\documentclass[5p,longtitle]{elsarticle}

\usepackage{epsfig}
\usepackage{graphicx}
\usepackage{dcolumn}
\usepackage{bm}
\usepackage{subfigure}
\usepackage{amsmath}
\usepackage{amssymb}
\usepackage{epstopdf}
\usepackage{hyperref}
\usepackage{lineno}
\usepackage[utf8]{inputenc}

\begin{document}
\begin{frontmatter}
\title{Technical Supplement to ``Polarization Transfer Observables in Elastic Electron-Proton Scattering at $Q^2= 2.5$, 5.2, 6.8 and 8.5 GeV$\mathbf{^2}$''\tnoteref{citeprc}}
\tnotetext[citeprc]{Technical supplement to original research published in~\cite{Puckett:2017flj}.}


\author[uconn]{A. J. R. Puckett\corref{cor1}} 
\ead{andrew.puckett@uconn.edu}
\author[cnu,jlab]{E. J. Brash}
\author[jlab]{M. K. Jones}
\author[lanzhou]{W. Luo}
\author[wm]{M. Meziane}
\author[jlab]{L. Pentchev}
\author[wm]{C. F. Perdrisat}
\author[nsu]{V. Punjabi}
\author[nsu]{F. R. Wesselmann}
\author[gwu]{A. Afanasev}
% ...

\cortext[cor1]{Corresponding author}
\fntext[dec]{Deceased.}

\address[uconn]{University of Connecticut, Storrs, CT 06269}
\address[cnu]{Christopher Newport University, Newport News, VA 23606}
\address[jlab]{Thomas Jefferson National Accelerator Facility, Newport News, VA 23606}
\address[lanzhou]{Lanzhou University, Lanzhou 730000, Gansu, Peoples Republic of China}
\address[wm]{College of William and Mary, Williamsburg, VA 23187}
\address[nsu]{Norfolk State University, Norfolk, VA 23504}
\address[gwu]{The George Washington University, Washington, DC 20052}
% ...

\date{\today}
\begin{abstract}
The GEp-III and GEp-2$\gamma$ experiments, carried out in Jefferson Lab's Hall C from 2007-2008, consisted of measurements of polarization transfer in elastic electron-proton scattering at momentum transfers of $Q^2 = 2.5, 5.2, 6.8,$ and $8.54$ GeV$^2$. These measurements were carried out to improve knowledge of the proton electromagnetic form factor ratio $R = \mu_p G_E^p/G_M^p$ at large values of $Q^2$ and to search for effects beyond the Born approximation in polarization transfer observables at $Q^2 = 2.5$ GeV$^2$. The final results of both experiments were reported in a recent archival publication. A full reanalysis of the data from both experiments was carried out in order to reduce the systematic and, for the GEp-2$\gamma$ experiment, statistical uncertainties. This technical note provides additional details of the final analysis omitted from the main publication, including the final evaluation of the systematic uncertainties.
\end{abstract}
\end{frontmatter}


\section{Introduction}
\label{sec:intro}
Experiments E04-108 and E04-019,  commonly known as ``GEp-III'' and ``GEp-2$\gamma$'', respectively, ran in Jefferson Lab's experimental Hall C from October 2007 to June 2008. The GEp-III experiment, the results of which were originally published in Ref.~\cite{Puckett:2010ac}, aimed at measuring the proton's electromagnetic form factor ratio $\mu_p \frac{G_E^p}{G_M^p}$ to the highest possible $Q^2$, given the maximum electron beam energy of 5.71 GeV available at the time. The objective of the GEp-2$\gamma$ experiment, originally published in Ref.~\cite{Meziane:2010xc}, was to perform precise ($\lesssim 1\%$ total uncertainty) measurements of the $\epsilon$ dependence of the ratio $P_t/P_\ell \propto G_E^p/G_M^p$, and the ratio $P_\ell/P_\ell^{Born}$ of the longitudinal polarization transfer component to its Born approximation value, at a fixed $Q^2$ of 2.5 GeV$^2$, with the goal of searching for experimental signatures of effects beyond the Born approximation at a $Q^2$ where the extractions of $G_E^p/G_M^p$ from Rosenbluth separations and polarization observables disagree. For a recent overview of nucleon electromagnetic form factors, see, e.g., Ref.~\cite{Punjabi:2015bba} and references therein. The experiments used a combination of baseline Hall C equipment and new detectors that were constructed for the express purpose of facilitating the measurements in question. A final analysis of the data from both experiments was carried out to reduce the systematic and statistical uncertainties of the data. The final results of both experiments were recently reported in an archival publication~\cite{Puckett:2017flj}. The kinematics and results of the measurements, the details of the apparatus, the theoretical formalism of elastic electron-proton scattering, and the major aspects of the data analysis are described in detail in the main body of the archival publication~\cite{Puckett:2017flj}. The purpose of this document is to provide additional details of the data analysis that go beyond the scope of the main publication, including significant improvements to the analysis since the original publication of both experiments. This technical document is intended to be read as a companion to the archival publication~\cite{Puckett:2017flj}, and as such, assumes a basic familiarity with the background material presented in Ref.~\cite{Puckett:2017flj} on the part of the reader. It is organized as follows:
\begin{itemize}
\item Section~\ref{sec:EventRecon} gives an overview of the reconstruction of events, including calibrations of the detectors and the spectrometer optics, emphasizing the reconstruction algorithms, calibration procedures, and performance of the detector systems that were newly constructed for these experiments. 
\item Section~\ref{sec:elastic} provides additional details of the elastic event selection procedure.
\item Section~\ref{sec:dataqualitycheck} details several data quality checks for the maximum-likelihood estimators that confirm the validity of the extraction method. 
\item Section~\ref{sec:systematics} details the evaluation of the final systematic uncertainties of the main physics results.
\item Section~\ref{sec:conclusion} presents a summary and conclusion of this work.
\end{itemize}

\section{Overview of Event Reconstruction}
\label{sec:EventRecon}
In both experiments, the polarized electron beam of Jefferson Lab's Continuous Electron Beam Accelerator Facility (CEBAF)~\cite{Leemann:2001dg,Chao:2011za} was scattered from a liquid hydrogen target in experimental Hall C. Elastically scattered electrons were detected in a large-acceptance electromagnetic calorimeter called ``BigCal'' in coincidence with elastically scattered protons detected by the Hall C High Momentum Spectrometer (HMS), equipped with a double focal plane polarimeter (FPP) to measure the polarization of the recoiling protons. The decoding of the raw data is described in Ref.~\cite{Puckett:2015soa}. A brief overview of the event reconstruction procedures is given in this document, including detector calibrations, reconstruction algorithms, and a summary of the detector performance, including the new detectors that were constructed for these measurements, particularly in areas where the final analysis differs from the original analysis. More detailed descriptions of the event reconstruction algorithms and calibration procedures for the original analysis can be found in the Ph.D. thesis~\cite{Puckett:2015soa}.

Event reconstruction for the proton arm includes determination of the ``start time'' of an event from the analysis of the fast signals from the HMS trigger scintillators, pattern recognition and track reconstruction in the drift chambers of the HMS, FPP1 and FPP2, reconstruction of the proton kinematics from the known transport matrix of the HMS, and computation of the proton spin transport matrix through the HMS magnetic elements from the reconstructed proton kinematics. For the electron arm, event reconstruction involves determination of the detected electron's energy, its impact coordinates at the surface of BigCal, and its timing relative to the event start time, defined for real coincidence events by the HMS trigger. Combined with the reconstructed position of the interaction vertex from the HMS, the measured electron coordinates at the surface of BigCal are used to reconstruct the electron scattering angles. This section presents some details of the calibration procedures and reconstruction algorithms for the main detector systems.

\subsection{TRANSPORT coordinate system} 
\label{subsubsec:transport}
The proton's trajectory as it exits the hydrogen target and as it is measured by the HMS drift chambers is described in a coordinate system that is fixed with respect to the HMS optical axis, hereafter referred to as the TRANSPORT coordinate system. 
\begin{figure*}
  \begin{center}
    \includegraphics[width=0.98\textwidth]{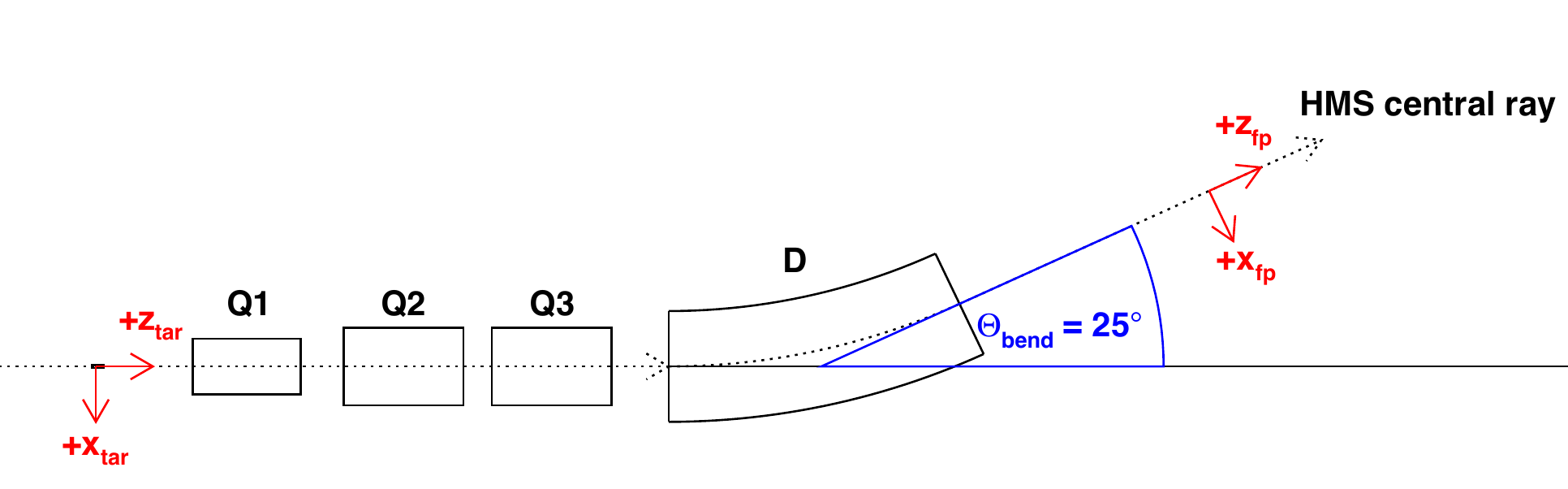}
  \end{center}
  \caption{\label{fig:HMS_centray} Sketch of the HMS magnet layout illustrating the definition of the HMS central ray (black dotted curve) and the relation between ``target'' and ``focal plane'' TRANSPORT coordinate systems. Magnet sizes, shapes, and positions are not to scale, and are merely drawn for illustrative purposes. The $+y$ axis of both the ``target'' and ``focal plane'' coordinate systems points into the page in this figure.}
\end{figure*}
In this coordinate system, the $+z$ axis is along the HMS optical axis in the direction of particle motion, the $+x$ axis lies in the dispersive plane in the direction of increasing particle momentum (vertically downward), and the $+y$ axis lies in the non-dispersive plane such that the $(x,y,z)$ axes form a right-handed, Cartesian coordinate system, as shown in Fig.~\ref{fig:HMS_centray}. Since the HMS is on the right side of the beam, the $+y$ axis of the TRANSPORT system at the target points in the direction of decreasing scattering angles in the horizontal plane, as shown in Fig.~\ref{fig:yztransport}. Figure~\ref{fig:xztransport} illustrates the definition of $x_{tar}$.
\begin{figure}
  \begin{center}
    \includegraphics[width=0.98\columnwidth]{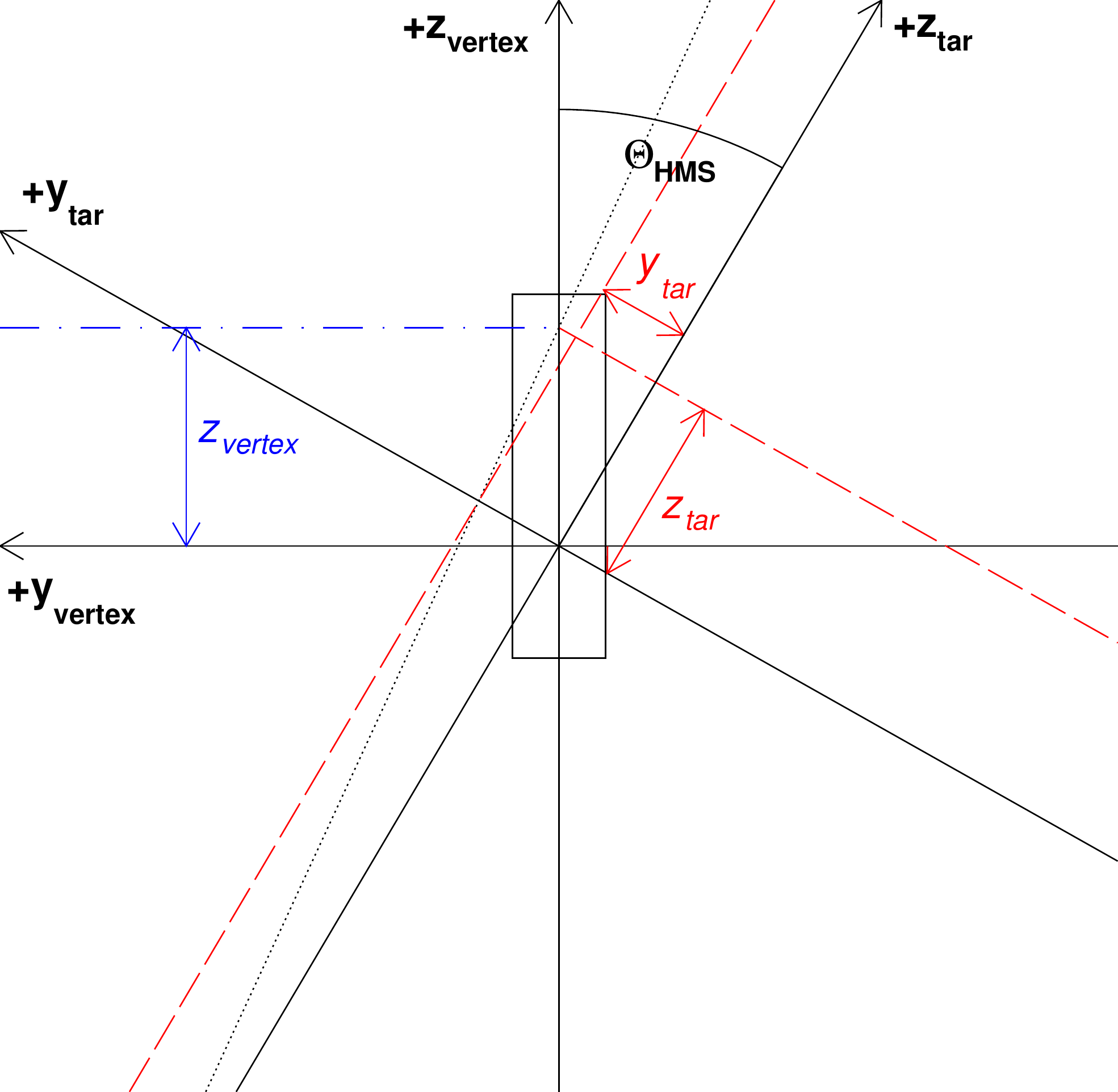}
  \end{center}
  \caption{\label{fig:yztransport} Horizontal ($yz$) plane projection of the TRANSPORT coordinate system at the target, as viewed from above. $\Theta_{HMS}$ is the HMS central angle. The $+x_{tar}$ axis points into the page in this figure. The box (not to scale) indicates the cylindrical liquid hydrogen target cell with downstream offset of the target center with respect to the origin. The $+z_{vertex}$ axis indicates the (nominal) beam direction. The black dotted line is the $yz$-plane projection of a trajectory originating at $z_{vertex}$ with a non-zero $y'_{tar}$. Red dashed lines illustrate the definitions of $y_{tar} = z_{vertex} \sin \Theta_{HMS} - y'_{tar} z_{tar}$ and $z_{tar} = z_{vertex} \cos \Theta_{HMS}$. The blue dot-dashed line illustrates the definition of $z_{vertex}$. The beam in this example is assumed to be horizontally centered with respect to the origin ($y_{vertex} = 0$).}
\end{figure}
\begin{figure}
  \begin{center}
    \includegraphics[width=0.98\columnwidth]{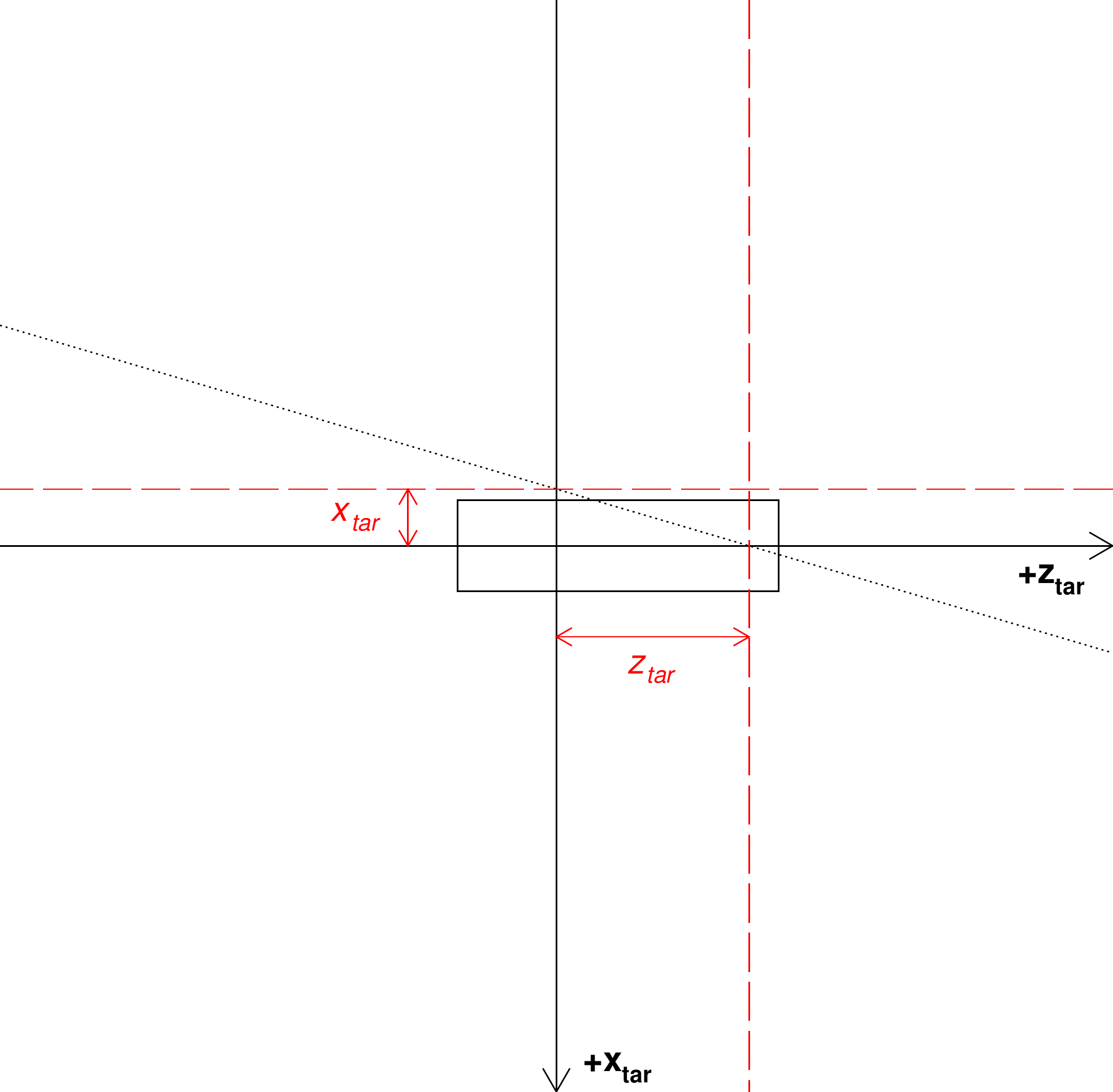}
  \end{center}
  \caption{\label{fig:xztransport} Vertical ($xz$) plane projection of the TRANSPORT coordinate system at the target, viewed from the side, perpendicular to the HMS optical axis. The $+y_{tar}$ axis points into the page in this figure. The box indicates the projection of the target length along the $z_{tar}$ axis. The black dotted line is the $xz$-plane projection of a trajectory originating from a scattering event occurring at $z_{tar}$ with a non-zero $x'_{tar}$. Red dashed lines indicate the definitions of $z_{tar}$ and $x_{tar} = -x'_{tar} z_{tar}$ (assuming a vertically centered beam; i.e., $x_{vertex} = 0$). }
\end{figure}
In the ``target'' coordinate system, the $z = 0$ plane is perpendicular to
the HMS optical axis and contains the origin of Hall C (the center of
the spectrometer pivot). The proton's trajectory at the target ($z_{tar} =
0$) is described by five parameters $(x_{tar}, y_{tar}, x'_{tar},
y'_{tar}, \delta)$, where $x_{tar}$ and $y_{tar}$ are the coordinates,
$x'_{tar} = \frac{dx}{dz}$ and $y'_{tar} = \frac{dy}{dz}$ are the
track slopes and $\delta \equiv 100 \times \frac{p-p_0}{p_0}$ is the
percentage deviation of the particle momentum from the HMS central
momentum setting. The origin of the HMS detector or ``focal plane''
coordinate system lies approximately 25 meters downstream of the
origin of Hall C along the HMS central ray, as depicted in
Fig.~\ref{fig:HMS_centray}. 

The $z = 0$ plane of the detector coordinate system lies between the two HMS drift chambers and approximately coincides with the focal point of the HMS when operated in its standard tune. The focal-plane coordinate system is rotated vertically upward by the 25-degree central bend angle of the HMS relative to the target coordinate system, as shown in Fig.~\ref{fig:HMS_centray}. The proton's trajectory as measured by the HMS drift chambers is described by the four parameters $(x_{fp}, y_{fp}, x'_{fp}, y'_{fp})$, where $x_{fp}$ and $y_{fp}$ are the track coordinates at $z = 0$, and $x'_{fp} \equiv dx/dz$ and $y'_{fp} \equiv dy/dz$ are the track slopes. Hereafter, the term ``TRANSPORT'' will be used generically to refer to both the ``target'' and ``focal-plane'' coordinate systems of the HMS, and the subscripts ``tar'' and ``fp'' will be used to distinguish between the two. We will also occasionally refer to trajectory angles $\theta = \arctan (x') \approx x'$ and $\phi = \arctan (y') \approx y'$ instead of the slopes $x'$ and $y'$ for either the ``focal-plane'' or ``target'' trajectories. Because $\left|x'\right|$ and $\left|y'\right|$ are small within the HMS acceptance, the small-angle approximation is valid and the trajectory slopes and angles can be used more or less interchangeably. 

\subsection{HMS scintillator reconstruction}
\label{sec:HMShodo}
The time at which the proton track crossed the HMS focal plane is reconstructed from the fast timing signals provided by the trigger scintillator planes. Only the ``S1X'' and ``S1Y'' planes (see Ref.~\cite{Puckett:2017flj} for definitions) were used in the timing reconstruction, as the timing resolution of the ``S0'' plane installed upstream of the HMS drift chambers was too poor to meaningfully improve the resolution. The reconstruction proceeds in two iterations. In the first iteration, which occurs prior to tracking, the scintillator signals are analyzed assuming that the detected particle is a proton moving along the central trajectory of the HMS at the central momentum. The results from the first iteration define a reference time for the measurement of drift times in the HMS and FPP drift chambers. In the second iteration, the results are refined using the reconstructed track information, again assuming that the detected particle is a proton.

\begin{figure}
  \begin{center}
    \includegraphics[width=0.9\columnwidth]{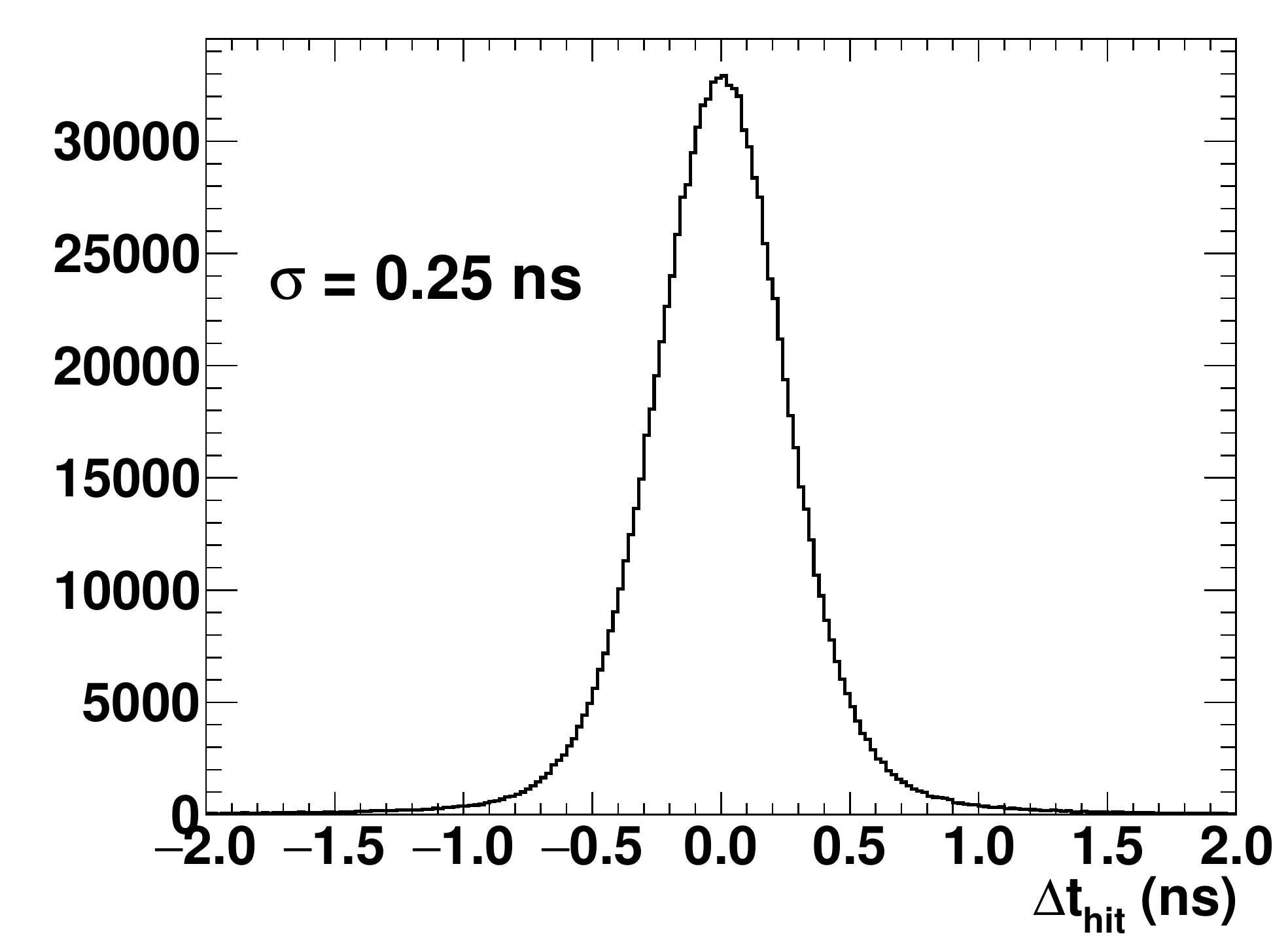}
  \end{center}
  \caption{\label{fig:timeres} Difference between individual corrected PMT hit times and the average corrected time of all other PMT hits on scintillators pointed to by the same track, for elastically scattered protons at a central momentum of 3.59 GeV ($Q^2 = 5.2$ GeV$^2$ setting).}
\end{figure}
For each scintillator paddle pointed to by the best proton track reconstructed from the HMS drift chamber signals, a final corrected time at the HMS focal plane was determined by correcting the raw TDC signals from the PMTs for the effective average light propagation delay in the paddles\footnote{For paddles with PMTs at both ends firing, }, the time walk due to the pulse height dependence of the time at which the signal crossed the fixed discriminator threshold, a constant offset to account for channel-to-channel variations in cable and electronic delays, and the particle time-of-flight from the HMS focal plane to the scintillator paddle, again assuming the particle is a proton, a good assumption in the context of this analysis. The parameters describing each correction for each PMT/paddle were determined in a calibration procedure described in Ref.~\cite{Puckett:2015soa}.
As shown in Fig.~\ref{fig:timeres}, the typical per-PMT timing resolution achieved after all corrections was approximately 250 ps, implying $\sim$125 ps resolution for the average of all four PMTs attached to the two paddles pointed to by the track\footnote{Approximately $1/8$ of tracks pass through three scintillator paddles, because the S1X (S1Y) paddles are staggered and interleaved such that they overlap by approximately 1/8$^{\text{th}}$ of their width along the $x$ ($y$) direction. An even smaller fraction pass through four paddles, firing eight PMTs.}. The effectively realized timing resolution varied slightly with experiment conditions, but never exceeded 350 ps per PMT in any configuration. The average of all focal plane times measured by PMTs on paddles pointed to by the reconstructed track was then corrected for the variation in time of flight of the proton from the target to the focal plane as a function of the reconstructed kinematics using a standard parametrization based on the HMS COSY model~\cite{Blok:2008jy} for comparison to the timing of the scattered electron shower in BigCal. More details of the scintillator reconstruction can be found in Ref.~\cite{Puckett:2015soa}.
\subsection{HMS drift chamber tracking}

The design of the HMS drift chamber pair is discussed in detail in Ref.~\cite{Baker:1995ky}. The HMS tracking system consists of two identical planar drift chambers, spaced approximately 80 cm apart along the HMS optical axis. Each chamber consists of six wire planes with four different wire orientations, with a stacking order along $z$ of $XYUVYX$. The $X (Y)$ planes, of which there are two in each chamber, measure the $x (y)$ coordinate, while the $U$ and $V$ planes measure the coordinates at $\pm 15^\circ$ angles relative to the $x$ axis. The HMS drift chamber signals were read out by LeCroy model 1877 Fastbus multihit TDCs operated in common stop mode~\cite{Puckett:2015soa}. Potentially useful hits for tracking were selected by rejecting hits with raw TDC values outside a broad window encompassing the allowed range of arrival times for hits caused by the primary track responsible for the HMS trigger. To the extent that there are multiple hits on the same wire in the same event within the allowed window (a relatively rare occurence under the conditions of these experiments), the time of the earliest hit is retained for further analysis. During pattern recognition, the drift time and drift distance are computed independently for each wire in each potentially valid track combination in which it appears. The track-independent contribution to the measured drift time for each wire is obtained from the raw TDC value by subtracting the ``start time'' determined on the first iteration of the HMS hodoscope reconstruction described in Section~\ref{sec:HMShodo} and a per-wire zero offset that aligns the drift time spectra of all individual wires in a common window for the time-to-distance conversion. The small correction to the drift time for particle time of flight between different planes is effectively absorbed into the zero offset for each wire, which is calibrated separately for each kinematic setting. Using the approximate position of each candidate track at each plane based on the fit to wire positions alone with no timing information, drift times are then corrected for the signal propagation delay from the position along the wire where the track crossed the plane to the front-end electronics. The drift distance is determined from the corrected drift time by mapping the observed corrected drift time spectrum of hits included in final tracks onto a uniform drift distance distribution within a drift cell. The time-to-distance calibration was performed separately for each wire plane for each data acquisition run. 

A detailed description of the HMS drift chamber pattern recognition and tracking algorithm specific to the GEp-III and GEp-2$\gamma$ experiments can be found in~\cite{Puckett:2015soa}. Several modifications relative to the ``standard'' HMS tracking algorithm were implemented for this analysis (and the analysis leading to the originally published results). These included fixing a bug in the existing tracking code that increased the probability of an incorrect solution of the left-right ambiguity at high rates, and adding an improved method for solving the left-right ambiguity by considering one-dimensional projections of the track along the $xz$ and $yz$ planes separately. 

Following pattern recognition, all potentially valid wire combinations in each drift chamber are fitted individually with straight lines referred to as ``track stubs'', and all combinations of one ``stub'' from each chamber whose fitted track parameters agree to within tolerances chosen to optimize the tracking efficiency and accuracy within the interesting range of track parameters are considered as candidates for full track fitting. A ``full track'' candidate consists of a combination of 10-12 hits in unique wire planes, with 5-6 hits from each drift chamber. In the ``standard'' HMS tracking algorithm, the wire positions, drift distances and left-right combinations of the hits are taken from the pattern recognition/stub fit results, and a straight line fit to all the hit positions is performed, assuming the left-right combinations from the ``stub'' fits are correct. In the GEp-III/GEp-2$\gamma$ analyses, the determination of the best left-right combination of the hits was further refined for full track candidates with hits in at least 3 of 4 planes in both the $x$ and $y$ directions. For full track candidates satisfying this condition, the projection of the track along the $xz$($yz$) plane was fitted to the $x$($y$) hits considered in isolation and used to fix the left-right combinations of these hits. Then, the $x$ and $y$ hits were combined and re-fitted, and the resulting track used to fix the left-right combination of the $u$ and $v$ planes (if applicable). Finally, the full track was re-fitted using all available hits, with the left-right combination of all hits fixed by this procedure. If the new left-right solution improved the $\chi^2/ndf$ of the track compared to the initial solution from the ``stub'' fits, it was kept. Otherwise, the original solution was retained. This procedure significantly improved the tracking resolution compared to the ``standard'' HMS tracking algorithm, especially under high-rate conditions. 


The final per-plane coordinate resolution, as measured by the tracking residuals, was approximately 280 $\mu$m for 2 GeV protons~\cite{Puckett:2015soa}. This corresponds to a per-drift chamber spatial resolution of approximately 140 $\mu$m (200 $\mu$m) in $x_{fp} (y_{fp})$ and a resulting resolution of 0.24 (0.35) mrad in the track slopes $x'_{fp}$ and $y'_{fp}$. The resolution of the reconstructed proton trajectory angles at the target depends additionally on the optical magnification of the resolution of the drift chambers and the additional smearing of the proton trajectory by multiple scattering in the 1-cm-thick ``S0'' trigger scintillator installed upstream of the HMS drift chambers. Compared to the standard HMS configuration with no extra materials between the exit window of the HMS vacuum and the drift chambers, multiple scattering in S0 made the HMS angular resolution roughly a factor of 3 worse at the lowest momentum setting corresponding to $Q^2 = 2.5$ GeV$^2$ (proton momentum $p_p = 2.07$ GeV), and about a factor of 1.4 worse at the highest momentum setting corresponding to $Q^2 = 8.5$ GeV$^2$ ($p_p = 5.41$ GeV). On the other hand, the effect of S0 on the HMS momentum resolution was negligible, since the momentum reconstruction is mainly sensitive to the position of the proton at the HMS focal plane, rather than the slope of its trajectory.

It is worth noting that the exclusivity cuts applied to select elastic events (see Ref.~\cite{Puckett:2017flj} and section~\ref{sec:elastic} for detailed discussions) reject events in which the proton scatters by large angles in S0 prior to being tracked, suppressing any significant false asymmetry arising from spin-dependent scattering in S0, since any such asymmetry must vanish in the limit $\vartheta \rightarrow 0$ by definition. All applied exclusivity cuts are symmetric about the elastic peak and sufficiently loose ($\pm 3\sigma$) to prevent the introduction of any significant left-right or up-down bias of the selection of elastic events by scattering direction in S0, such that any residual false asymmetry arising from spin-dependent scattering in S0 is strongly suppressed. Since the asymmetry of interest is that of the secondary scattering in the CH$_2$ analyzers of the FPP, the only observable effect of any spin-dependent scattering in S0 would be an asymmetry in the number of protons incident on the FPP for positive and negative beam helicities, which does not noticeably affect the extraction of the polarization transfer observables in any case. In practice, no statistically significant asymmetry in the number of incident protons between positive and negative beam helicities was observed for any of the kinematics after all exclusivity cuts were applied, confirming that any effects of spin-dependent scattering in S0 were negligible.
\subsection{HMS optics calibration}
\label{subsubsec:HMSoptics}
The precise measurement of the proton's coordinates and trajectory at the HMS focal plane is combined with the knowledge of the transport matrix of the HMS to reconstruct the proton kinematics at the target. In principle, the problem of reconstructing the target coordinates from the focal-plane coordinates requires solving a system of four equations in five unknowns, and is therefore underdetermined (see section~\ref{subsubsec:transport} for parameter definitions). In practice, a one-to-one mapping between target coordinates and focal-plane coordinates exists when one of the target coordinates is fixed. For a thin target located at the central pivot of Hall C, the vertical spectrometer coordinate $x_{tar}$ is fixed by the vertical beam position on target. For extended targets such as the 20-cm liquid hydrogen cell used in this experiment, $x_{tar}$ varies significantly with the position of the interaction vertex and the proton trajectory slope in the dispersive plane (see Fig.~\ref{fig:xztransport}):
\begin{eqnarray}
  x_{tar} &=& -y_{beam} - x'_{tar} z_{vertex} \cos(\Theta), \label{eq:xtar}
\end{eqnarray}
where $y_{beam}$ is the vertical beam position on target (in the ``Hall C'' coordinate system with $+y$ vertically upward), $z_{vertex}$ is the position of the interaction vertex along the beam direction, and $\Theta$ is the central scattering angle of the HMS. The vertical angular acceptance of the HMS is approximately $\pm 70$ mrad when used with the larger of its two acceptance-defining octagonal collimators as in this experiment. The center of the 20-cm hydrogen target cell used for most kinematic settings was offset by 3.84 cm downstream from the origin to accommodate electron scattering angles up to 120 degrees using the standard Hall C scattering chamber exit window. In the most extreme case, at $Q^2 = 8.5$ GeV$^2$ with $\Theta = 11.6$ deg., $x_{tar}$ can differ from $-y_{beam}$ by up to 1 cm for extreme rays.
This uncertainty in $x_{tar}$ significantly affects the reconstruction of both $x'_{tar}$ and $\delta$ for an extended target, as the first-order sensitivities are $(dx'_{tar}/dx_{tar}, d\delta/dx_{tar}) \approx (1$ mrad/mm$, 0.08\%$/mm$)$. The optical design of the HMS largely decouples the measurement of $z_{vertex}$ from the measurement of $x'_{tar}$, such that the accuracy of the reconstruction can be significantly improved with a small number of subsequent iterations in which the knowledge of $x_{tar}$ is refined using the reconstructed values of $x'_{tar}$ and $z_{vertex}$ from the previous iteration.

The transport matrix of the HMS consists of an independent polynomial expansion of each target coordinate to be reconstructed in terms of the four measured focal-plane coordinates and the ``known'' value of $x_{tar}$:
  \begin{eqnarray}
    \left[\begin{array}{c} x'_{tar} \\ y'_{tar} \\ y_{tar} \\ \delta \end{array}\right] &=& \sum_{i,j,k,\ell,m=0}^{i+j+k+\ell+m \le n} \left[\begin{array}{c} C_{x'}^{ijk\ell m} \\ C_{y'}^{ijk\ell m} \\ C_{y}^{ijk\ell m} \\ C_{\delta}^{ijk\ell m} \end{array}\right] T_{ijk\ell m}, \label{eq:recon_coeff} \\
    T_{ijk\ell m} &=& (x_{fp})^i (y_{fp})^j (x'_{fp})^k (y'_{fp})^\ell (x_{tar})^m \nonumber
  \end{eqnarray}
  The order $n$ of the expansion is arbitrary, but is typically chosen to be either 5 or 6 depending on experimental requirements for accuracy and acceptance. For this experiment, a sixth-order expansion was used for the reconstruction of $x'_{tar}$, $y'_{tar}$ and $y_{tar}$, while an existing fifth-order expansion was used for the reconstruction of $\delta$. The main advantage of the polynomial expansion~\eqref{eq:recon_coeff} is that $\chi^2$ is a linear function of the parameters, such that the least-squares solution for the expansion coefficients can be found using computationally inexpensive linear-algebra techniques such as singular-value decomposition.

 While many calibrations and optimizations of the HMS optics have been performed in the past (see~\cite{Blok:2008jy} for a representative example), no previous experiment had used the HMS with such a long target, particularly with the large downstream offset of the 20-cm liquid hydrogen cell used in this experiment. For this reason, pre-existing versions of the HMS transport matrix, optimized for experiments with much thinner targets, did a relatively poor job of describing the HMS optics in the full phase space coverage of this experiment. The difficulty is exacerbated by the tendency of polynomial fits to diverge uncontrollably when extrapolated outside regions where they are directly constrained by data. For this reason, a new set of optics calibration data was collected by measuring inelastic electron scattering at a beam energy of 4.109 GeV on several multi-foil targets with the HMS ``sieve slit'' collimator installed. Dedicated optics runs included:
\begin{itemize}
\item Three-foil aluminum target with \emph{nominal} $z$ positions\footnote{The ``nominal'' target foil positions are the design values. The actual target foil positions can deviate slightly from the nominal positions due to, e.g., small misalignments of the target ladder with respect to the spectrometer pivot, motion of the target ladder associated with the cooldown procedure, etc.} of $z = \{-7.5,0,7.5\}$ cm, with central HMS angle $\Theta = 22^\circ$ and central momentum $p_0 = 2.4$ GeV. 
\item Two-foil aluminum target with \emph{nominal} $z$ positions of $z = \{-3.8,3.8\}$ cm, with $\Theta = 22^\circ$, $p_0 = 2.4$ GeV.
\item Two-foil carbon target with \emph{nominal} $z$ positions of $z = \{-2,2\}$ cm, with $\Theta = 22^\circ$ and $p_0 = 2.4$ GeV. 
\item 20-cm aluminum ``dummy'' target with \emph{nominal} $z$ positions of $z = \{-6.16, 13.84\}$ cm (also used to measure the target endcap contribution to the hydrogen elastic production data). Data were collected at angles $\Theta = 22^\circ$ and $26^\circ$ at $p_0 = 2.4$ and 2.15 GeV, respectively.

\item 15-cm aluminum ``dummy'' target with \emph{nominal} $z$ positions of $z = \{-7.5,7.5\}$ cm. Data collected at angles of $\Theta = 22$, 26, and 30$^\circ$ with $p_0 = 2.4$, 2.15 and 1.9 GeV, respectively.
\end{itemize} 
\begin{figure*}
  \begin{center}
    \includegraphics[width=0.98\textwidth]{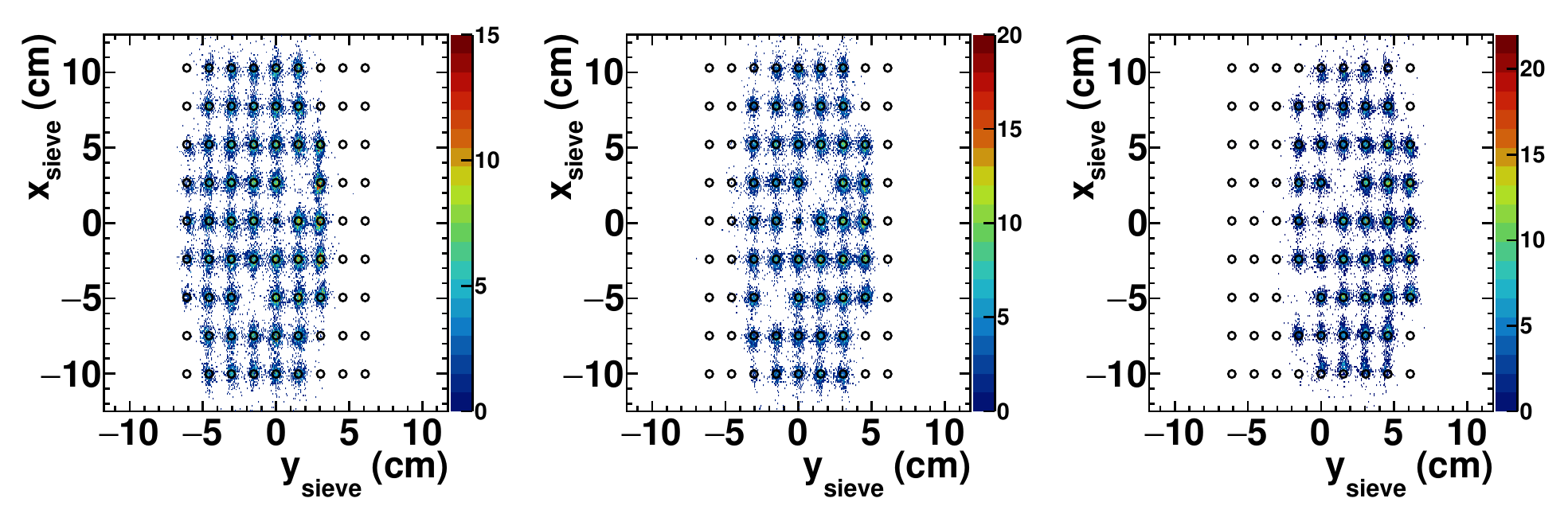}
  \end{center}
  \caption{\label{fig:sieveslit} Sieve slit reconstruction for the three-foil aluminum target, after optimization, for the foils at $z_{foil} = \{-7.5,0,7.5\}$ cm, from left to right. $x_{sieve}$ and $y_{sieve}$ are the projected coordinates of the reconstructed trajectory at the surface of the sieve slit collimator. The circles mark the sieve hole positions and diameters. The color scale represents the number of events. The shifting pattern of sieve holes populated by scattering from different target foils reflects the HMS angular acceptance.} 
\end{figure*}
The sieve slit collimator is a 3.175 cm-thick slab of \emph{densimet} ($\rho \simeq 17$ g/cm$^3$) with a regular rectangular grid of circular ``sieve'' holes, as described in Ref.~\cite{Puckett:2015soa}. 
Two of the holes are blocked in order to verify the correct up-down and left-right orientation of the reconstructed angles. When used with multiple thin target foils at known locations along the beamline, the rays from beam-foil intersection points to the known sieve hole positions determine the target coordinates $(x_{tar}, y_{tar}, x'_{tar}, y'_{tar})$ with a high degree of precision and accuracy. 

Figure \ref{fig:sieveslit} illustrates the quality of the reconstruction of the HMS sieve hole pattern for the three-foil aluminum target. For a \emph{point} target, the in-plane angular acceptance of the HMS is approximately $\pm 28$ mrad. For an extended target, a wider range of in-plane angles is accepted, because the center of the in-plane angle acceptance of the HMS is shifted to smaller (larger) angles for particle trajectories originating from points upstream (downstream) of the origin. Because of the significant corrections to $x'_{tar}$ and $\delta$ arising from the variation of $x_{tar}$ as a function of $x'_{tar}$ and $z_{vertex}$, all $x_{tar}$-independent matrix elements were optimized up to sixth order while fixing all $x_{tar}$-dependent matrix elements at values calculated from a detailed sixth-order COSY~\cite{Makino:2006sx} model of the HMS. It is worth remarking that the S0 scintillator was removed from the HMS during the optics calibration, because the effect of multiple-scattering in S0 on the angular resolution of the HMS made it impossible to isolate tracks passing through individual sieve holes with S0 in place.

\begin{figure}
  \begin{center}
    \includegraphics[width=0.85\columnwidth]{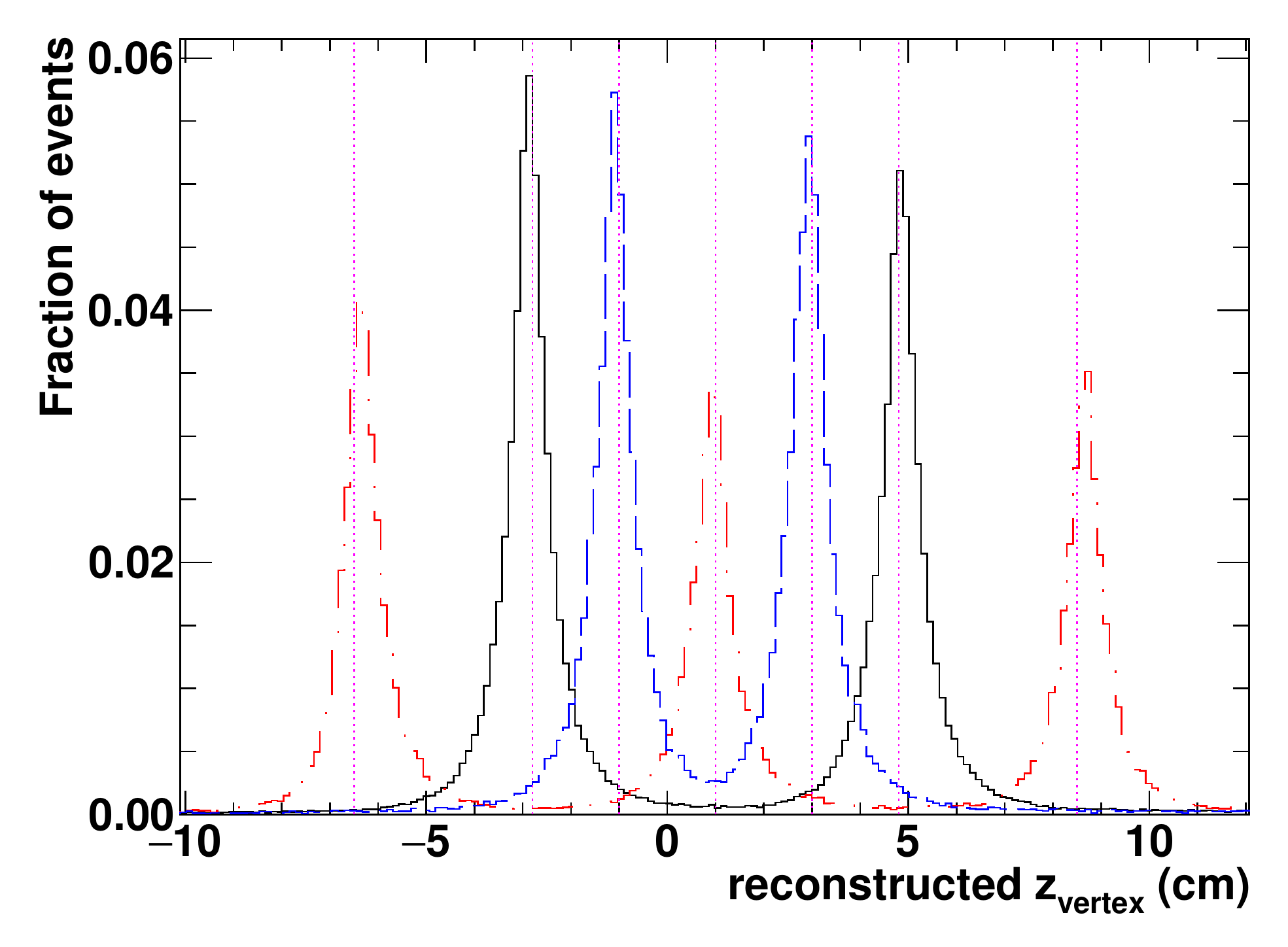}
  \end{center}
  \caption{\label{fig:zvertex}  Vertex reconstruction for the three-foil aluminum target (red dot-dashed), the two-foil aluminum target (black solid), and the two-foil carbon target (blue dashed). ``True'' foil positions are indicated by the pink dotted vertical lines.}
\end{figure}
Figure~\ref{fig:zvertex} shows the reconstructed vertex coordinate along the beamline for several of the optics targets, after optimization. The vertex $z$ coordinate is defined in this context as the intersection of the horizontal projection of the scattered particle's trajectory with the \emph{ideal} beamline, and is related to the TRANSPORT coordinates $y_{tar}$ and $y'_{tar}$ by:
\begin{eqnarray}
  y_{tar} &=& z_{vertex} \left[ \sin(\Theta) - y'_{tar} \cos(\Theta) \right]. \label{eq:ytar_ideal}
\end{eqnarray}
While the positions of the various target foils relative to each other are known with a high degree of certainty, the position of the beam-foil intersection point relative to the HMS optical axis is subject to considerable uncertainty. The reconstructed $z_{vertex}$ values are displaced by approximately 1 cm downstream of the nominal foil locations, independently of which target foil is analyzed. Most of the apparent offset in the location of the interaction vertex is attributable to an offset in the (average) horizontal beam position of approximately 3 mm to the left of the ideal beamline position above the central pivot of Hall C. 
For scattering from a thin foil at position $z_{foil}$, the physical interaction vertex is located at the point $(x_{beam}, y_{beam}, z_{foil})$. In a beamline coordinate system with $+x$ to beam left, $+y$ vertically upward, and $+z$ along the nominal beam direction, the global vertex coordinates are related to the TRANSPORT coordinates $y_{tar}$ and $z_{tar}$ by:
\begin{eqnarray}
  z_{tar} &=& z_{foil} \cos(\Theta) - x_{beam} \sin(\Theta) \nonumber \\
  y_{tar} &=& z_{foil} \sin(\Theta) + x_{beam} \cos(\Theta) - y'_{tar} z_{tar} \nonumber \\
          &=& z_{foil} \left[\sin(\Theta) - y'_{tar} \cos(\Theta) \right] + \nonumber \\
          & &  x_{beam}\left[\cos(\Theta) + y'_{tar} \sin(\Theta) \right]. \label{eq:ytar_real}
\end{eqnarray}
For a central trajectory ($y'_{tar} = 0$), it follows from Eqs.~\eqref{eq:ytar_ideal}-\eqref{eq:ytar_real} that $z_{vertex} = z_{foil} + x_{beam} \cot(\Theta)$. For example, at $\Theta = 22^\circ$, the angle at which most of the optics data (and all the data shown in Fig.~\ref{fig:zvertex}) were collected, $x_{beam} = +3$ mm corresponds to an offset $z_{vertex} - z_{foil} \approx 7.4$ mm. The rest of the observed offset can be accounted for by a possible error in the assumed horizontal beam position (the uncertainty $\Delta x_{beam} \approx 1$ mm, corresponds to $\Delta (z_{vertex}-z_{foil}) \approx 2.5$ mm at $\Theta = 22^\circ$), and a possible global offset of the actual target foil positions relative to their nominal positions, not expected to exceed 1 mm. Since the Hall C beam position monitors (BPMs) were not calibrated relative to the Hall C superharp~\cite{Yan:1995nc} system for the beam conditions specific to the optics data taking, a more accurate determination of the beam position on target was not possible. In light of the fact that the horizontal beam position during the optics calibration was not known with sufficient accuracy to improve on the pre-existing determinations of the zero offsets in $y_{tar}$ and $y'_{tar}$, the effective $z$ positions of all the target foils were shifted by $+1$ cm downstream of the nominal foil positions during the optimization, in order to match the 1-cm offset observed in the data for the foil nominally located at $z = 0$. This procedure amounts to assuming that the initial zero offsets in $y_{tar} $ and $y'_{tar}$ are correct, and absorbing all systematic effects contributing to the effective $z$ position of the foils into a single global $z$ offset applied to all foils in the optimization. In other words, the goal of the optimization was not to improve the knowledge of the optics of the central ray (see, however, Section~\ref{sec:systematics}), which would have required more accurate knowledge of the beam position and the absolute target foil positions relative to the HMS optical axis, but to improve the behavior of the expansion of Eq.~\eqref{eq:recon_coeff} for extreme rays by obtaining a set of calibration data populating as much as possible of the wider phase space acceptance at the HMS focal plane for the extended, 20-cm target. 

\begin{figure*}
  \begin{center}
    \includegraphics[width=0.8\textwidth]{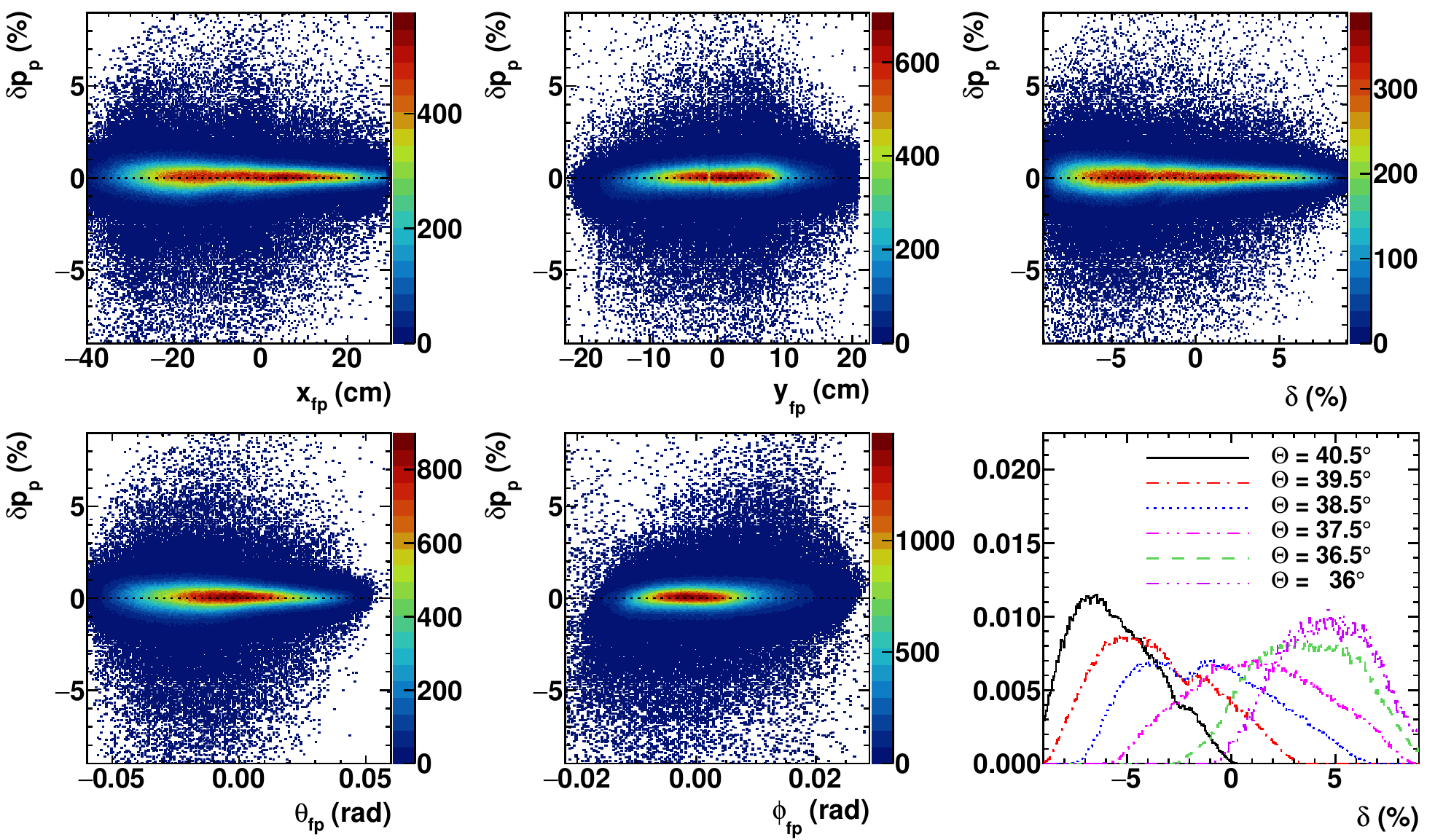}
  \end{center}
  \caption{\label{fig:deltarecon}  Difference $\delta p_p \equiv 100\times \frac{p_p -p_p(\theta_p)}{p_0}$ between the measured proton momentum $p_p$ and the momentum $p_p(\theta_p)$ required by elastic kinematics at the measured proton scattering angle, expressed as a percentage of the HMS central momentum $p_0$, plotted as a function of the focal plane track parameters $x_{fp}$ (top left), $y_{fp}$ (top middle), $\theta_{fp} \equiv \arctan(x'_{fp})$ (bottom left) and $\phi_{fp} \equiv \arctan(y'_{fp})$ (bottom middle), and as a function of the percentage deviation $\delta$ (top right) of the proton momentum from the HMS central momentum. Elastic $ep$ scattering was measured at a fixed $p_0 = 2.02$ GeV for six HMS central angles. The bottom right panel shows the $\delta$ distribution at each HMS central angle. }
\end{figure*}
The quality of the HMS momentum reconstruction was also checked by measuring elastic $ep$ scattering, with elastically scattered protons (electrons) detected in the HMS (BigCal). At a fixed HMS central momentum of 2.02 GeV, elastic scattering was measured at six different HMS central angles from 40.5$^\circ$ to 36$^\circ$. The beam energy was fixed at 4.109 GeV. As the central angle of the HMS is varied at a fixed beam energy and central momentum, different regions of the HMS acceptance are populated by elastically scattered protons. Inelastically scattered protons were rejected by placing cuts on the angular correlations between the measured proton track and the electron detected by BigCal as described in Ref.~\cite{Puckett:2017flj} and Section~\ref{sec:elastic}. The beam energy was corrected for the average energy loss along the target length prior to scattering, and the measured proton momentum was corrected for the average energy loss in materials along its path from the interaction vertex to the entry window of the HMS vacuum. Figure~\ref{fig:deltarecon} shows the quality of the momentum reconstruction achieved using a pre-existing fifth-order expansion of the $\delta$ matrix elements, including $x_{tar}$-dependent terms in the expansion~\eqref{eq:recon_coeff}. In elastic $ep$ scattering, the scattered proton's momentum is related to its scattering angle by:
\begin{eqnarray}
  p_p(\theta_p) &=& \frac{2M_pE_e(M_p+E_e)\cos(\theta_p)}{M_p^2+2M_pE_e + E_e^2 \sin^2 \theta_p }\label{eq:pp_ptheta},
\end{eqnarray}
where $M_p$ is the mass of the proton and $E_e$ is the beam energy. The difference $\delta p_p \equiv \frac{p_p-p_p(\theta_p)}{p_0}$ between the reconstructed proton momentum and the expected momentum of an elastically scattered proton exhibited no significant correlations with any of the focal plane track parameters or with $\delta$, indicating that no further optimization of the transport matrix elements for $\delta$ was needed. 

The phase space coverage of the optics calibration data in terms of $(x'_{tar}, y'_{tar}, \delta)$ equaled or exceeded that of the elastic $ep$ production data for all of the GEp-III and GEp-2$\gamma$ kinematics. The $y_{tar}$ coverage of the optics calibration data exceeded that of the elastic $ep$ production data for all but the two highest-$\epsilon$ kinematics of GEp-2$\gamma$, for which the $y_{tar}$ acceptance slightly exceeded that of the optics calibration data due to the larger HMS central angles involved (recall $y_{tar} \approx z_{vertex} \sin \Theta$), requiring a modest extrapolation outside the phase space coverage of the calibration for these kinematics. The small fraction of the data lying outside the $y_{tar}$ coverage of the fit were nonetheless included in the final analysis, because the overdetermined two-body kinematics of the elastic $ep$ reaction (see section~\ref{sec:elastic}) and the data quality checks described in section~\ref{sec:dataqualitycheck} showed that the proton kinematic reconstruction and the spin transport calculation were both sufficiently well-behaved throughout the HMS acceptance. This is also indirectly demonstrated by Fig.~\ref{fig:deltarecon}, the data for which were obtained at larger HMS central angles than any of the production kinematics. 

\subsection{FPP drift chamber tracking}
The FPP drift chamber tracking algorithm is similar to the tracking algorithm used for the HMS drift chambers, but differs in several important respects due to differences in the design and function of each detector. The FPP consists of two CH$_2$ analyzer blocks, each followed by a pair of two drift chambers. Each chamber contains three planes of parallel wires oriented at $+45^\circ$ (V), $90^\circ$ (X) and $-45^\circ$ (U) with respect to the $x$ direction of TRANSPORT coordinates, in order of increasing $z$. Within each FPP, the total number of wire planes, and therefore the largest possible number of coordinate measurements to define a track, is six. The roughly 21-cm separation in $z$ between the two chambers within each FPP is large compared to the 1.8-cm $z$ spacing of planes within a chamber, so that each chamber can be thought of as measuring essentially one point along the track in three-dimensional space, to a good approximation.

Owing to the lack of redundancy of coordinate measurements and the relatively high multiplicity of tracks in the FPP chambers, and the fact that the interesting range of track angles and positions was much wider than for the HMS drift chamber tracks since the angular distribution of the secondary scattering was the observable of interest, the strategy for pattern recognition and track fitting in the FPP required a more exhaustive consideration of possible wire combinations. As in the HMS tracking, the FPP hits were filtered through a loose cut on their raw TDC values to suppress noise and accidental background, and rough drift times were computed from the TDC values and the ``start time'' determined from the hodoscope analysis. Individual $t_0$ offsets were determined for each wire to align the drift time spectra in a window\footnote{Because the FPP chamber wire spacing was twice that of the HMS chambers and both sets of chambers used the same gas mixture and operated in a similar high-voltage regime, the drift time window for useful hits for FPP tracking was roughly twice as wide as that for HMS tracking.} from approximately zero to 200 ns. The drift time calculation was refined at a later stage using the track information. 

\begin{figure}
  \begin{center}
    \includegraphics[width=0.75\columnwidth]{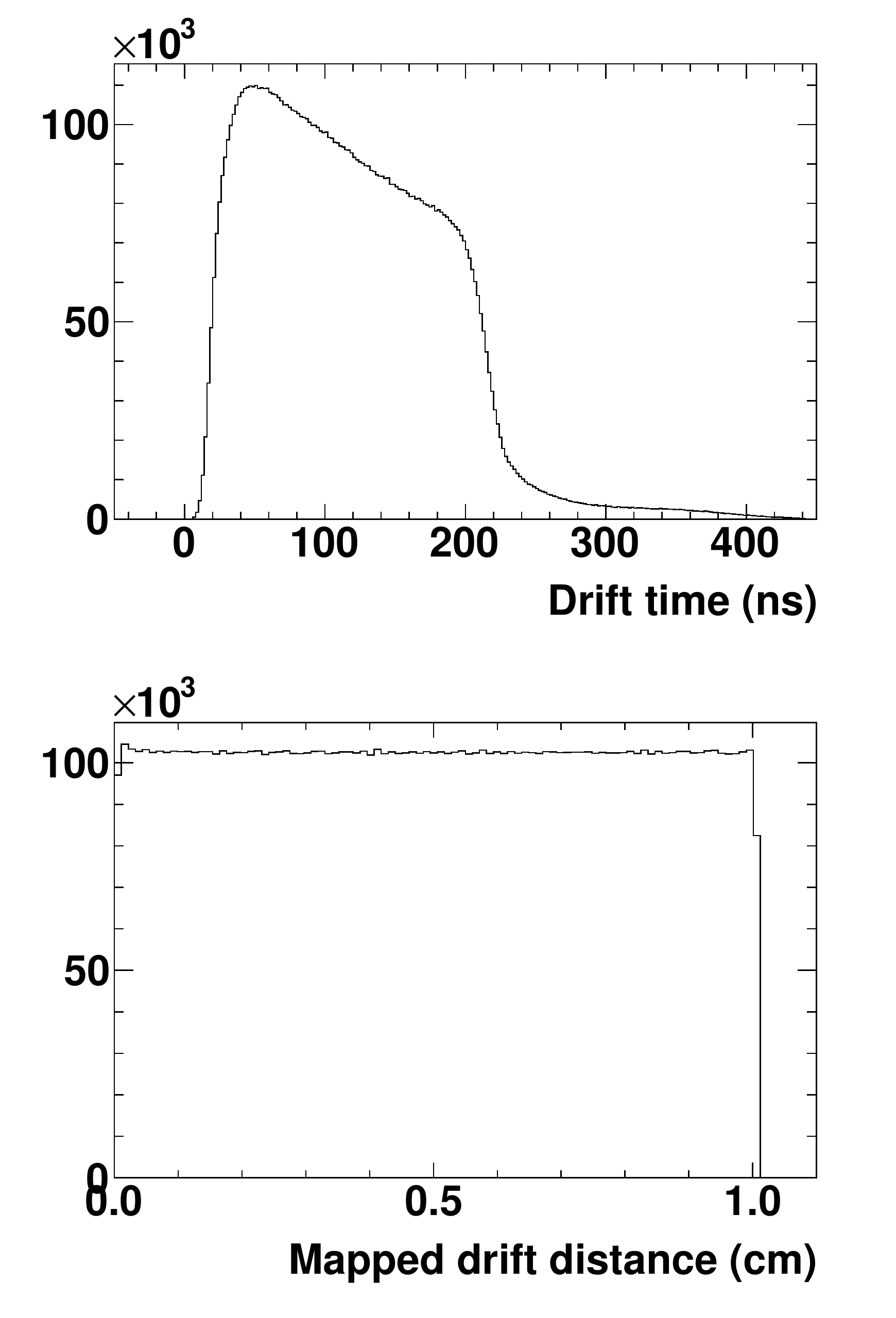}
  \end{center}
  \caption{\label{fig:FPPdriftmapping} Top: example FPP drift time
    spectrum for hits included in final tracks, after all
    corrections. Bottom: Mapped drift distance distribution. Both
    spectra are averaged over all wires in all planes for the run in question.}
\end{figure}

The FPP pattern recognition algorithm tests all possible combinations of one hit wire per plane as potential track candidates. If and only if no valid wire combinations are found with all six planes firing, wire combinations with five out of six planes firing are considered\footnote{The reasons for the preferential treatment of six-plane tracks on the first iteration of pattern recognition and track reconstruction are that (a) the selection of hits and tracks based on $\chi^2$ strongly favors five-plane tracks over six-plane tracks due to the small number of degrees of freedom (only 1(2) for tracks with 5(6) hits), and (b) the greater ambiguity in the determination of the correct left-right combination of the hits in the case of five-plane tracks. Events with five-plane tracks reconstructed in the FPP were nonetheless retained in the final analysis to minimize $\varphi$-dependent variations of tracking efficiency.}. For each potentially valid wire combination, a straight line is fitted to the wire positions only without considering drift distance information, and if the $\chi^2$ per d.o.f. of the fit to wire positions is less than an upper limit corresponding to a maximum in-plane track-wire distance of $\pm$1.4 cm (the FPP in-plane wire spacing is 2 cm), the wire combination is marked as potentially valid. To choose the best wire combination to construct the first track from among all potentially valid combinations, the drift distance information is also used. For each candidate hit combination, the drift time for each hit is corrected for the propagation delay from the point along the wire where the track crossed the wire plane (based on the fit to wire positions only) to the front-end electronics and then used to compute the drift distance. As in the HMS drift chamber tracking algorithm, the time-to-distance conversion is performed by mapping the observed drift time spectrum onto a uniform drift distance distribution within the cell, as shown in Fig.~\ref{fig:FPPdriftmapping}. 

Once the drift distances are computed from the corrected drift times for all hits in a candidate track, straight-line tracks are fitted to all $2^6 = 64$ (or $2^5 = 32$ in the case of five-plane tracks) possible combinations of wire position $\pm$ drift distance, and the combination resulting in the smallest $\chi^2/ndf$ is chosen as the ``best'' left-right combination of the candidate hits. Candidate tracks are required to have $\chi^2/ndf < 100$, corresponding to a maximum tracking residual for any individual hit of 1.2 mm for six-plane tracks ($ndf = 2$)\footnote{The assumed intrinsic coordinate resolution in the $\chi^2$ calculation is $\sigma \approx 250\ \mu$m.}. The maximum $\chi^2$ imposed on drift-based track candidates was chosen to be as small as possible without artificially reducing the efficiency to reconstruct tracks firing all six planes. From among all potentially valid wire combinations, the combination with the smallest $\chi^2/ndf$ of the fit to wire positions $\pm$ drift distances is chosen as the first track in a given chamber pair. The hits used to construct the first track are then marked as used and the pattern recognition/track fitting is repeated until no additional tracks are found. If more than one track is found in either FPP, the track resulting in the smallest polar scattering angle $\vartheta_{fpp}$ relative to the incident proton track reconstructed by the HMS is chosen as the ``best'' track for further analysis, although only the single-track events were ultimately used in the final analysis.

After calibration of the time-to-distance conversion run-by-run, the final RMS tracking residuals in the FPP for elastically scattered protons averaged about 125 $\mu$m for tracks firing all six planes\footnote{The tracking residuals for five-plane tracks were generally much smaller since straight-line fits to these tracks have only one degree of freedom.}, roughly independent of proton momentum. However, this is not a true measure of the coordinate resolution because the residuals are obtained by comparing the in-plane coordinate of each hit to the projected coordinate at each plane of the fitted track, including the hit in question. Since the fitted track is defined by only six coordinate measurements, each hit significantly influences the fitted track. According to Monte Carlo simulations of tracking in the FPP drift chambers, the observed residuals correspond to an intrinsic per-plane coordinate resolution of about 270 $\mu$m, which closely matches the tracking residuals of the HMS drift chambers, for which the tracking residuals more nearly approximate the intrinsic coordinate resolutions due to the larger number of degrees of freedom of the fitted tracks. This is not surprising since the FPP and HMS drift chambers shared the same gas mixture, had similar electric field/drift velocity characteristics, and used very similar front-end and readout electronics. 
\begin{figure}
  \begin{center}
    \includegraphics[width=0.75\columnwidth]{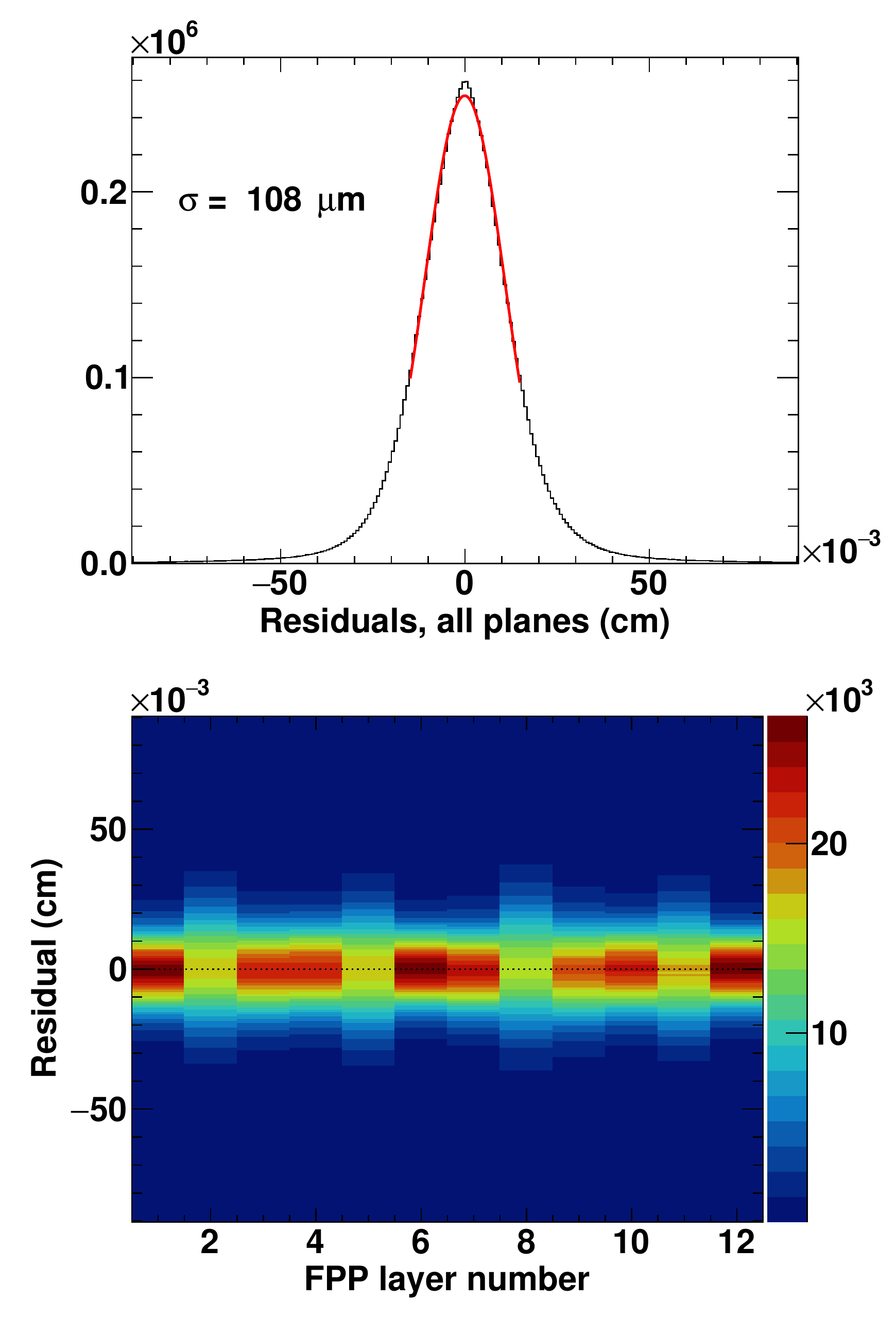}
  \end{center}
  \caption{\label{fig:FPPresolution} FPP drift chamber tracking residuals, averaged over all planes (top) and as a function of plane number (bottom). The red curve in the top panel is a Gaussian fit, resulting in $\sigma = 108\ \mu$m, as indicated. In the bottom panel, planes 1-6 correspond to FPP1, while planes 7-12 correspond to FPP2. Residuals shown are for ``straight-through'' electron tracks at a central momentum of 2.4 GeV. See text for details.}
\end{figure}
Figure~\ref{fig:FPPresolution} shows a typical example of FPP tracking residuals for ``straight-through'' tracks of electrons at a central momentum of 2.4 GeV. Electron tracks give slightly smaller \emph{rms} tracking residuals than protons, mainly due to reduced multiple-scattering in the drift chamber materials and the somewhat greater amount of ionization of the chamber gas mixture by electrons than protons in the momentum range of this experiment. All coordinate measurements are weighted equally in the $\chi^2$ calculation during the fitting of tracks, reflecting the fact that the intrinsic coordinate resolution is the same for all planes. The observed pattern of different widths of tracking residuals in different planes emerges as an artifact of the geometric layout (wire orientations and plane ordering) of the FPP chambers and the limited number of coordinate measurements, and is reproduced by Monte Carlo simulations. 
\subsection{FPP straight-through data and alignment}
\label{subsubsec:FPPalignment}
The FPP drift chambers were surveyed in place after installation. However, the absolute accuracy of the surveyed positions was not better than about $\pm$1 mm. A more accurate determination of the position and orientation of the FPP drift chambers relative to the HMS drift chambers was achieved by collecting dedicated ``straight-through'' data with the CH$_2$ analyzers retracted from the HMS acceptance. The support structure for the CH$_2$ analyzers, including the insertion/retraction mechanism, was separate from that of the FPP drift chambers, ensuring that the drift chambers could not move during insertion/removal of the analyzers. Although several dedicated straight-through runs were taken with elastically scattered protons (including at least one for each kinematic setting), the final alignment of the FPP drift chambers was actually performed using straight-through data collected simultaneously with optics calibration data on multi-foil Aluminum and Carbon targets, with the HMS set to detect inelastically scattered electrons at a central momentum of 2.4 GeV. The advantage of using these data for alignment of the FPP drift chambers was that the inelastically scattered electrons populated a wider region of the HMS acceptance at the focal plane than did elastically scattered protons for any of the production kinematics, thus providing greater sensitivity to the small rotational offsets of the FPP chambers relative to the HMS. The coordinate and angular resolution for electrons at 2.4 GeV was also better than for protons at 2.07 GeV. In the following discussion, the subscripts '$HMS$', '$FPP1$', and '$FPP2$' refer to the set of drift chambers measuring the track; unless otherwise noted, all track parameters in the following discussion are expressed at $z = 0$ in the TRANSPORT coordinate system at the HMS focal plane (see section \ref{subsubsec:transport}).

The goal of the software alignment procedure was to determine the set of translational ($x_0, y_0, z_0$) and rotational $(\alpha_x, \alpha_y, \alpha_z)$ offsets of each FPP drift chamber pair that minimized the sum of squared differences between HMS tracks and FPP tracks in terms of the track slopes $(x', y')$ and the track coordinates $(x,y)$ projected to the HMS focal plane. The rotation angles were assumed to be sufficiently small that a linearized approximation to the rotation matrices was adequate; i.e., $\sin(\alpha_{x,y,z}) \approx \alpha_{x,y,z}$ and $\cos(\alpha_{x,y,z}) \approx 1$. Several iterations of the alignment procedure were carried out, using the results from the previous iteration of the fit as the starting point for the next iteration, until the translational and rotational offsets did not change appreciably on subsequent iterations of the fit. The straight-through data were then reconstructed using the final global alignment parameters, and the correlations of the FPP-HMS track parameter differences with the HMS track parameters were examined. The correlation study showed that some small residual correlations remained even after the geometric alignment. 
\begin{figure}
  \begin{center}
    \includegraphics[width=0.7\columnwidth]{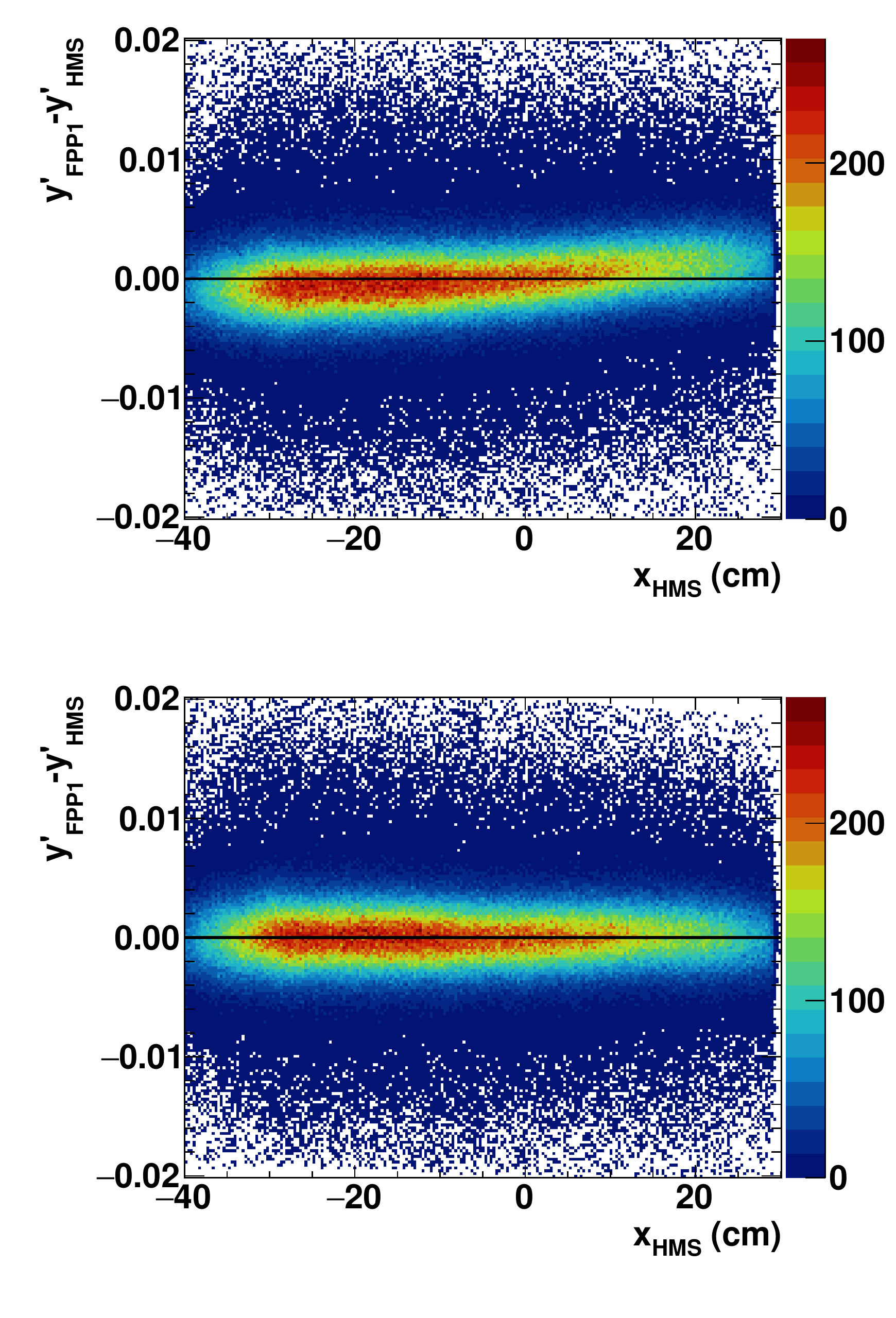}
  \end{center}
  \caption{\label{fig:dypx_correlation} Difference $\Delta y' = y'_{FPP1} - y'_{HMS}$ between the FPP1 and HMS track slopes in the non-dispersive direction as a function of the dispersive-plane coordinate $x_{HMS}$ of the HMS track at the focal plane, \emph{after} geometric alignment, before (top) and after (bottom) applying \emph{ad hoc} corrections represented by Eq.~\eqref{eq:fpp_track_correction} to the FPP track.}
\end{figure}

\begin{figure*}
  \begin{center}
    \includegraphics[width=0.98\textwidth]{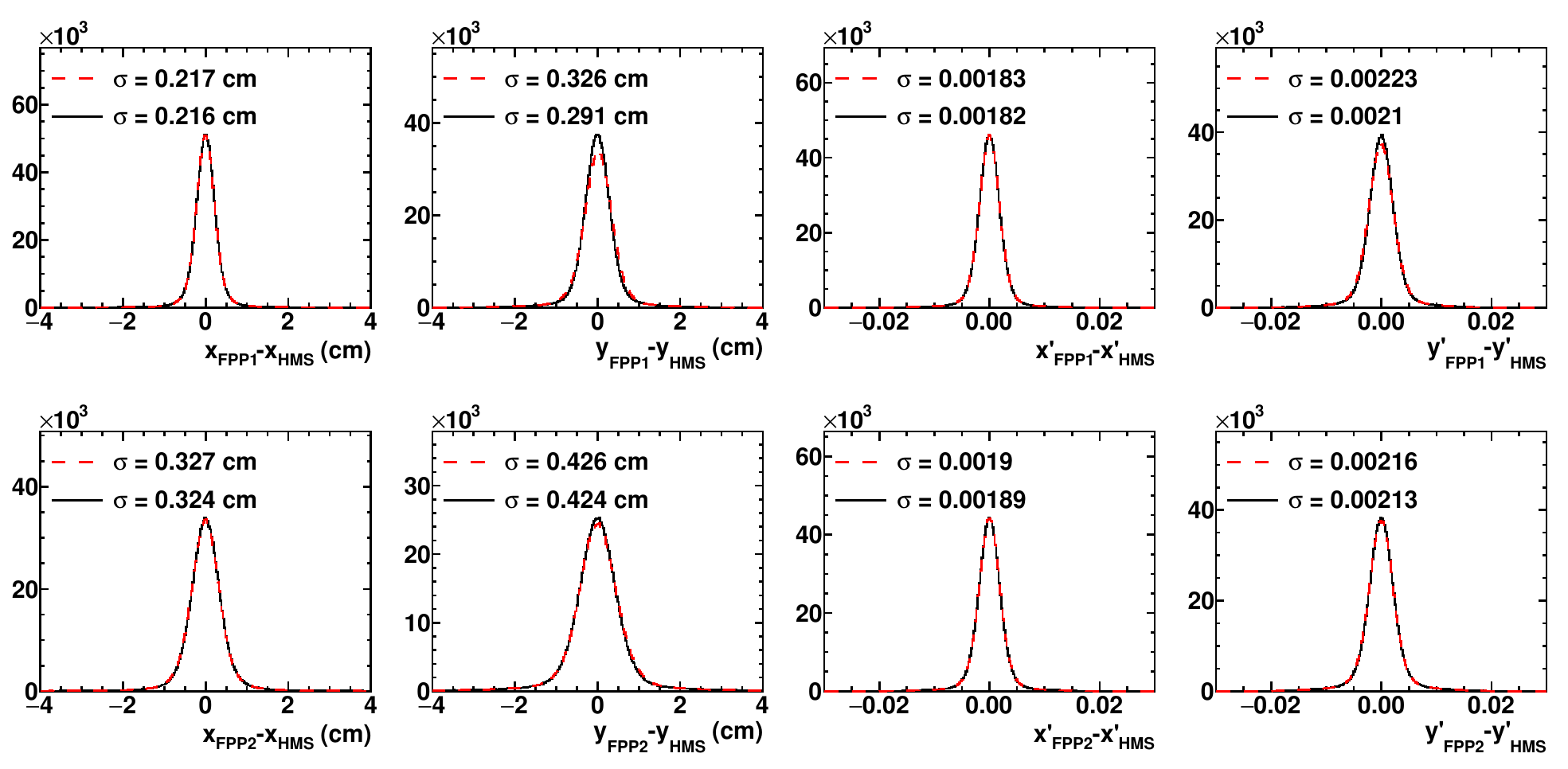}
  \end{center}
  \caption{\label{fig:fppalign1} Distributions of the differences between reconstructed FPP and HMS track parameters from ``straight-through'' data, after geometric alignment, before (red dashed) and after (black solid) applying the \emph{ad hoc} corrections of Eq.~\eqref{eq:fpp_track_correction} to the FPP tracks on an event-by-event basis. Straight-through tracks in this figure are electrons scattered inelastically from multi-foil carbon and aluminum targets at a central momentum of 2.4 GeV. Top (bottom) row shows FPP1 (FPP2) differences. From left to right, parameter differences are $\Delta x$, $\Delta y$, $\Delta x'$ and $\Delta y'$. Coordinate differences include the error in projecting the FPP tracks from the chamber locations where they are measured back to the HMS focal plane.}
\end{figure*}
For example, as shown in Fig.~\ref{fig:dypx_correlation}, the difference $\Delta y' = y'_{FPP1} - y'_{HMS}$ of the track slope in the non-dispersive direction exhibited a correlation with the dispersive-plane coordinate $x_{HMS}$ of $\frac{d(\Delta y')}{dx_{HMS}} \approx 3.9$ mrad/m, implying an error in the \emph{relative} $y'$ between the FPP and HMS tracks of up to 2 mrad at the extremes of the HMS acceptance. These kinds of residual correlations are symptomatic of internal offsets and/or misalignments of the HMS and/or FPP drift chambers. No attempt was made to further refine the parameters describing the global alignment (position and orientation) of the HMS and/or FPP drift chambers. Instead, the effect of the residual correlations on the reconstruction of the secondary scattering angles $\vartheta$ and $\varphi$ was minimized by applying small, \emph{ad hoc} corrections to the parameters of each reconstructed FPP track. The parameters of the correction were determined by fitting the straight-through data with the following second-order expansion of the FPP-HMS track parameter differences $\Delta x = x_{FPP} - x_{HMS}$, $\Delta y = y_{FPP} - y_{HMS}$, $\Delta x' = x'_{FPP} - x'_{HMS}$ and $\Delta y' = y'_{FPP} - y'_{HMS}$ in terms of the \emph{HMS} track parameters:
\begin{eqnarray}
  \label{eq:fpp_track_correction}
  \Delta u &=& C^{(u)}_0 + C^{(u)}_{x} x + C^{(u)}_{y} y + C^{(u)}_{x'} x' + C^{(u)}_{y'} y' + \nonumber \\
           & & C^{(u)}_{xx} x^2 + C^{(u)}_{xy} xy + C^{(u)}_{xx'} xx' + C^{(u)}_{xy'} xy' + \nonumber \\
           & &  C^{(u)}_{yy} y^2 + C^{(u)}_{yx'} yx' + C^{(u)}_{yy'} yy' + C^{(u)}_{x'x'} {x'}^2 + \nonumber \\
           & & C^{(u)}_{x'y'} x'y' + C^{(u)}_{y'y'} {y'}^2,
\end{eqnarray}
where $\Delta u = \{\Delta x, \Delta y, \Delta x', \Delta y'\}$ is the
track parameter difference in question, and the $C^{(u)}$'s are the
coefficients of each term in the expansion. The track parameters in
Eq.~\eqref{eq:fpp_track_correction} refer to the \emph{HMS} track. 
Because the residual correlations were small to begin with, a second-order expansion easily suppressed them to a level well below the intrinsic resolution of track coordinates and angles. Moreover, fitting the correction terms using straight-through data populating a wider region of the HMS acceptance than that occupied by elastically scattered protons for any of the production kinematic settings guarantees that the correction applied to any given FPP track will be small, and that no extra error will be introduced by extrapolating the correction outside the phase space region where it is constrained by straight-through data. 

The full analysis was carried out with and without the \emph{ad hoc} correction of Eq.~\eqref{eq:fpp_track_correction}, and the effect of the correction on the polarization transfer observables was found to be negligible. For the final analysis, the correction \emph{was} applied. Figure~\ref{fig:fppalign1} summarizes the results of the software alignment, comparing all four track parameter differences between FPP1/2 and HMS, after geometric alignment, before and after applying the \emph{ad hoc} correction from Eq.~\eqref{eq:fpp_track_correction}. Small reductions in the peak widths are seen for all parameters. The most significant improvements are seen in $\Delta y$ and $\Delta y'$ for FPP1 (the latter shown in Fig.~\ref{fig:dypx_correlation}). The straight-through data also provide an estimate of the FPP angular resolution; for 2.4-GeV electrons, the angular resolutions are $(\sigma_{x'}, \sigma_{y'}) \approx (1.8, 2.1)$ mrad. The resolution asymmetry between the $x$ and $y$ directions simply reflects the fact that only four of six wire planes in each FPP chamber pair have sensitivity to the $y$ coordinate, while all six planes have some sensitivity to the $x$ coordinate. The quality of the geometric alignment of the FPP chambers and \emph{ad hoc} track corrections was checked by reconstructing straight-through data from several of the elastically scattered proton kinematics using the alignment parameters determined from the optics calibration data. The small differences among the various settings were used to set an upper limit on the systematic uncertainty in the reconstructed secondary scattering angles $\vartheta$ and $\varphi$. Specifically, based on the repeatability of the alignment for straight-through runs taken in different kinematic settings, a conservatively estimated upper systematic uncertainty limit of 0.1 mrad in $\Delta x'$ and $\Delta y'$, which translates into a $\vartheta$-dependent systematic uncertainty $\Delta \varphi \approx 0.14\mbox{ mrad}/\sin(\vartheta)$ in the azimuthal scattering angle $\varphi$, was assigned for the scattering angle reconstruction in the FPP. 

\subsection{FPP event selection criteria, angular distributions and closest approach parameters}
\begin{table}
  \caption{\label{tab:FPPselections} FPP event selection criteria as a function of $Q^2$. Only single-track events passing the ``cone test'' were included in the analysis. No explicit $\vartheta$ cuts were applied. Instead, the $\vartheta$ ranges shown are the effective ranges resulting from the $p_T$ cuts. The same criteria were applied to all three $\epsilon$ values at $Q^2 = 2.5$ GeV$^2$. $s_{close}$ and $z_{close}$ are defined, respectively, as the distance of closest approach between the incident and scattered tracks, and the $z$-coordinate of the point of closest approach between incident and scattered tracks, with $z = 0$ at the HMS focal plane. See text for details.}
  \begin{center}
    \begin{tabular}{lcccc}
      \hline \hline 
      $Q^2$ (GeV$^2$) & 2.5 & 5.2 & 6.8 & 8.5 \\ \hline 
      $p_T^{min}$ (GeV/c) & 0.06 & 0.05 & 0.05 & 0.05 \\
      $p_T^{max}$ (GeV/c) & 1.2 & 1.5 & 1.5 & 1.5 \\
      FPP1 $\vartheta_{min}^{eff} (^\circ)$ & 1.71 & 0.81 & 0.65 & 0.53 \\
      FPP1 $\vartheta_{max}^{eff} (^\circ)$ &  36.7 & 25.1 & 19.9 & 16.3 \\
      FPP2 $\vartheta_{min}^{eff} (^\circ)$ & 1.82 & 0.84 & 0.67 & 0.55 \\
      FPP2 $\vartheta_{max}^{eff} (^\circ)$ & 39.5 & 26.0 & 20.4 & 16.6 \\
      FPP1 $s_{close}^{max}$ (cm) & 2.2 & 1.7 & 1.4 & 1.2 \\
      FPP2 $s_{close}^{max}$ (cm) & 6.5 & 5.1 & 4.1 & 3.3 \\
      FPP1 $z_{close}^{min}$ (cm) & 108 & 108 & 108 & 108 \\
      FPP1 $z_{close}^{max}$ (cm) & 168 & 168 & 168 & 168 \\
      FPP2 $z_{close}^{min}$ (cm) & 207 & 207 & 207 & 207 \\
      FPP2 $z_{close}^{max}$ (cm) & 267 & 267 & 267 & 267 \\ \hline \hline 
    \end{tabular}
  \end{center}
\end{table} 

Table~\ref{tab:FPPselections} summarizes the event selection criteria for the FPP. Cuts are applied to the ``transverse momentum'' $p_T \equiv p_p \sin \vartheta$, the distance of closest approach $s_{close}$ between incident (HMS) and scattered (FPP) tracks, and the coordinate $z_{close}$ of the point of closest approach between incident and scattered tracks. As described in Ref.~\cite{Puckett:2017flj}, a ``cone test'' was also applied to the reconstructed FPP tracks to minimize acceptance-related azimuthal asymmetries. For events reconstructed in the second polarimeter (FPP2) drift chambers, it is possible to choose either the HMS track or any track reconstructed in the first polarimeter (FPP1) as the ``reference'' track with respect to which the scattering angles $\vartheta, \varphi$ and the closest-approach parameters $s_{close}, z_{close}$ are reconstructed. For the final analysis, the scattering angles and closest approach parameters of the FPP2 track were always reconstructed relative to the HMS track\footnote{Because of the significant probability of mistracking in either set of FPP drift chambers, the scattering parameters of the FPP2 track relative to the FPP1 track were unreliable in a small but significant fraction of events with good track reconstruction in FPP2.}. Any event reconstructed in FPP2 with scattering parameters relative to the HMS track consistent with a single scattering in the second analyzer was counted in the analysis, regardless of the results of tracking in FPP1. This approach to the analysis of the FPP2 data was found to give the best overall figure-of-merit, and is also the most logically consistent and unbiased way to analyze the data. In the analysis of the GEp-III kinematics, with statistics-limited uncertainties, single-track events in FPP2 consistent with a single scattering in the \emph{first} analyzer were also counted, provided the same events had not already been counted in FPP1, due to e.g., mistracking, detection inefficiency and/or FPP1 data quality issues. These events were \emph{not} included in the analysis of the GEp-2$\gamma$ data, except during data acquisition runs for which all of the FPP1 data were rejected due to data quality issues. This is because the accuracy of the GEp-2$\gamma$ data was not statistics-limited, and because the analyzing power, the accurate description of which is essential for the reliable extraction of $P_\ell/P_\ell^{Born}$, is subject to greater uncertainty for this event topology. 

\begin{figure}
  \begin{center}
    \includegraphics[width=0.7\columnwidth]{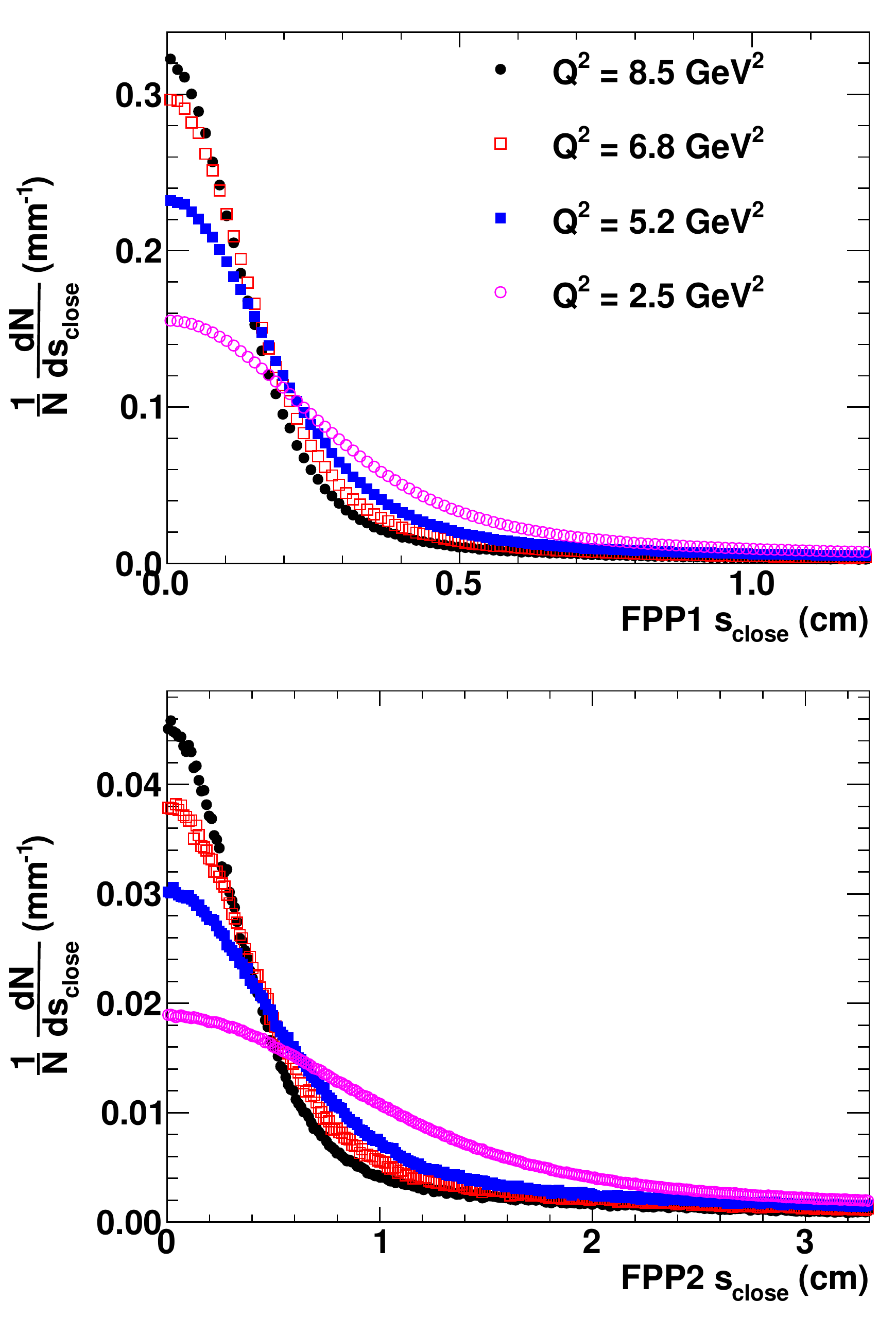}
  \end{center}
  \caption{\label{fig:sclose}  Distance of closest approach $s_{close}$ between incident and scattered tracks, for FPP1 (top) and FPP2 (bottom), for single-track events passing the cone test and with point of closest approach $z_{close}$ reconstructed within the region corresponding to the physical extent of the analyzer, shown in Fig.~\ref{fig:zclosetheta}. Note the different horizontal scales between the top and bottom panels.}
\end{figure}

The distribution of $s_{close}$ is shown for all four $Q^2$ values in Fig.~\ref{fig:sclose}. At each $Q^2$, the $s_{close}$ distribution is normalized to the total number of elastic events producing exactly one track in the polarimeter in question (see Fig.~\ref{fig:FPPtrackmult} for the FPP track multiplicities per event.). The resolution of $s_{close}$ improves with increasing proton momentum, as the width of the multiple-scattering distribution in the fixed thickness of analyzer material decreases. The $s_{close}$ distribution of FPP2 events is nearly three times as wide as that of FPP1 events, because protons detected in FPP2 traverse approximately three times the average path length in CH$_2$ as those detected in FPP1 prior to scattering. The effective $\vartheta$ ranges for FPP1 and FPP2 differ for the same reason;  the same $p_T$ corresponds to a slightly larger $\vartheta$ in FPP2 due to the additional energy losses prior to scattering by protons detected in FPP2.

\begin{figure}
  \begin{center}
    \includegraphics[width=0.75\columnwidth]{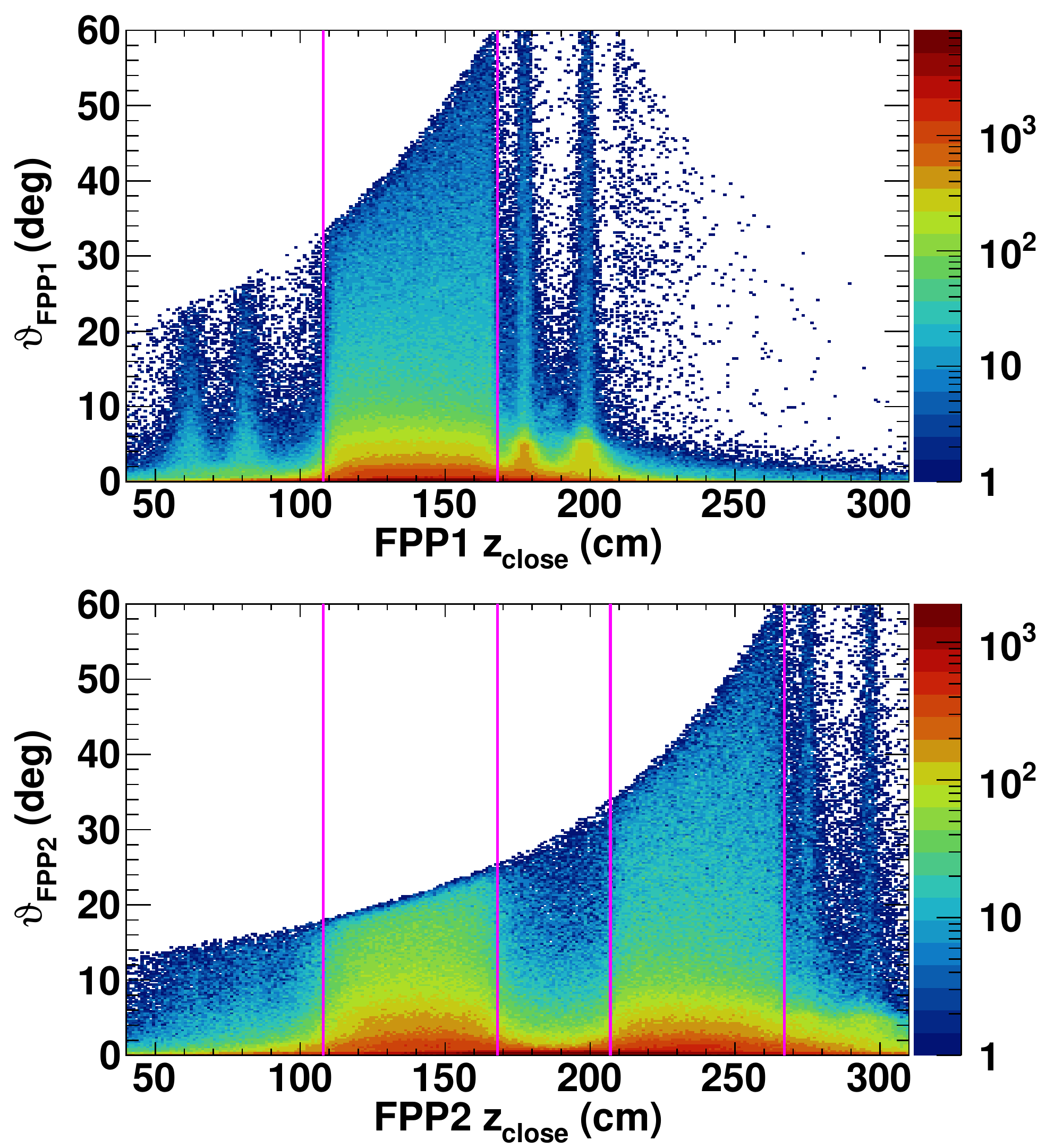}
  \end{center}
  \caption{\label{fig:zclosetheta}  Correlation between $z_{close}$,
    the $z$ coordinate of the point of closest approach between
    incident and scattered tracks, and the polar scattering angle
    $\vartheta_{fpp}$, for incident elastically scattered protons at $Q^2 = 8.5$ GeV$^2$. Single-track events passing the ``cone test'' with distance of closest approach $s_{close} \le s_{close}^{max}$ are shown for FPP1 (top panel) and FPP2 (bottom panel). Vertical lines indicate the region(s) of $z_{close}$ included in the analysis. In the bottom panel, the distribution of FPP2 events is shown regardless of the FPP1 tracking results. See text for details.}
\end{figure}
Figure~\ref{fig:zclosetheta} shows the correlation between $z_{close}$ and $\vartheta_{FPP}$ for events with an incident elastically scattered proton and a single track reconstructed in FPP1 and/or FPP2 passing the cone test and the $s_{close}$ cut, at $Q^2 = 8.5$ GeV$^2$. The $z_{close}$-dependent maximum $\vartheta$ cutoff reflects the acceptance of the cone test. For tracks reconstructed in FPP1 (FPP2) with $z_{close}$ values corresponding to scattering in the first (second) analyzer, the cone test is essentially 100\% efficient for $\vartheta \le 30^\circ$, regardless of $z_{close}$. For events reconstructed in FPP2 with $z_{close}$ corresponding to scattering in the \emph{first} analyzer, the cone test is efficient regardless of $z_{close}$ for $\vartheta \le 16^\circ$. The distribution of FPP2 events in Fig.~\ref{fig:zclosetheta} is shown regardless of the tracking results in FPP1, and therefore includes many events that would have already been counted in FPP1 (and thus not counted in FPP2 to avoid double-counting). The narrow stripes at $z_{close} \approx 60$ cm and $z_{close} \approx 80$ cm correspond to scattering in the S1X and S1Y scintillator planes. 

The ``stripes'' at the locations of the FPP drift chambers result to some extent from tracks that actually scatter in the chambers, but mainly from tracks with incorrect solutions of the left-right ambiguity. As discussed in Refs.~\cite{Puckett:2017flj} and \cite{Puckett:2015soa}, the design of the FPP drift chambers minimized the number of wire planes in order to minimize the cost of chamber construction, the number of readout channels, and the space occupied by the drift chambers in the HMS shield hut. The efficiency of the design comes at the price of an irreducible left-right ambiguity for a subset of tracks passing through a chamber pair at close to normal incidence near the geometric center of the chambers, where the $X$, $U$, and $V$ wires share a common intersection point. For this subset of tracks, two mirror-image solutions exist, that are essentially indistinguishable in terms of $\chi^2$, with the hits placed on opposite sides of all three wires that fired in a given chamber. These mistracked events are particularly prominent for $\vartheta \lesssim 6^\circ$. The peaks at the drift chamber locations with $\vartheta \lesssim 6^\circ$ correspond to tracks in the Coulomb peak of the $\vartheta$ distribution for which all three hits in one of the two drift chambers in the pair are placed on the wrong side of the wires that fired in that chamber. Given the 2-cm FPP drift cell size in each wire plane, the incorrect left-right assignment displaces the position of the track at that drift chamber by up to 2 cm. Since the $z$ separation between the two chambers in a pair is approximately 21 cm, a 2-cm displacement of one of the two measured points along a track with $\vartheta \approx 0$ is $\Delta \vartheta \approx \arctan (2$ cm$/21$ cm$) = 5.4^\circ$. This is why the number of events in the ``stripes'' at the drift chamber locations decreases sharply for $\vartheta \gtrsim 6^\circ$.

\begin{figure}
  \begin{center}
    \includegraphics[width=0.75\columnwidth]{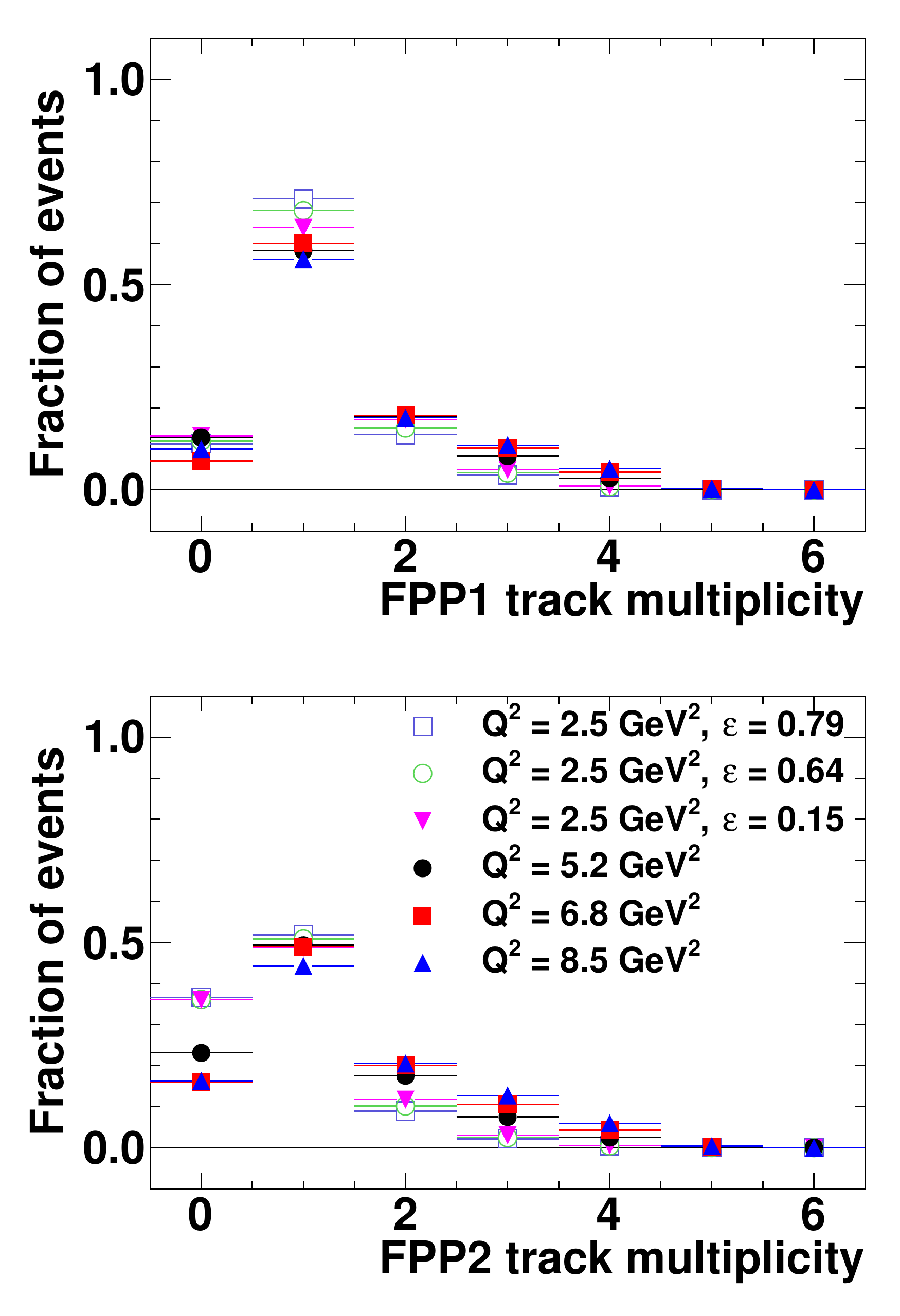}
  \end{center}
  \caption{\label{fig:FPPtrackmult} FPP track multiplicity per event for all six kinematic settings. The fraction of all identified elastic events as a function of track multiplicity is shown for FPP1 (top) and FPP2 (bottom). }
\end{figure}
Figure~\ref{fig:FPPtrackmult} shows the multiplicity of reconstructed tracks per incident elastically scattered proton for FPP1 and FPP2. In FPP1 (FPP2), the fraction of single-track events ranges from 55-70\% (45-50\%). The single-track fraction decreases somewhat as $Q^2$ increases, as the available phase space for multi-particle production increases. 
The fraction of events with zero tracks, which reflects detection
inefficiencies, large-angle scatterings in which the proton escapes
detection, and/or proton absorption/capture/charge-exchange reactions
that don't produce any charged tracks, ranges from 8-13\% (16-36\%)
for FPP1 (FPP2). In the case of FPP2, the fraction of events with zero
tracks depends more strongly on the proton momentum, which is expected
given the higher probability of large angle scattering for lower momentum protons and the greater analyzer thickness the protons must pass through before detection in FPP2. 

\begin{figure}
  \begin{center}
    \includegraphics[width=0.7\columnwidth]{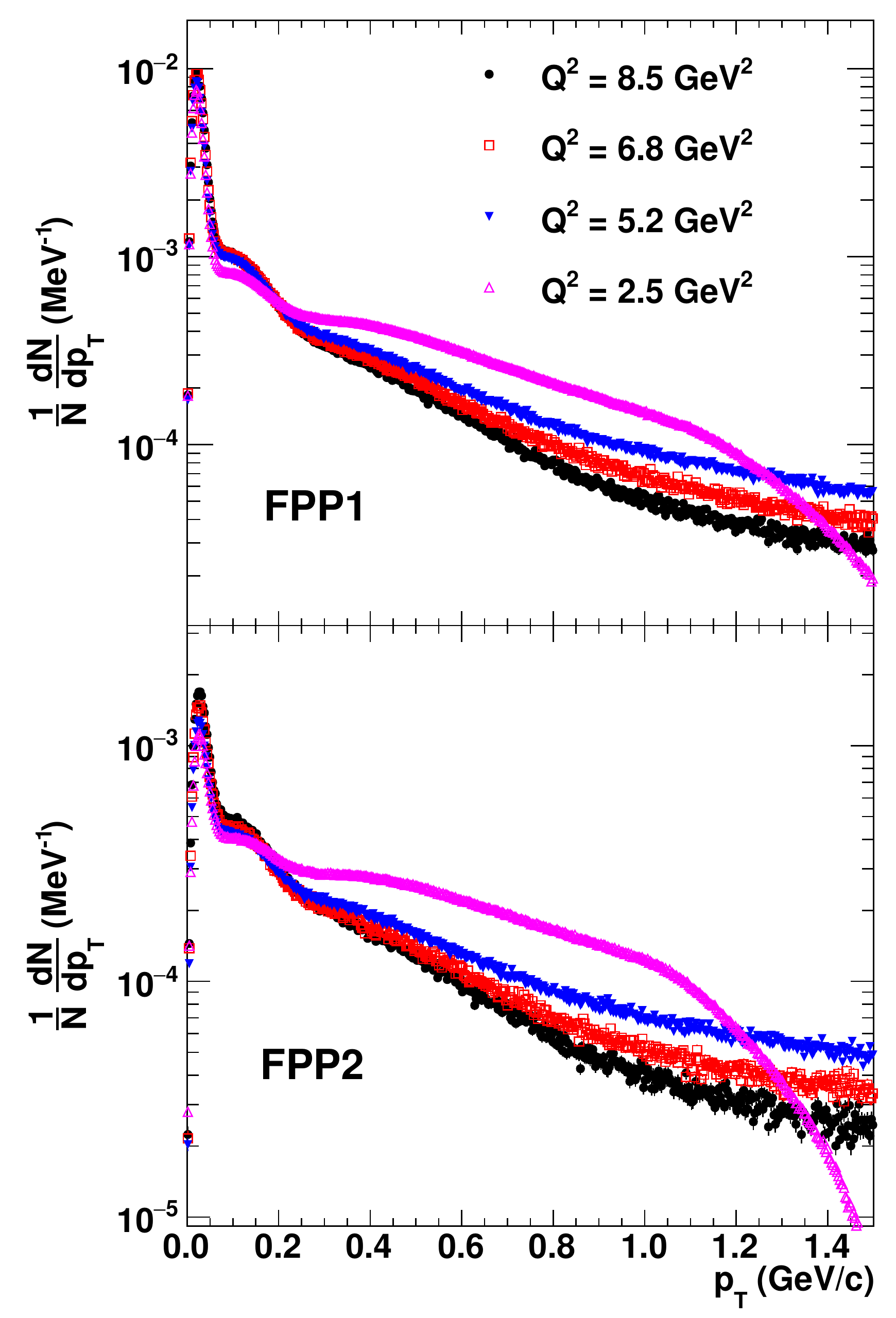}
  \end{center}
  \caption{\label{fig:pTgep3} FPP angular distributions for all four $Q^2$ values, plotted in terms of the ``transverse momentum'' $p_T \equiv p_p \sin \vartheta_{fpp}$, illustrating the approximate scaling of the angular distribution of nuclear scattering events with momentum. Single-track events passing the cone test as well as the $s_{close}$ and $z_{close}$ cuts in Table~\ref{tab:FPPselections} are shown for FPP1 (top panel) and FPP2 (bottom panel). See text for details.}
\end{figure}

Figure~\ref{fig:pTgep3} shows the $p_T$ distributions of single-track events in both polarimeters for all four $Q^2$ values. For each kinematic setting, the distribution is normalized to the total number of elastic events producing exactly one track in the polarimeter in question. The distributions are qualitatively similar, but clearly not identical. In particular, the shape of the $p_T$ distribution at $Q^2 = 2.5$ GeV$^2$ differs significantly from its shape at higher $Q^2$. The global features of the distribution are well understood. The small-angle peak corresponds to multiple-Coulomb scattering; the analyzing power vanishes in the $p_T \rightarrow 0$ limit. The width of the Coulomb peak in the $p_T$ distribution is independent of the incident proton momentum, to a good approximation, but is slightly wider in FPP2 than FPP1 due to the greater thickness of analyzer traversed by the proton before detection in FPP2. The vanishing yield as $p_T \rightarrow 0$ is an effect of the vanishing solid angle in the $\vartheta \rightarrow 0$ limit. For events outside the Coulomb peak, the angular distribution shifts gradually toward smaller $p_T$ values as the incident proton momentum increases, thus showing that the scaling of the width of the angular distribution with proton momentum is not exact. At $Q^2 = 2.5$ GeV$^2$, the steep drop-off for $p_T \gtrsim 1$ GeV/c is caused by the detector acceptance; $p_T = 1$ GeV corresponds to $\vartheta \approx 29^\circ$ at this $Q^2$. As shown in Fig.~\ref{fig:zclosetheta}, the cone test starts to cut off the acceptance at about 30 degrees at the upstream edge of the analyzer closest to each drift chamber pair. The total probability for an incident proton to produce a single-track event within the useful range $0.06 \le p_T \mbox{(GeV/c)} \le 1.2$ decreases slowly as a function of momentum for a given analyzer thickness, a fact relevant to the planning of future experiments at higher $Q^2$. 

\begin{figure}
  \begin{center}
    \includegraphics[width=0.7\columnwidth]{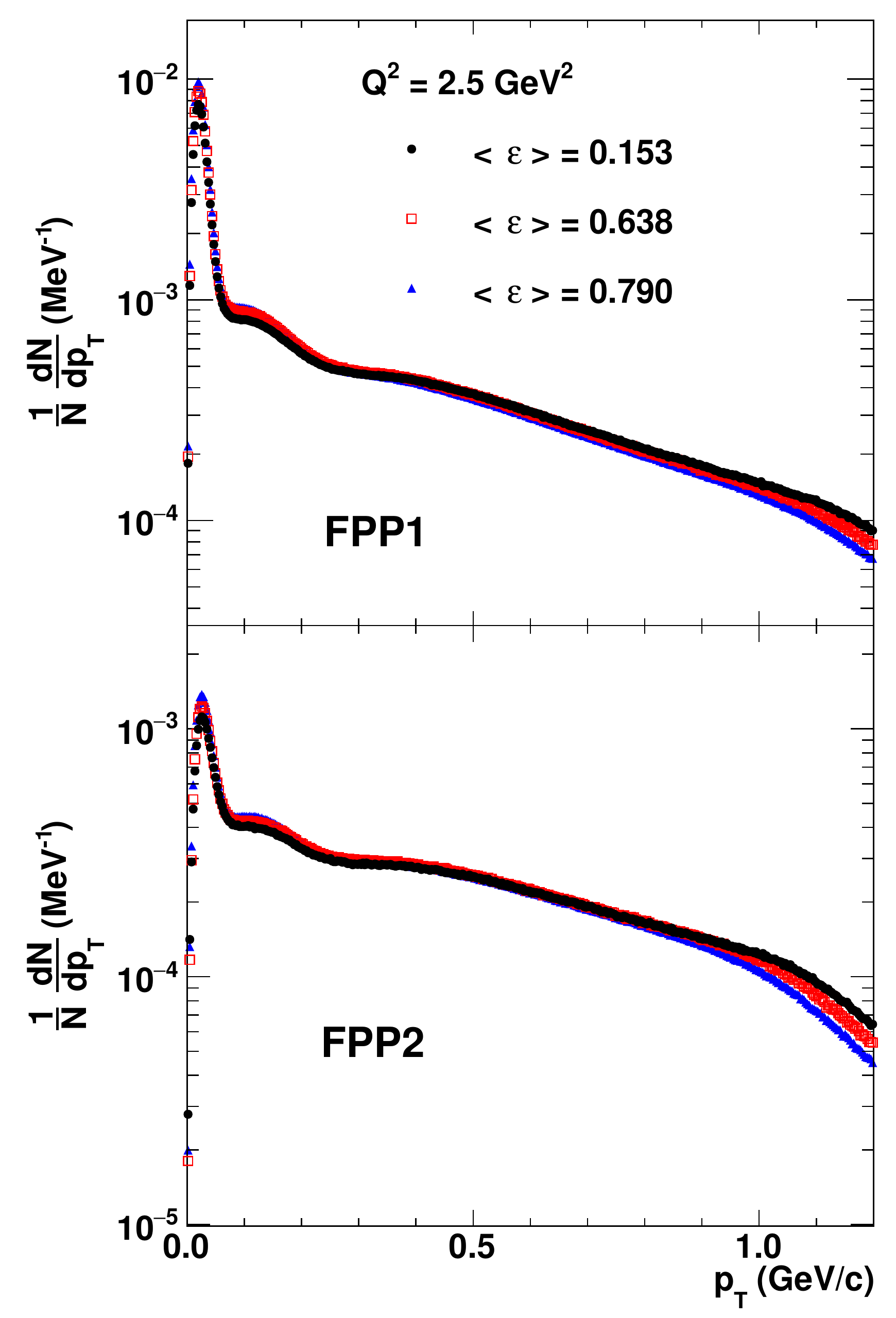}
  \end{center}
  \caption{\label{fig:pTgep2g} FPP angular distributions in terms of $p_T \equiv p_p \sin \vartheta_{fpp}$ for the GEp-2$\gamma$ kinematics, for single-track events passing the cone test as well as the $s_{close}$ and $z_{close}$ cuts in Table~\ref{tab:FPPselections}, for FPP1 (top) and FPP2 (bottom). The shape of the angular distributions is the same for all three $\epsilon$ values at the few-percent level within the useful $p_T$ range.}
\end{figure}
Figure~\ref{fig:pTgep2g} shows the $p_T$ distributions for each of the three $\epsilon$ values at $Q^2 = 2.5$ GeV$^2$. Because the measurements are at the same fixed $Q^2$, the angular distributions should be the same, in first approximation. In both polarimeters, the $p_T$ distributions are indeed observed to be the same at the few-percent level within the useful $p_T$ range. However, even for the same central momentum setting, there are some slight differences resulting from the different momentum and phase space distributions of incident protons for the different $\epsilon$ values. At the lowest $\epsilon$, corresponding to the most forward proton scattering angle, the fixed HMS angular acceptance corresponds to a very small range of $Q^2$ (see Eq.~\eqref{eq:pp_ptheta}), meaning that the envelope of elastically scattered protons at $\epsilon = 0.153$ is confined to a narrow region at the center of the HMS focal plane. At large $\epsilon$, the smaller reaction Jacobian leads to a much wider $Q^2$ acceptance, and the envelope of elastically scattered protons is spread out over a much wider region of the HMS focal plane. For this reason, the $p_T$ distribution falls off slightly faster at large angles for the two higher-$\epsilon$ kinematics than at the lowest $\epsilon$, because the probability of an event failing the cone test is greater at a given $p_T$ when the incident protons are spread out over a wider region of the HMS acceptance. 

\subsection{BigCal Event Reconstruction}
The reconstruction of the scattered electron's energy and scattering angles begins by grouping adjacent lead-glass blocks with large signals into ``clusters'' of hits representing the electromagnetic showers initiated by (presumably) single electrons (or high-energy photons). The raw signals from each block were recorded by charge-integrating ADCs with a gate width of 150-250 ns, chosen based on the kinematics\footnote{Longer gate widths were used for kinematics with greater elastically scattered electron energies, producing larger pulse amplitudes and somewhat longer pulse durations.}. The raw ADC values were then converted to deposited energies by subtracting the mean ``pedestals''\footnote{The ``pedestal'' is defined as the mean ADC value for events with no signal; i.e., the baseline.} from the digitized signals and multiplying the pedestal-subtracted ADC values by calibration constants (specific to each channel) relating the charge and the energy deposition. Periodic calibration and gain-matching of BigCal was performed \emph{in situ} using elastically scattered electrons, the energies of which were precisely determined by the measured proton kinematics in the HMS. Details of the calibration procedure are given in Ref.~\cite{Puckett:2015soa}. As the overall signal amplitude in BigCal dropped due to radiation-induced darkening of the lead-glass, the PMT high voltages were periodically increased to maintain a roughly constant average signal amplitude even as the energy resolution worsened significantly due to reduced photoelectron statistics. Additionally, a database of time-dependent calibration constants was developed for the offline analysis. 

\begin{figure}
  \begin{center}
    \includegraphics[width=0.98\columnwidth]{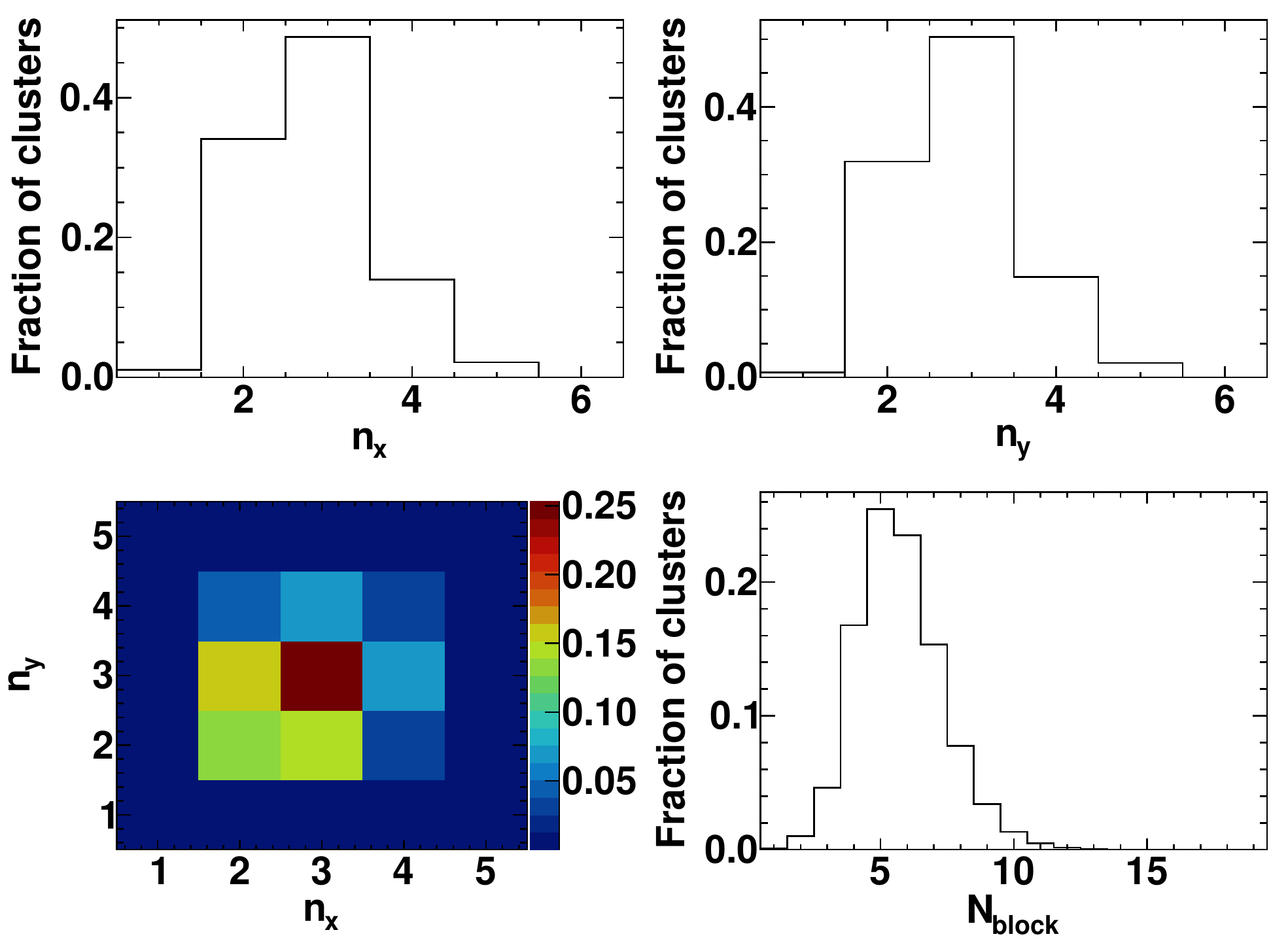}
  \end{center}
  \caption{\label{fig:ClusterSize} BigCal cluster size distributions for elastically scattered electrons at $Q^2 = 2.5$ GeV$^2$, $\left<E'_e\right> = 1.52$ GeV. The horizontal (vertical) cluster size is denoted $n_x$ ($n_y$). $N_{block}$ is the total number of blocks per cluster with a signal above (software) threshold. 98.3\% of elastically scattered electron clusters in this example are at least two blocks wide in both the vertical and horizontal directions. The mean horizontal (vertical) cluster size is 2.82 (2.86) blocks, and the most probable cluster size is $3\times 3$. The mean (most probable) total number of hits per cluster is 5.73 (5).}
\end{figure}
Figure \ref{fig:ClusterSize} shows the cluster size distribution in BigCal for elastically scattered electrons at an average energy of $E'_e \approx 1.5$ GeV. These distributions are typical of all the kinematics except for $Q^2 = 2.5$ GeV$^2$, $E'_e \approx 0.54$ GeV, for which the average cluster size was smaller owing to the much lower scattered electron energy. The Moli\`{e}re radius of the TF1-0 lead-glass used in BigCal is about 4.7 cm. Given the roughly 4-cm transverse size of the individual lead-glass blocks, the typical electromagnetic shower initiated by an elastically scattered electron at normal incidence deposits about 90\% of its energy in a $3 \times 3$-block area, and about 99\% of its energy in a $5 \times 5$-block area. Details of the clustering algorithm are given in Ref.~\cite{Puckett:2015soa}. 
For events with multiple clusters, the ``best'' cluster was chosen as the cluster with the minimum squared difference $(E_{clust} - E'_e(\theta_{clust}))^2$ between the cluster energy sum and the expected energy of an elastically scattered electron at the measured scattering angle $\theta_{clust}$, after filtering the clusters through several additonal criteria, as detailed in~\cite{Puckett:2015soa}. 

\begin{figure}
  \begin{center}
    \includegraphics[width=0.98\columnwidth]{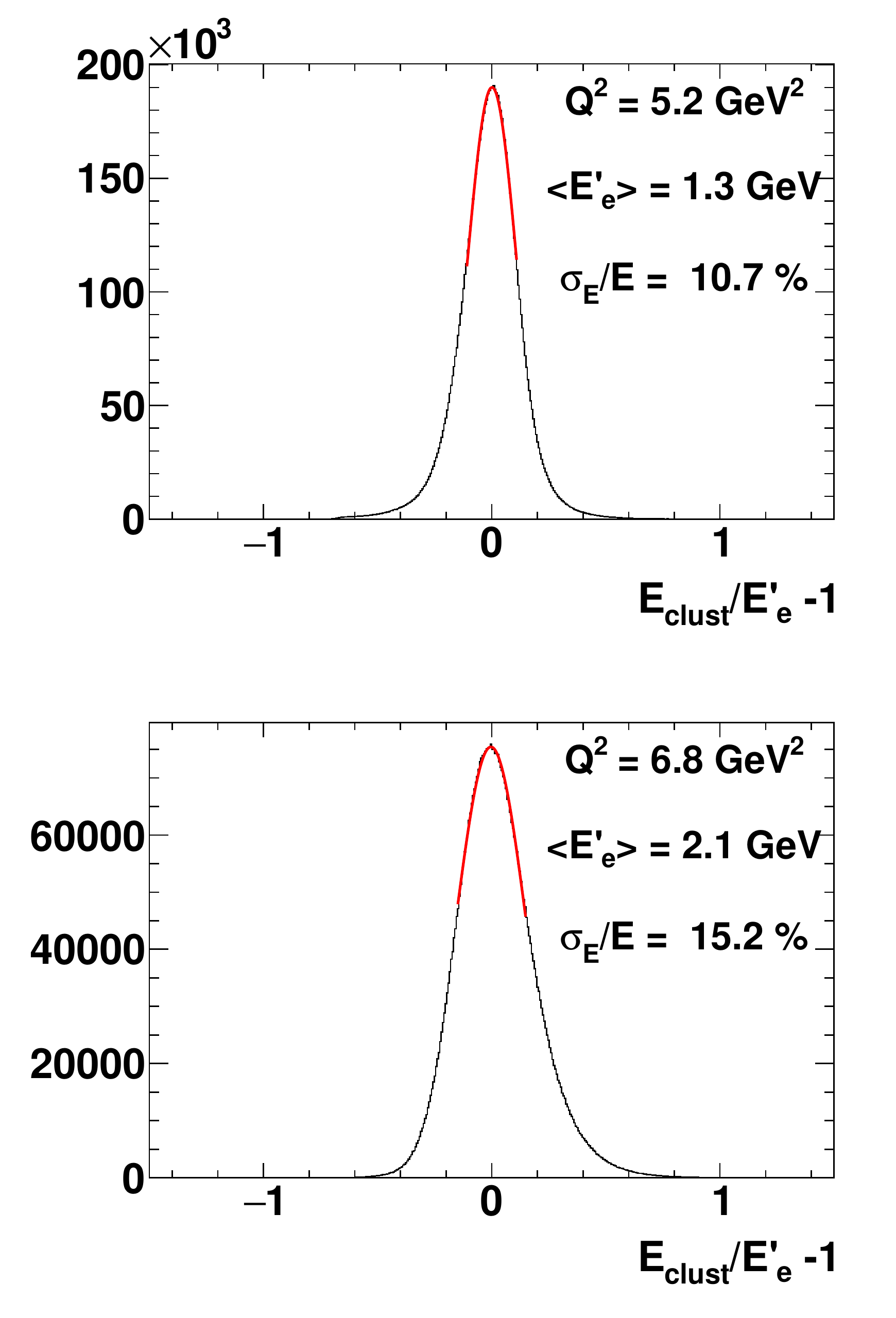}
  \end{center}
  \caption{\label{fig:BigCalEres} Fractional energy resolution of BigCal after calibration, averaged over all elastic events, for the $Q^2 = 5.2$ GeV$^2$ setting (top), collected near the beginning of the experiment, and for the $Q^2 = 6.8$ GeV$^2$ setting (bottom), collected at the end of the experiment. Assuming dominance of the stochastic contribution $\frac{\sigma_E}{E} \propto \frac{1}{\sqrt{E}}$, the scaled energy resolution at 1 GeV worsened from 12\% to 22\% during the experiment. }
\end{figure}
Figure~\ref{fig:BigCalEres} shows the average energy resolution of BigCal achieved using the final calibration database, during the $Q^2 = 5.2$ GeV$^2$ and $Q^2 = 6.8$ GeV$^2$ kinematics, taken at the beginning and the end of the experiment, respectively. Following the initial calibration and gain matching before the start of production data taking, the energy resolution of BigCal with the 4-inch thick aluminum absorber in place was 10.4\% at 1.1 GeV, compared to an expected resolution of $\sim 9$\% from Monte Carlo simulations. The difference is attributable to effects not included in the simulation, including electronics noise, calibration uncertainties, and possible differences in light collection efficiency and PMT quantum efficiency compared to the assumptions used in the simulation. The 4-inch absorber thickness was used for all kinematics except $Q^2 = 2.5$ GeV$^2$, $E'_e \approx 0.54$ GeV, for which a 1-inch thick absorber was used to improve the energy resolution (and trigger efficiency) for the lower-energy electrons. Since the radiation dose rate was much lower at the very large electron scattering angles of this setting, the signal loss rate due to radiation damage was slower than for the other settings, even with a factor of 4 thinner absorber. The average energy resolution scaled to 1 GeV energy with the thinner absorber was 8.0\% (10.9\%) for the data collected at this setting in 2007 (2008), as the two run periods at this setting bookended the two higher-$\epsilon$ kinematics with much higher dose rates. By the end of the experiment, radiation damage had worsened the energy resolution of BigCal by roughly a factor of two relative to the start of the experiment, even after the partial UV curing of the glass during the February-March 2008 accelerator shutdown.

The shower coordinate reconstruction procedure used for the final analysis starts with the calculation of shower ``center of gravity'' coordinates, defined as energy-weighted average block positions in a cluster:
\begin{eqnarray}
  \bar{x} &\equiv& \frac{\sum_{i=1}^{N_{block}}x_i E_i}{\sum_{i=1}^{N_{block}} E_i} \nonumber \\
  \bar{y} &\equiv& \frac{\sum_{i=1}^{N_{block}}y_i E_i}{\sum_{i=1}^{N_{block}} E_i} 
\end{eqnarray}
The electron impact coordinates at the surface are then reconstructed under the assumption that the shower center-of-gravity coordinates are monotonically increasing functions of the electron impact coordinates. The ``true'' electron impact coordinates are assumed to be uniformly distributed within the cell with the largest energy deposition.
\begin{figure}
  \begin{center}
    \includegraphics[width=0.7\columnwidth]{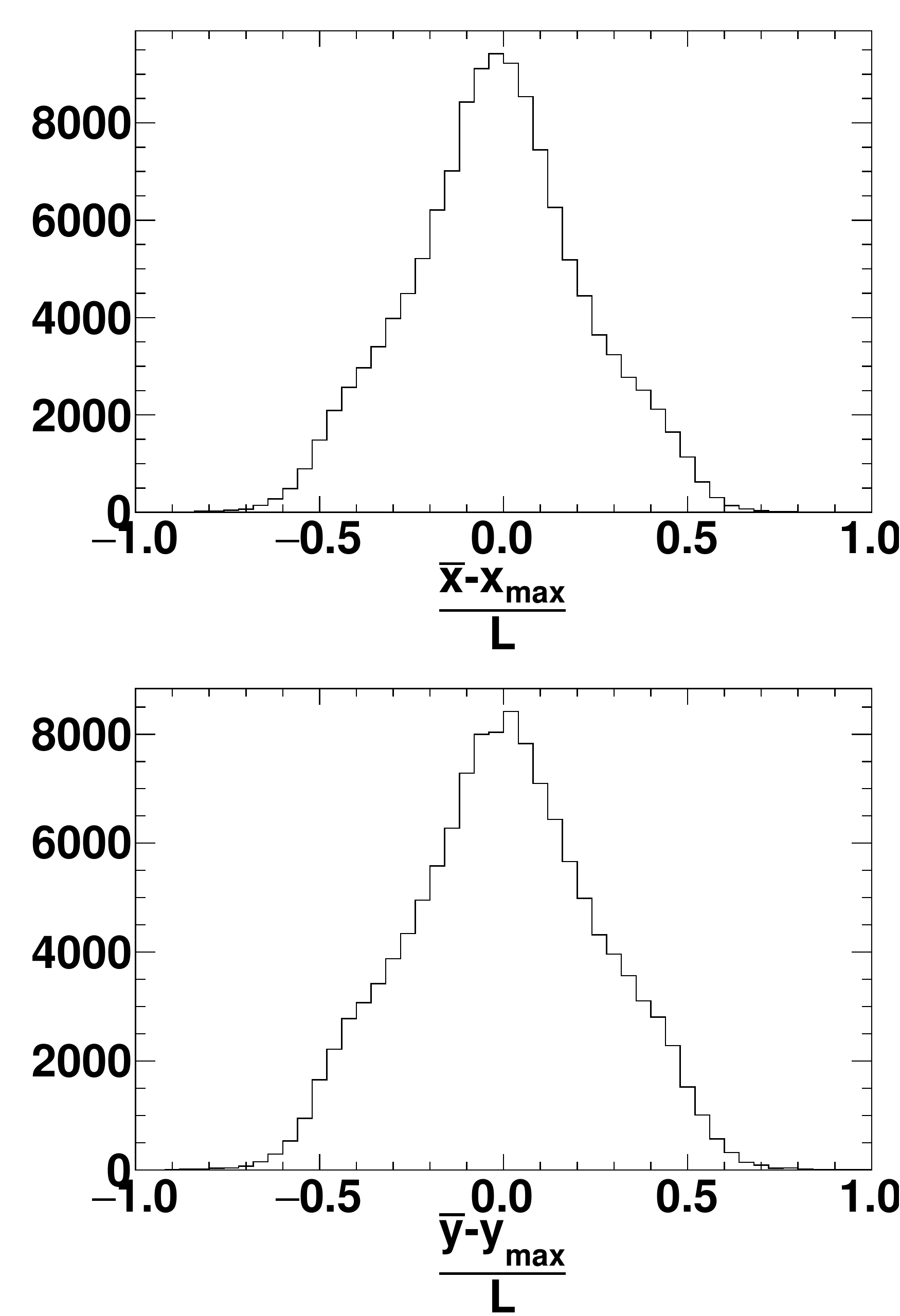}
  \end{center}
  \caption{\label{fig:showerprofile} Difference between horizontal (top) and vertical (bottom) shower center of gravity coordinates $(\bar{x}, \bar{y})$ and the coordinates $(x_{max}, y_{max})$ of the center of the cell with maximum energy deposition, divided by the block transverse size $L_x = L_y \equiv L$. The distributions are averaged over the entire surface of the calorimeter, for the $Q^2 = 6.8$ GeV$^2$, $E'_e = 2.1$ GeV kinematics. See text for details.}
\end{figure}
Figure~\ref{fig:showerprofile} shows a representative example of the distributions in the $x$ (horizontal) and $y$ (vertical) directions of the difference between the shower center-of-gravity coordinates $(\bar{x},\bar{y})$ and the coordinates of the cell with maximum energy deposition $(x_{max},y_{max})$, normalized to the cell size $L$. Both distributions are peaked near zero, and mostly contained within $\pm L/2$. The shape of the distribution reflects the fact that the energy deposition-weighted average block position tends to overweight the central maximum. The central block contains a larger fraction of the total shower energy when the electron impacts near the center of the block than when it impacts closer to the edge of a block, sharing more of the total shower energy with neighboring blocks. 

BigCal was divided into four horizontal sectors and seven vertical sectors, and the distributions of Fig.~\ref{fig:showerprofile} were formed separately for elastically scattered electron clusters in each sector for each kinematic setting. The division into sectors is required because the shape of the shower profiles varies across the surface of BigCal, becoming wider and more asymmetric near the horizontal and vertical extremes of the calorimeter surface due to the variation of the average incident angle of the electron trajectory relative to the surface normal. Within each sector, the distributions of $(\bar{x} - x_{max})$ and $(\bar{y} - y_{max})$ were mapped onto uniform distributions within the cell with the largest energy deposition. The assumption that the shower impact coordinate is uniformly distributed within the cell with the largest energy deposition is a reasonably good approximation, according to Monte Carlo simulations of electromagnetic showers in BigCal, but is violated with larger probability by tracks with large incident angles.   

A position and energy-dependent correction was applied to the shower coordinates resulting from the aforementioned procedure to account for the average incident angles of the electron trajectory, under the assumption that the transverse displacement of the shower maximum with respect to its impact coordinates at the surface of BigCal is proportional to the transverse displacement of the point of maximum energy deposition along the primary electron's trajectory in the lead-glass, with the constant of proportionality fixed by the results of detailed Monte Carlo simulations of BigCal. More details of the coordinate reconstruction procedure can be found in~\cite{Puckett:2015soa}. The ``ideal'' coordinate resolution of BigCal predicted by the Monte Carlo simulation, which again does not include the effects of electronics noise and calibration uncertainties, was $\sigma_{x,y} \approx 0.54\mbox{ cm}/\sqrt{E (\mbox{GeV})}$, using the coordinate reconstruction procedure described above. While the chosen coordinate reconstruction procedure is not unique, the achieved resolution in Monte Carlo is close to the intrinsic limiting resolution of the calorimeter based on fluctuations in shower development, photoelectron statistics, and the transverse size of the blocks relative to that of the shower. A more sophisticated neural-network-based approach to coordinate reconstruction in BigCal was developed for the analysis of the SANE experiment~\cite{Maxwell:2017mhs}, in which electrons were detected over a wider range of energies and angles after deflection in a magnetic field. The neural network approach achieved comparable resolution to the simple approach used in this analysis.

\begin{figure}
  \begin{center}
    \includegraphics[width=0.98\columnwidth]{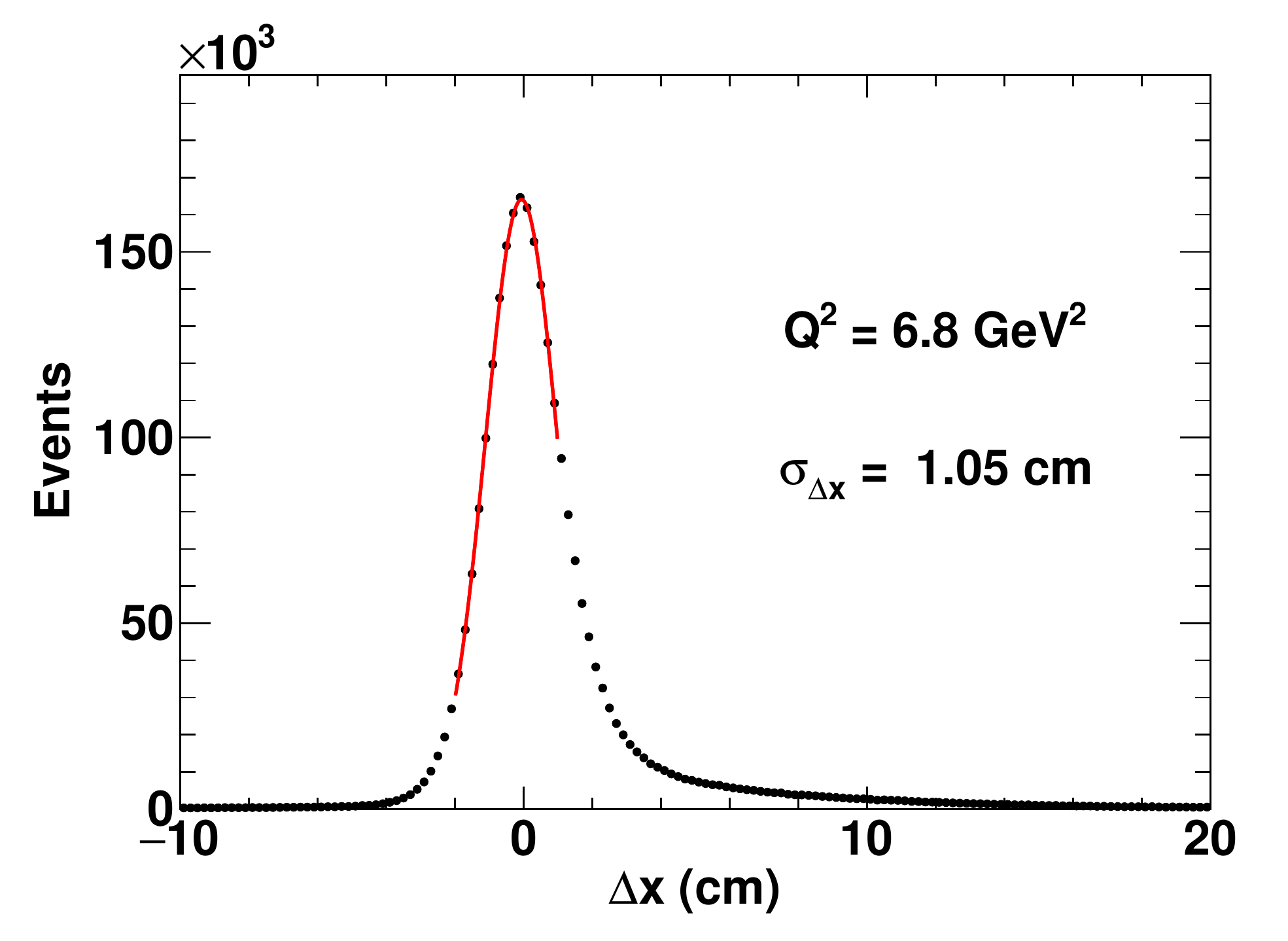}
  \end{center}
  \caption{\label{fig:deltax_q68} Distribution of the difference $\Delta x$ between the horizontal shower coordinate reconstructed in BigCal and the value predicted from the reconstructed proton kinematics assuming elastic scattering, at $Q^2 = 6.8$ GeV$^2$. The red curve is a Gaussian fit to the elastic peak, with a resulting $\sigma$ of 1.05 cm.}
\end{figure}
The experimentally realized coordinate resolution is somewhat difficult to quantify
because the angle, momentum, and vertex resolution of the HMS dominate
the resolution of the exclusivity variables used to select elastic
events. Any estimate of the BigCal coordinate resolution is highly
sensitive to the assumed momentum and vertex resolution of the
HMS. The most favorable kinematic setting to estimate the coordinate
resolution of BigCal was $Q^2 = 6.8$ GeV$^2$. The high proton momentum
$p_p \approx 4.46$ GeV reduced the effect of multiple scattering on
the resolution of the interaction vertex coordinate and the proton
momentum, and the central electron scattering angle of $44.2^\circ$
was relatively favorable in terms of the resolution of the electron
polar scattering angle $\theta_e$ predicted from the measured proton
momentum $p_p$ and the beam energy. Figure~\ref{fig:deltax_q68} shows
the distribution of the difference $\Delta x$ between the
reconstructed horizontal shower coordinate and the value predicted
from the measured proton kinematics, at 6.8 GeV$^2$. The contributions
to the resolution of the measured $\theta_e$ include the coordinate
resolution of BigCal, the vertex resolution of the HMS, and multiple
scattering of the electron in air. Assuming a momentum resolution
$\sigma_p/p$ of 0.1\% and vertex resolution $\sigma_{y_{tar}}$ of 1.7
mm (based on HMS optics calibration data), the coordinate resolution
of BigCal was estimated by subtracting in quadrature the contributions
of $\sigma_p$, $\sigma_{y_{tar}}$, and multiple scattering from the
observed width of the elastic peak in the distribution of $\Delta x$,
the difference between the measured and predicted horizontal shower
coordinates at BigCal. Based on these assumptions, the coordinate
resolution at 2.1 GeV was estimated to be $\sigma_{x} \approx 6$ mm at
the end of the experiment, \emph{after} most of the total radiation
dose had been absorbed. This estimate is consistent with an estimate
based on the $\Delta x$ distribution for the  $Q^2 = 2.5$ GeV$^2$, $E'_e = 2.35$ GeV setting, using identical assumptions for the HMS resolution. For all the other kinematics, the contribution of the BigCal coordinate resolution to the width of the elastic peak in $\Delta x$ was too small for a meaningful estimate of $\sigma_x$. In any case, the coordinate resolution was significantly better than required for a clean selection of elastic events, despite the significant degradation of the energy resolution by radiation damage.

\subsection{BigCal timing and coincidence}
Timing information was not recorded for each individual channel of BigCal. Instead, copies of the 224 ``first-level'' sums of (up to) eight channels, that were subsequently combined in the ``second-level'' sums of (up to) 64 channels used to define the BigCal trigger (see Ref.~\cite{Puckett:2015soa} for details), were sent to discriminators and then to LeCroy Fastbus 1877 model TDCs. The fixed discriminator threshold applied to the ``first-level'' sums was equivalent to about 100 MeV of energy deposition. For all first-level sums with TDC hits, a corrected hit time was computed by subtracting a constant zero offset and applying a time-walk correction based on the sum of all recorded ADC values in the channels corresponding to that sum. A ``cluster time'' was computed for each individual cluster as the energy-weighted average corrected hit time of all unique first-level sums with a TDC hit containing ADC channels included in the cluster. For clusters with good timing information, the achieved timing resolution of BigCal was approximately 1.5 ns, as determined by the width of the real coincidence peak in the time difference between HMS and BigCal after correcting for variations in particle time-of-flight within the acceptance of each detector. The contamination of elastic events by random coincidences was found to be negligible after applying the exclusivity cuts described in Ref.~\cite{Puckett:2017flj} and Section~\ref{sec:elastic}. 

A small fraction of the first-level sums failed to produce reliable timing information during a significant fraction of the experiment, due to malfunctions in the electronics chain involving either individual discriminator channels, summing modules or TDC channels. While these malfunctions did not affect the individual ADC signals from the BigCal PMTs, the second-level sums, or the BigCal trigger, they did affect the BigCal timing information for roughly 2\% of clusters identified as elastic by their angular correlations with elastically scattered protons detected by the HMS. Analysis of the distributions of the exclusivity variables used to select elastic events showed no significant differences between events with good timing information for the chosen cluster and those without first-level timing information. Nonetheless, a loose cut of $\left|\Delta t\right| \le 10$ ns was applied to the time-of-flight-corrected HMS-BigCal time difference in order to minimize the systematic uncertainty associated with the subtraction of the inelastic background asymmetry from that of the elastic signal. 

\begin{figure}
  \begin{center}
    \includegraphics[width=0.98\columnwidth]{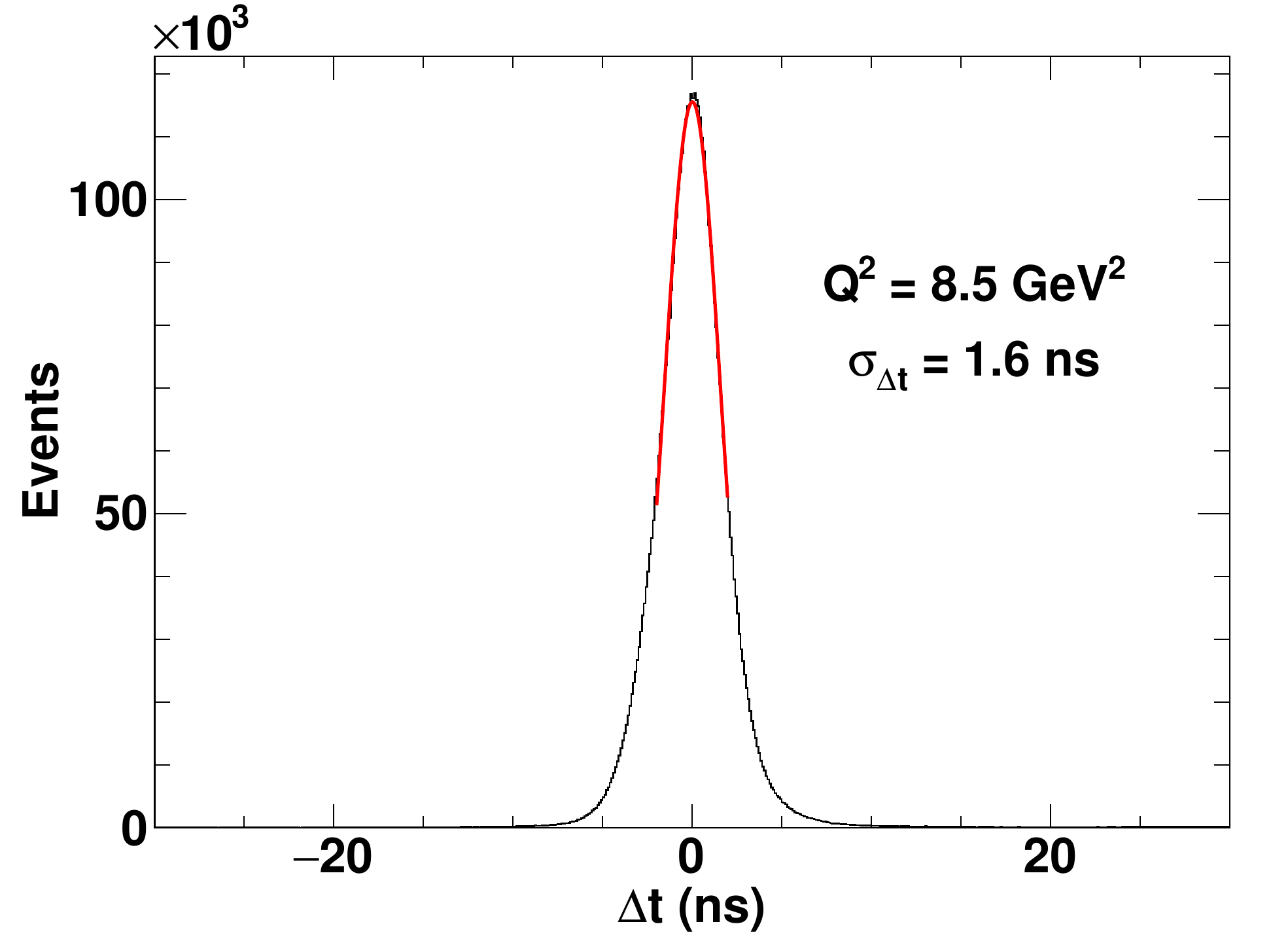}
  \end{center}
  \caption{\label{fig:deltat} Time-of-flight corrected difference $\Delta t$ between HMS and BigCal timing signals for $Q^2 = 8.5$ GeV$^2$, for events identified as elastic. The timing resolution in this example is $\sigma_{\Delta t} \approx 1.6$ ns, from a Gaussian fit represented by the red curve. A loose cut of $\left|\Delta t\right| \le 10$ ns was applied for all kinematics. }
\end{figure}
Figure~\ref{fig:deltat} shows the distribution of the difference $\Delta t$ between the HMS and BigCal timing signals for events identified as elastic at $Q^2 = 8.5$ GeV$^2$, the measurement with the largest backgrounds from inelastic processes and accidental coincidences, after applying all other exclusivity cuts. Given the relatively poor timing resolution of BigCal (and the lack of valid timing information for a small fraction of the BigCal sums-of-eight), the coincidence timing resolution was not, generally speaking, sufficient to resolve the 2-ns beam-bunch structure of CEBAF. As described in Ref.~\cite{Puckett:2015soa}, the arrival timing of the HMS and BigCal triggers at the logic unit used to define the coincidence trigger was configured such that the BigCal trigger arrived approximately 20-30 ns before the HMS trigger for elastic $ep$ events, near the center of the 50-ns coincidence window defined by the width of the HMS trigger logic pulse. This arrangement ensured that the HMS trigger, with timing resolution on the order of a few hundred picoseconds, defined the timing of the coincidence trigger for real elastic $ep$ events, as well as the timing of the gate/start/stop signals sent to the readout electronics for all detectors, including BigCal. 

The CEBAF bunch length during the GEp-III/GEp-2$\gamma$ experiments
was approximately 0.3 ps~\cite{Leemann:2001dg,Chao:2011za}, which is
negligible compared to the timing resolution of any detector in Hall
C. A standard practice in coincidence measurements in Halls A and C
involving two precision spectrometers is to record a timing signal in
each spectrometer that is locked to a subharmonic of the accelerator
RF frequency of 499 MHz. This provides a reference time that is fixed relative to the event start time (the time at which the beam bunch responsible for the interaction crossed the target) that can be used for precise calibration of the absolute time-of-flight of the particles and the relative timing between the two spectrometers. The RF timing signals were not used in this analysis, however, because the timing resolution of BigCal was too poor for the RF corrections to the timing signals to meaningfully improve the suppression of accidentals. Indeed, the loose timing cut provides only marginal additional suppression of accidental coincidences relative to the exclusivity cuts defined in terms of the reconstructed particle kinematics.

\section{Additional details of elastic event selection procedure}
\label{sec:elastic}
As detailed in Ref.~\cite{Puckett:2017flj}, the definitions of the exclusivity cut variables used to select elastic events are:
\begin{enumerate}
\item $\delta p_p \equiv 100 \times \frac{p_p-p_p(\theta_p)}{p_0}$ is the difference between the measured proton momentum and the expected momentum of an elastically scattered proton at the measured $\theta_p$, expressed as a percentage of the HMS central momentum. This quantity depends only on the measured proton kinematics.
\item $\delta p_e \equiv 100 \times \frac{p_p-p_p(\theta_e)}{p_0}$ is defined the same way as $\delta p_p$, except in this case the expected proton momentum is computed using the measured \emph{electron} scattering angle $\theta_e$. 
\item $\delta \phi \equiv \phi_e - \phi_p - \pi$ is the ``acoplanarity'' defined in terms of the measured azimuthal scattering angles of the electron and proton. 
\end{enumerate}

\begin{figure*}
  \begin{center}
    \includegraphics[width=0.75\textwidth]{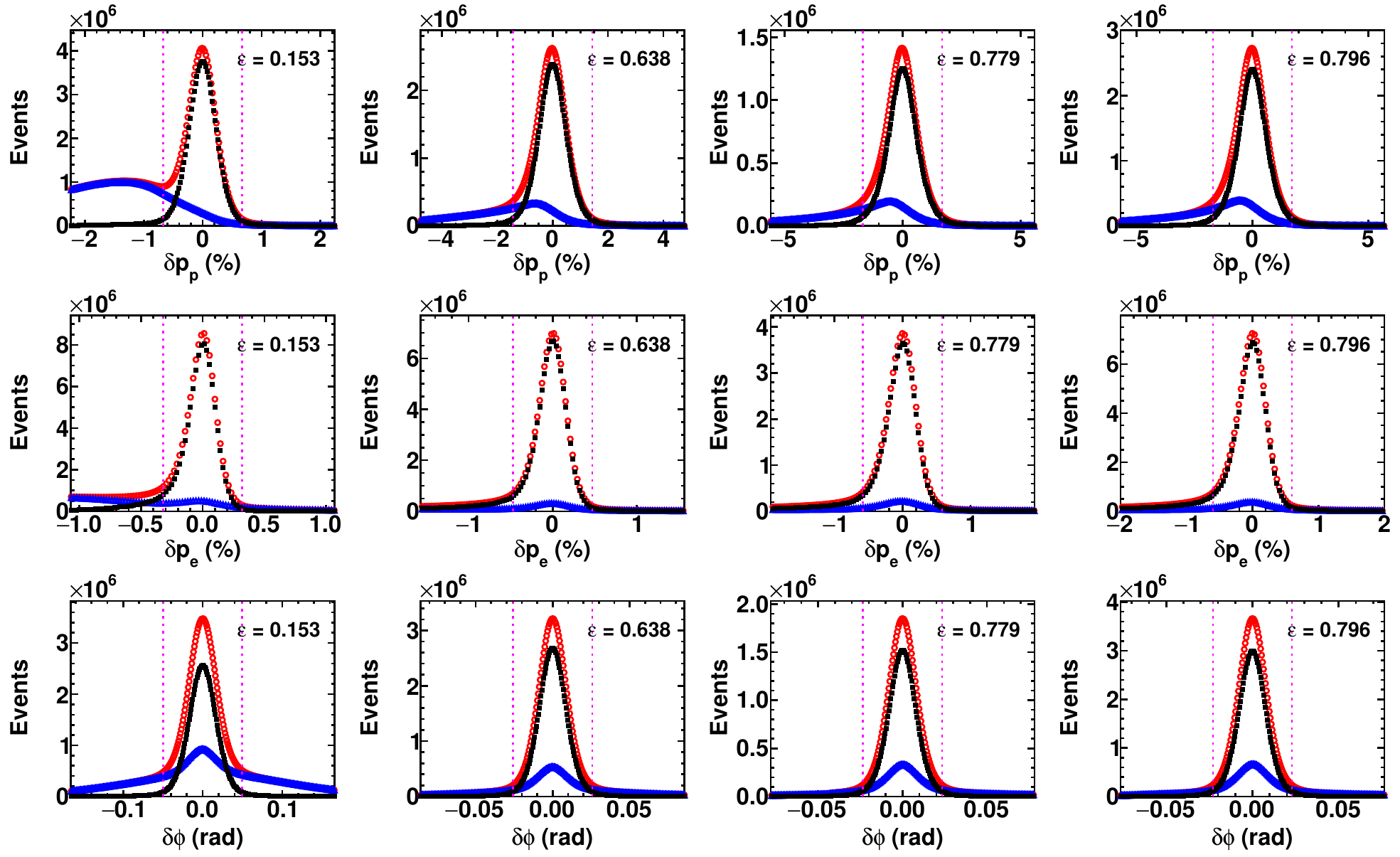}
  \end{center}
  \caption{\label{fig:exclusivitycuts_gep2gamma} Simplified illustration of elastic event selection for the GEp-2$\gamma$ kinematics at $Q^2 = 2.5$ GeV$^2$: $\left<\epsilon\right> = 0.153$ (left column), $\left<\epsilon\right> = 0.638$ (middle left column), $\left<\epsilon\right> = 0.779$ ($E_e = 3.548$ GeV, middle right column), $\left<\epsilon\right> = 0.796$ ($E_e = 3.680$ GeV, right column). Exclusivity cut variables are $\delta p_p \equiv 100 \times \frac{p_p - p_p(\theta_p)}{p_0}$ (top row), $\delta p_e \equiv 100 \times \frac{p_p - p_p(\theta_e)}{p_0}$ (middle row), and $\delta \phi \equiv \phi_e - \phi_p - \pi$ (bottom row). The distribution of each variable is shown for all events (red empty circles), events selected by applying $\pm 3\sigma$ cuts to \emph{both} of the other two variables (black filled squares), and events rejected by these cuts (blue empty triangles). Vertical dotted lines indicate the $\pm 3\sigma$ cut applied to each variable. Similar plots for the GEp-III kinematics can be found in the main publication~\cite{Puckett:2017flj}.}
\end{figure*}
Figure~\ref{fig:exclusivitycuts_gep2gamma} shows the same simplified illustration of the elastic event selection procedure for the GEp-2$\gamma$ kinematics as shown for the GEp-III kinematics in the main publication~\cite{Puckett:2017flj}. In contrast to the GEp-III case, the inelastic background levels in the vicinity of the elastic peak in the GEp-2$\gamma$ data are low even before applying exclusivity cuts, and are extremely low after applying the cuts. In fact, only the lowest-$\epsilon$ point has significant inelastic contamination after the cuts. As in the GEp-III case, the signal-to-background ratio before cuts is highest for the $\delta p_e$ distribution. 

The application of fixed-width, $\pm 3\sigma$ cuts centered at zero yields an efficient selection of elastic events with small inelastic contamination. However, it was found that the efficiency and signal-to-background ratio of the $\delta p_p$ and $\delta \phi$ cuts could be improved by applying variable cuts that account for observed variations of the width and/or position of the elastic peak within the acceptance. 
The correlations of $\delta p_p$, $\delta p_e$, and $\delta \phi$ with all reconstructed parameters of the proton trajectory (and the reconstructed electron angles) were examined for all kinematics, in order to verify the quality of the reconstruction and to optimize the event selection cuts. 
\begin{figure*}
  \begin{center}
    \includegraphics[width=0.75\textwidth]{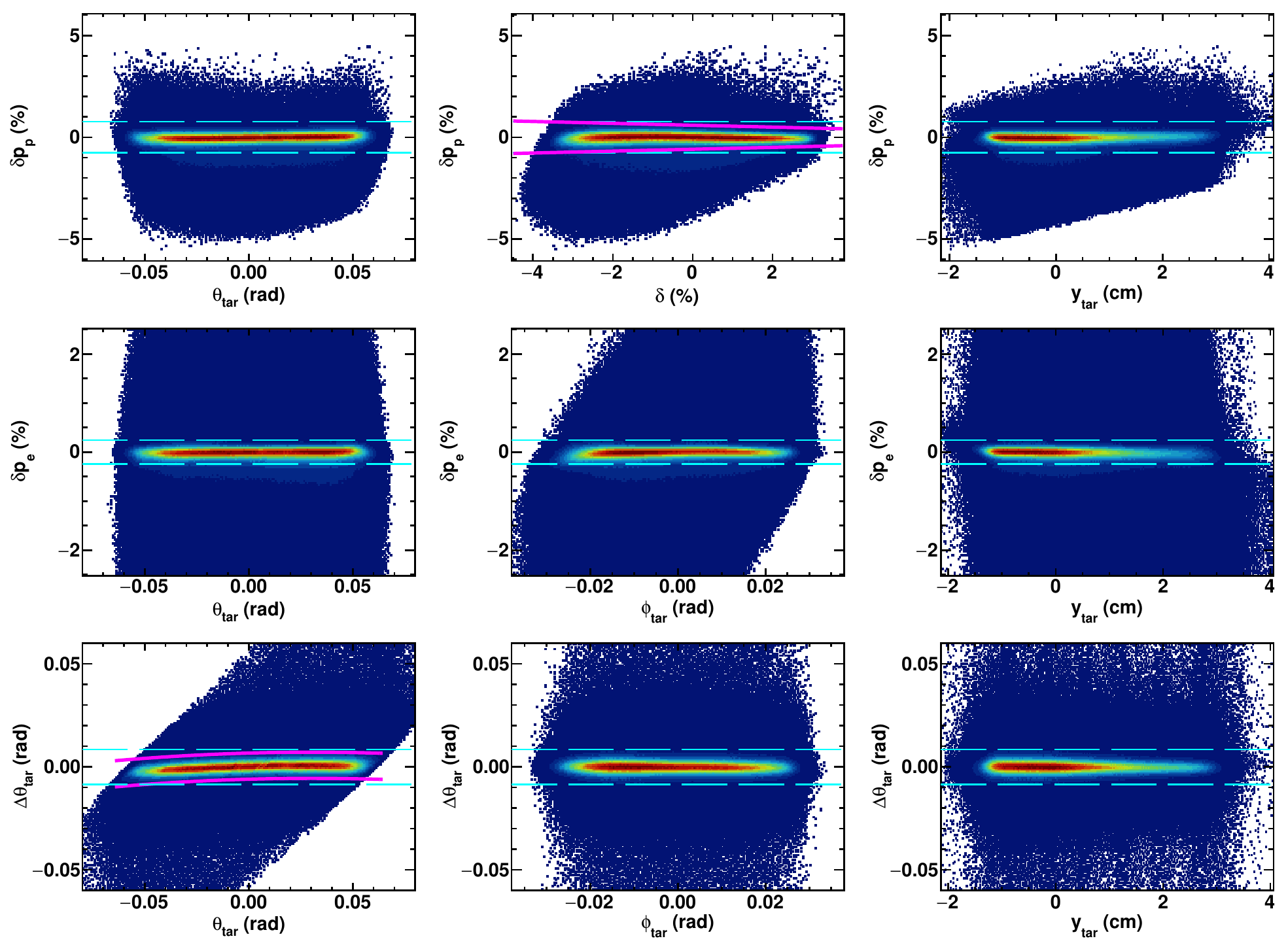}
  \end{center}
\caption{\label{fig:cutkine_q85} Correlations of exclusivity cut variables with reconstructed proton trajectory parameters for $Q^2 = 8.5$ GeV$^2$. $\delta p_p$ (top row), $\delta p_e$ (middle row), and $\Delta \theta_{tar}$ (bottom row) are shown as a function of $\theta_{tar}$ (left column), $\phi_{tar}$ or $\delta$ (middle column), and $y_{tar}$ (right column). The $\phi_{tar}$ and $\delta$ dependences are redundant due to the proton angle-momentum correlation in elastic $ep$ scattering. Pink solid curves represent variable $\pm 3\sigma$ cuts. Cyan dashed horizontal lines represent fixed, $\pm 4\sigma_{avg}$ cuts for $\delta p_p$ and $\Delta \theta_{tar}$, and $\pm 3\sigma$ cuts for $\delta p_e$. See text for details.}
\end{figure*}
Figure~\ref{fig:cutkine_q85} shows a typical example of the correlations of $\delta p_p$, $\delta p_e$, and $\Delta \theta_{tar} \equiv \theta_{tar} - \theta_{tar}(\theta_e, \phi_e)$ with the reconstructed proton kinematics, for $Q^2 = 8.5$ GeV$^2$. Recall that $\theta_{tar} \equiv \arctan x'_{tar}$ is the reconstructed dispersive-plane trajectory angle of the scattered proton as it enters the HMS. In first approximation, $\delta \phi \approx -\Delta \theta_{tar}/\sin(\Theta)$, with $\Theta$ the central angle of the HMS. In Figures~\ref{fig:cutkine_q85}, \ref{fig:varcuts_gep3}, and \ref{fig:varcuts_gep2g}, $\Delta \theta_{tar}$ is shown instead of $\delta \phi$, because the experimental resolution of $\Delta \theta_{tar}$ is dominated by the resolution of $\theta_{tar}$, which depends only on the HMS central momentum setting and not the HMS central angle, whereas $\delta \phi$ also varies strongly with the HMS central angle. The use of $\Delta \theta_{tar}$ instead of $\delta \phi$ also allows a more direct comparison between the size of the observed correlation and the systematic uncertainty assigned to $\theta_{tar}$. 

The deviations from zero of the elastic peak positions in the distributions of the exclusivity cut variables are uncorrelated with the reconstructed proton kinematics, to within the estimated systematic uncertainties. However, two significant effects were observed motivating the use of variable exclusivity cuts for $\delta p_p$ and $\delta \phi$, as shown in Figures~\ref{fig:varcuts_gep3} and \ref{fig:varcuts_gep2g}.
\begin{figure*}
  \begin{center}
    \includegraphics[width=0.9\textwidth]{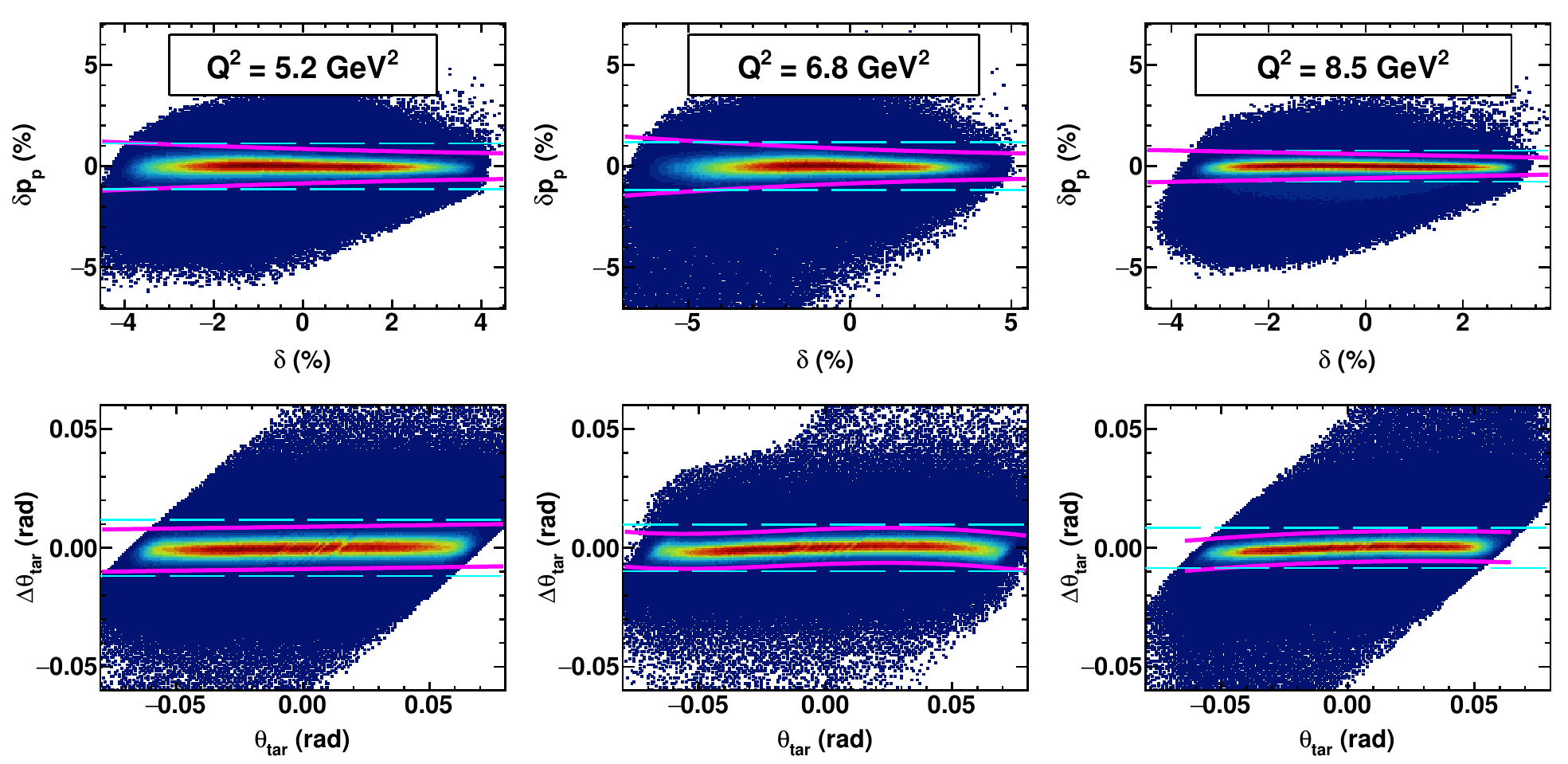}
  \end{center}
\caption{\label{fig:varcuts_gep3} Variable cuts for $\delta p_p$ as a function of $\delta$ (top row) and $\Delta \theta_{tar}$ as a function of $\theta_{tar}$ (bottom row) for the GEp-III kinematics. Pink solid curves represent variable, $\pm 3\sigma$ cuts, while the cyan dashed horizontal lines represent fixed-width, $\pm 4\sigma_{avg}$ cuts, with $\sigma_{avg}$ being the acceptance-averaged elastic peak width for the variable in question. The tighter of the two cuts is used throughout the acceptance.}
\end{figure*}
\begin{figure*}
  \begin{center}
    \includegraphics[width=0.9\textwidth]{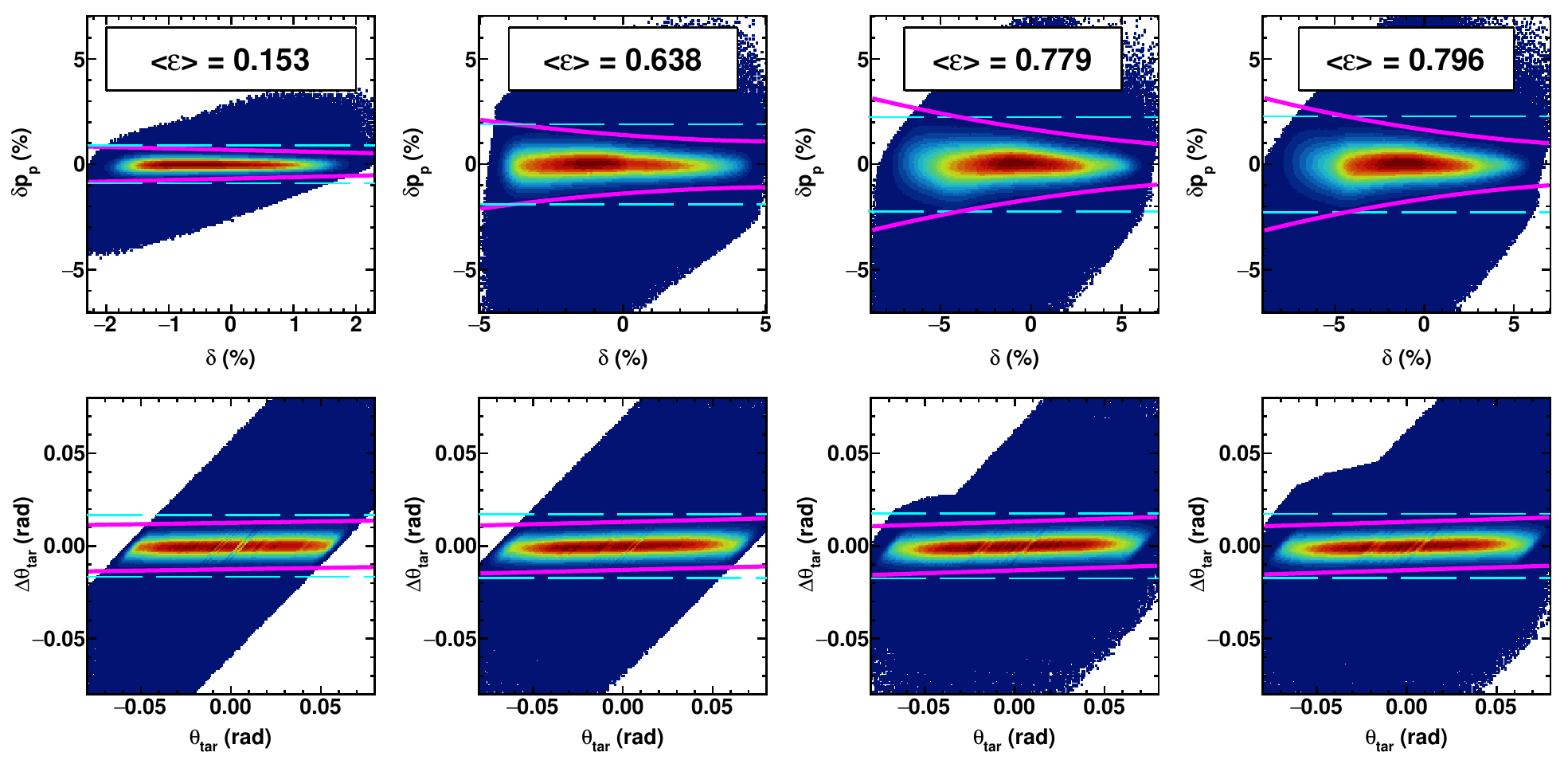}
  \end{center}
\caption{\label{fig:varcuts_gep2g} Variable cuts for $\delta p_p$ as a function of $\delta$ (top row) and $\Delta \theta_{tar}$ as a function of $\theta_{tar}$ (bottom row) for $Q^2 = 2.5$ GeV$^2$. Pink solid curves represent variable, $\pm 3\sigma$ cuts, while the cyan dashed horizontal lines represent fixed-width, $\pm 4\sigma_{avg}$ cuts, with $\sigma_{avg}$ being the acceptance-averaged elastic peak width for the variable in question. The tighter of the two cuts is used throughout the acceptance.}
\end{figure*}
First, the resolution of $\delta p_p$ varies significantly as a function of the proton momentum within the HMS acceptance, by more than a factor of two at $(Q^2,E_e) = (2.5\text{ GeV}^2, 3.68\text{ GeV})$, the setting with the largest $\delta$ acceptance. Second, the elastic peak position in the distribution of the ``acoplanarity'' $\delta \phi$, or, equivalently, $\Delta \theta_{tar}$, exhibits small correlations with $\theta_{tar}$. 

The observed $\delta$-dependence of the resolution of $\delta p_p$ is a combined effect of the intrinsic optical characteristics of the HMS, the reaction kinematics, and the exaggerated effect of multiple-scattering in ``S0'' on the HMS angular resolution, and is qualitatively similar for all six kinematics. The observed correlation between $\delta \phi$ (or, equivalently, $\Delta \theta_{tar}$) and $\theta_{tar}$ is a combined effect of all $\theta_{tar}$-dependent systematic errors in the reconstructed azimuthal angles $\phi_e$ and $\phi_p$. The proton azimuthal angle $\phi_p$ is mainly defined by $\theta_{tar}$, while the electron azimuthal angle $\phi_e$ is mainly defined by the vertical shower coordinate in BigCal, denoted $y_{clust}$. There are several different sources of systematic uncertainty in the reconstruction of $\theta_{tar}$ and/or $y_{clust}$ that can produce the observed correlations on their own or in combination. These include errors in the beam position on target, uncertainties in the HMS optics calibration, misalignments of BigCal, and $y_{clust}$-dependent distortions of the shower shape in BigCal that are not fully accounted for by the coordinate reconstruction procedure, largely due to the non-zero incident angle of the electron trajectory. Although the incident-angle distortion of the shower shape was corrected using an approximate formula based on Monte Carlo simulations~\cite{Puckett:2015soa}, no attempt was made to optimize the parameters of this correction using the real data, as it was not possible, generally speaking, to isolate this effect from other possible systematics affecting the polar and/or azimuthal angle correlation between the scattered electron and proton.

The magnitude of the observed deviation from zero of the elastic peak position in $\Delta \theta_{tar}$ does not exceed $2$ mrad anywhere within the limits of the $\theta_{tar}$ acceptance for any of the kinematics. For comparison, the global systematic uncertainty assigned to $\theta_{tar}$ in the evaluation of the systematic uncertainties in $R = \mu_p G_E^p/G_M^p$ and $P_\ell$ is $\pm 2.4$ mrad (see section~\ref{sec:systematics}). The slight non-linearity of the observed correlation between $\Delta \theta_{tar}$ and $\theta_{tar}$ for the measurements at $Q^2 = 6.8$ and 8.5 GeV$^2$ is caused by a vertical beam position offset of approximately $3$ mm above the HMS optical axis during the data collection with 5.71 GeV beam energy that distorts the $\theta_{tar}$ reconstruction in a non-linear fashion. This offset resulted from the procedure used to center the beam on target during this run period. The HMS optics calibration data were obtained with a beam position that was vertically centered with respect to the HMS, and approximately 3 mm below the beam position used during the high-$Q^2$ running. According to the HMS COSY model, the first-order sensitivity of $\theta_{tar}$ to a vertical beam offset for the HMS standard tune is about 1.1 mrad/mm. However, as described in Section~\ref{subsubsec:HMSoptics}, the higher-order $x_{tar}$-dependent matrix elements are taken at face value from the HMS COSY model and are not independently calibrated. Instead, their effect is absorbed into the calibration of the $x_{tar}$-independent matrix elements for the reconstruction of $\theta_{tar}$. Therefore, it is largely unsurprising that the reconstruction of $\theta_{tar}$ exhibits small, nonlinear distortions for a vertical beam position offset of this magnitude, given that no independent optimization of the $x_{tar}$-dependent HMS matrix elements exists.

For the final analysis, as shown in Figs.~\ref{fig:varcuts_gep3} and \ref{fig:varcuts_gep2g}, variable, $\pm 3\sigma$ cuts were applied to $\delta p_p$ as a function of $\delta$ and to $\delta \phi$ as a function of $\theta_{tar}$, up to a maximum of $\pm 4\sigma_{avg}$, with $\sigma_{avg}$ being the acceptance-averaged elastic peak width. In addition to optimizing the efficiency and purity of the elastic event selection, the application of variable cuts minimizes the potential cut-induced bias in the reconstructed proton kinematics. 
The relevance of such a bias is that the reconstructed parameters of the proton's trajectory at the target are the inputs to the calculation of the proton spin transport through the HMS. As discussed in Section~\ref{sec:systematics}, the ratio $P_t/P_\ell$ is highly sensitive to the proton's non-dispersive-plane (horizontal) trajectory angle $\phi_{tar}$, while the longitudinal polarization transfer component $P_\ell$ is more sensitive to the dispersive-plane (vertical) trajectory angle $\theta_{tar}$. The experimental resolutions $(\sigma_{\phi}, \sigma_{\theta})$ in $\phi_{tar}$ and $\theta_{tar}$, which are dominated by the effects of multiple-scattering in ``S0'' for most kinematics, ranged from $(3.5$ mrad$, 4.6$ mrad$)$ at $Q^2 = 2.5$ GeV$^2$ to $(1.9$ mrad$,2.7$ mrad$)$ at $Q^2 = 8.5$ GeV$^2$. For comparison, at $Q^2 = 8.5$ GeV$^2$, the first-order sensitivity $dR/d\phi_{tar} = -0.1/$mrad. 

Whereas the resolution of $\delta p_p$ ($\delta \phi$) is dominated by $\sigma_\phi$ ($\sigma_\theta$), the resolution of $\delta p_e$ is dominated by the HMS momentum resolution $\sigma_\delta$ and, to a lesser extent, the vertex resolution $\sigma_{y_{tar}}$, neither of which varies strongly within the HMS acceptance. Since the resolution of $\delta p_e$ is approximately constant throughout the HMS acceptance, and since the polarization transfer observables are less sensitive to the systematic errors in $\delta$ and $y_{tar}$ than those in $\theta_{tar}$ and $\phi_{tar}$, the use of fixed-width, $\pm 3\sigma_{avg}$ cuts for $\delta p_e$, which greatly simplifies the estimation of the residual inelastic contamination, was deemed appropriate. 
\begin{figure}
  \begin{center}
    \includegraphics[width=0.98\columnwidth]{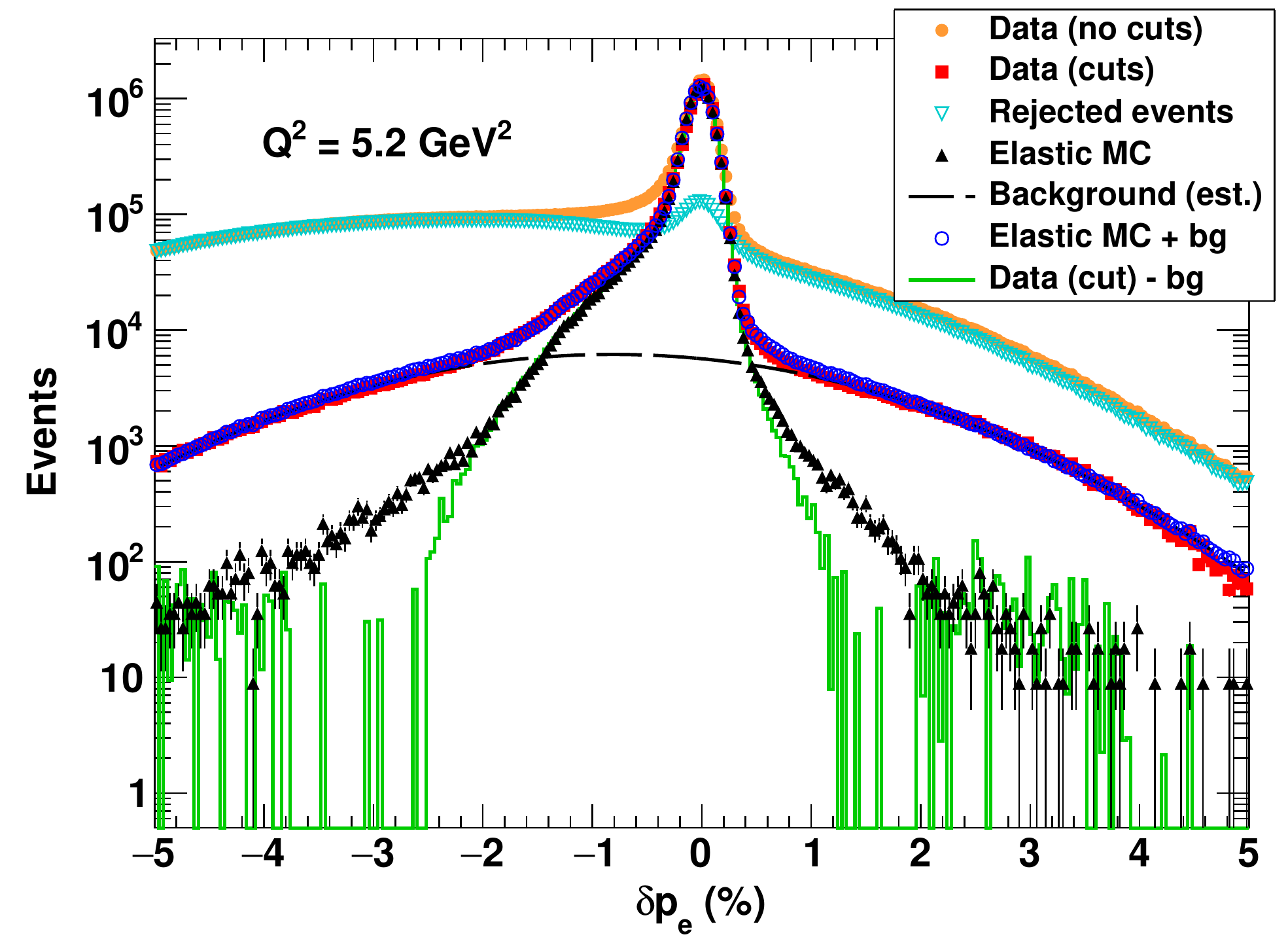}
    \includegraphics[width=0.98\columnwidth]{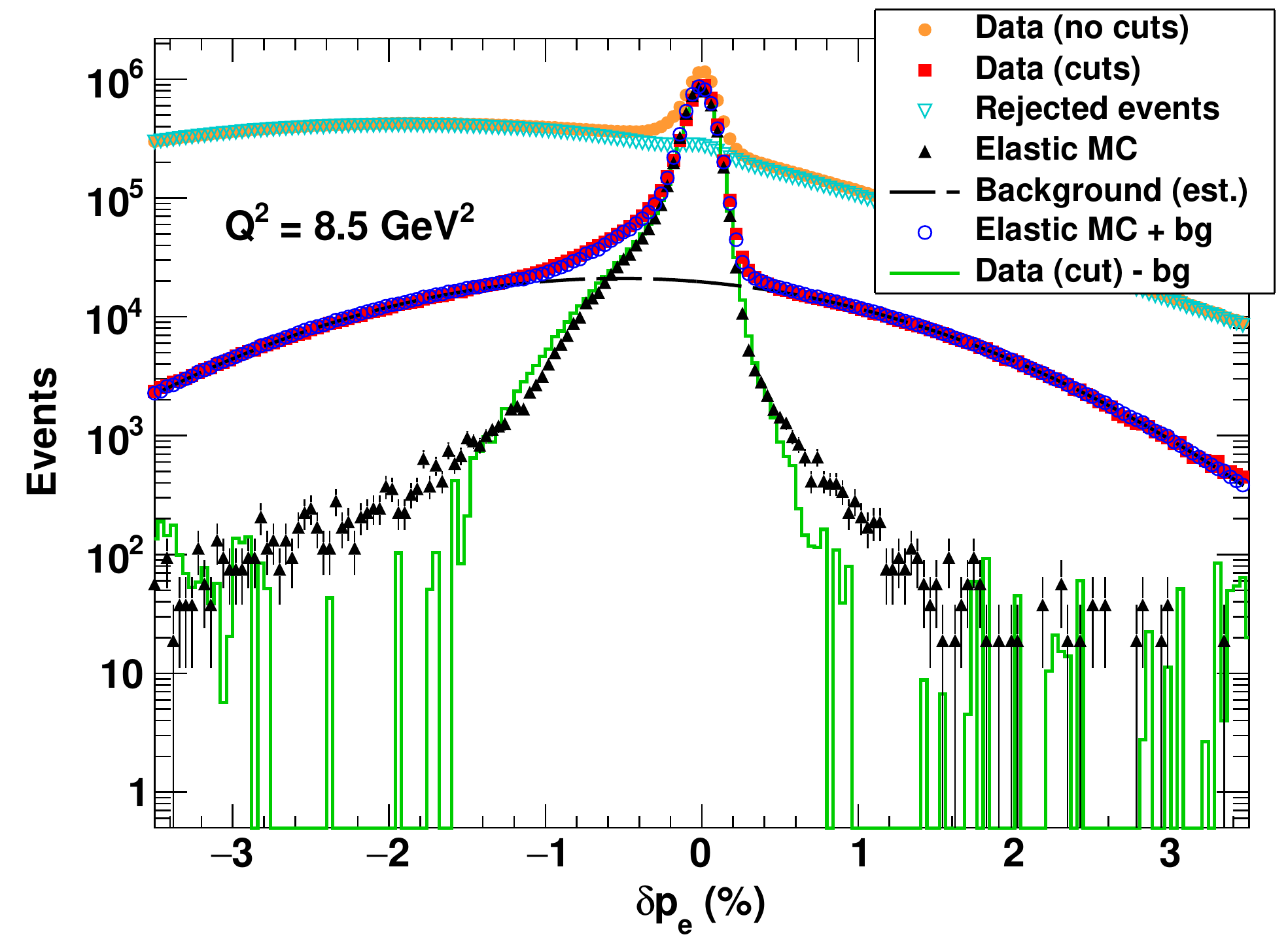}
  \end{center}
\caption{\label{fig:simdatacompare_dpe_q52_q85} $\delta p_e$ spectra at $Q^2 = 5.2$ GeV$^2$ (top) and $Q^2 = 8.5$ GeV$^2$ (bottom). Orange filled circles show the $\delta p_e$ distribution of all events. Red filled squares show the $\delta p_e$ distribution after applying the $\delta p_p$ and $\delta \phi$ cuts. Cyan empty triangles show the distribution of events rejected by the cuts. The dashed curve shows the inelastic background remaining after cuts, estimated using the Gaussian sideband method described in Ref.~\cite{Puckett:2017flj}. Black solid triangles show the simulated $\delta p_e$ distribution, including radiative effects, of elastic $ep$ events passing all exclusivity cuts other than $\delta p_e$. Blue empty circles show the sum of the simulated elastic events and the estimated background, while the green solid line shows the background-subtracted $\delta p_e$ distribution of events passsing the cuts.}
\end{figure}
Figure~\ref{fig:simdatacompare_dpe_q52_q85} compares the background-subtracted $\delta p_e$ distributions of the data to the simulated $\delta p_e$ distributions of elastic $ep \rightarrow ep$ scattering events including radiative effects, for $Q^2 = 5.2$ GeV$^2$ and $Q^2 = 8.5$ GeV$^2$, which are representative examples. The simulated elastic events in Fig.~\ref{fig:simdatacompare_dpe_q52_q85} were generated using the radiative Monte Carlo generator ``ESEPP'' described in Ref.~\cite{Gramolin:2014pva}, and convoluted with simplified, parametrized models for the acceptance and resolution of the detectors. The excellent agreement between the simulation and the data over roughly three orders of magnitude in relative yield as a function of $\delta p_e$ supports the validity of the Gaussian sideband method used to estimate the inelastic background, as described in Ref.~\cite{Puckett:2017flj}.

\section{Data quality checks for maximum-likelihood estimators}
\label{sec:dataqualitycheck}
\begin{figure*}
  \begin{center}
    \includegraphics[width=0.7\textwidth]{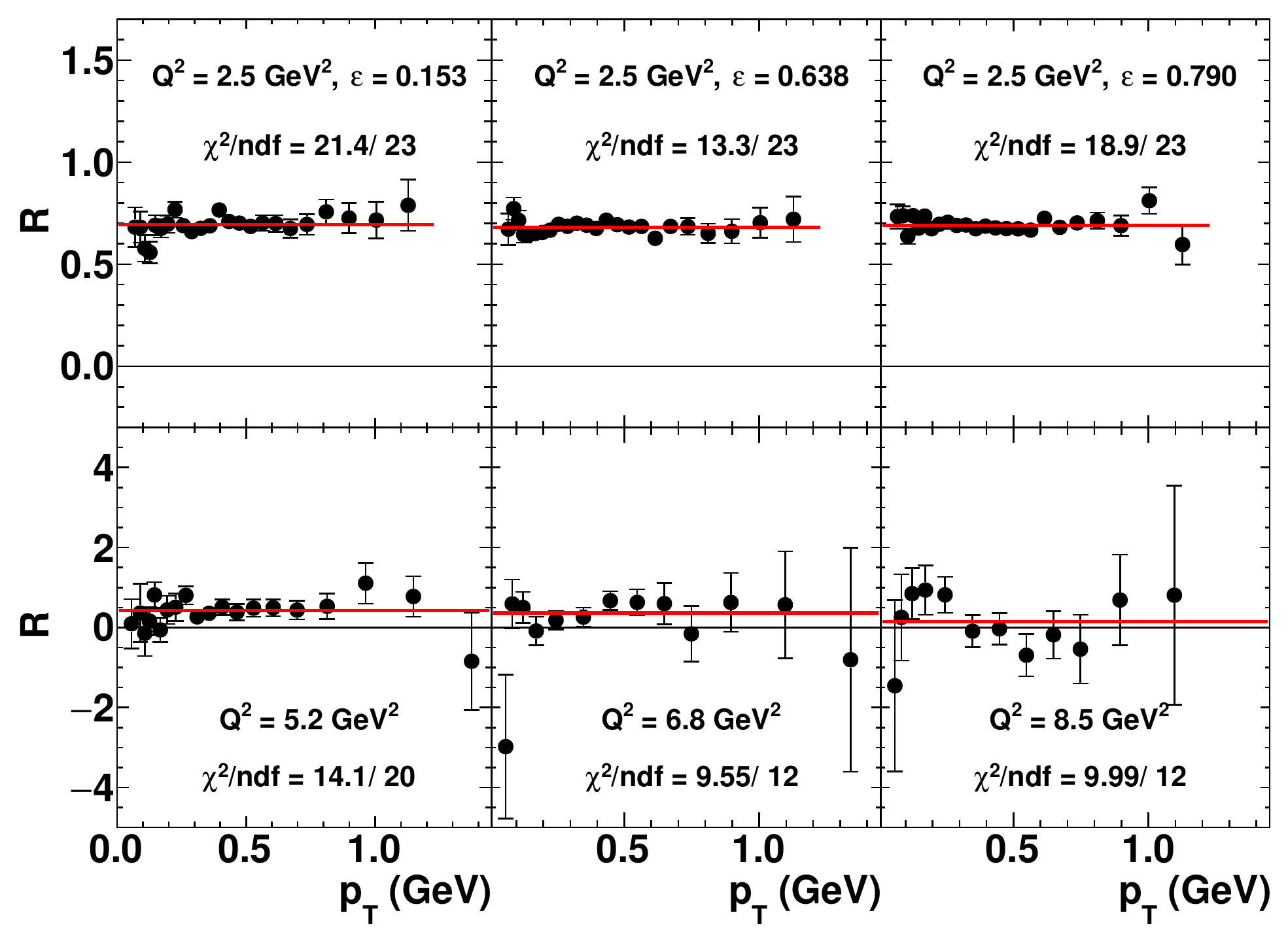}
  \end{center}
  \caption{\label{fig:R_pT} Dependence of the ratio $R \equiv -\mu_p \sqrt{\frac{\tau(1+\epsilon)}{2\epsilon}} \frac{P_t}{P_\ell} = \mu_p \frac{G_E^p}{G_M^p}$ on the ``transverse momentum'' $p_T \equiv p_p \sin(\vartheta_{FPP})$ for the combined data from FPP1 and FPP2, illustrating the cancellation of the analyzing power $A_y(p_T)$ in the ratio $P_t/P_\ell$. Red lines are constant fits to the data. See text for details. }
\end{figure*}
Figure~\ref{fig:R_pT} shows the dependence of the extracted ratio $R \equiv -\mu_p \sqrt{\frac{\tau(1+\epsilon)}{2\epsilon}} \frac{P_t}{P_\ell} \equiv -K P_t/P_\ell$, which equals $\mu_p G_E^p/G_M^p$ in the one-photon-exchange approximation, on the polar scattering angle in the FPP, expressed in terms of $p_T \equiv p_p \sin \vartheta$. The extracted form factor ratio shows no statistically significant $p_T$ dependence, according to the $\chi^2$ of a constant fit, confirming the cancellation of the analyzing power $A_y$ in the ratio $P_t/P_\ell$.
\begin{figure*}
  \begin{center}
    \includegraphics[width=0.7\textwidth]{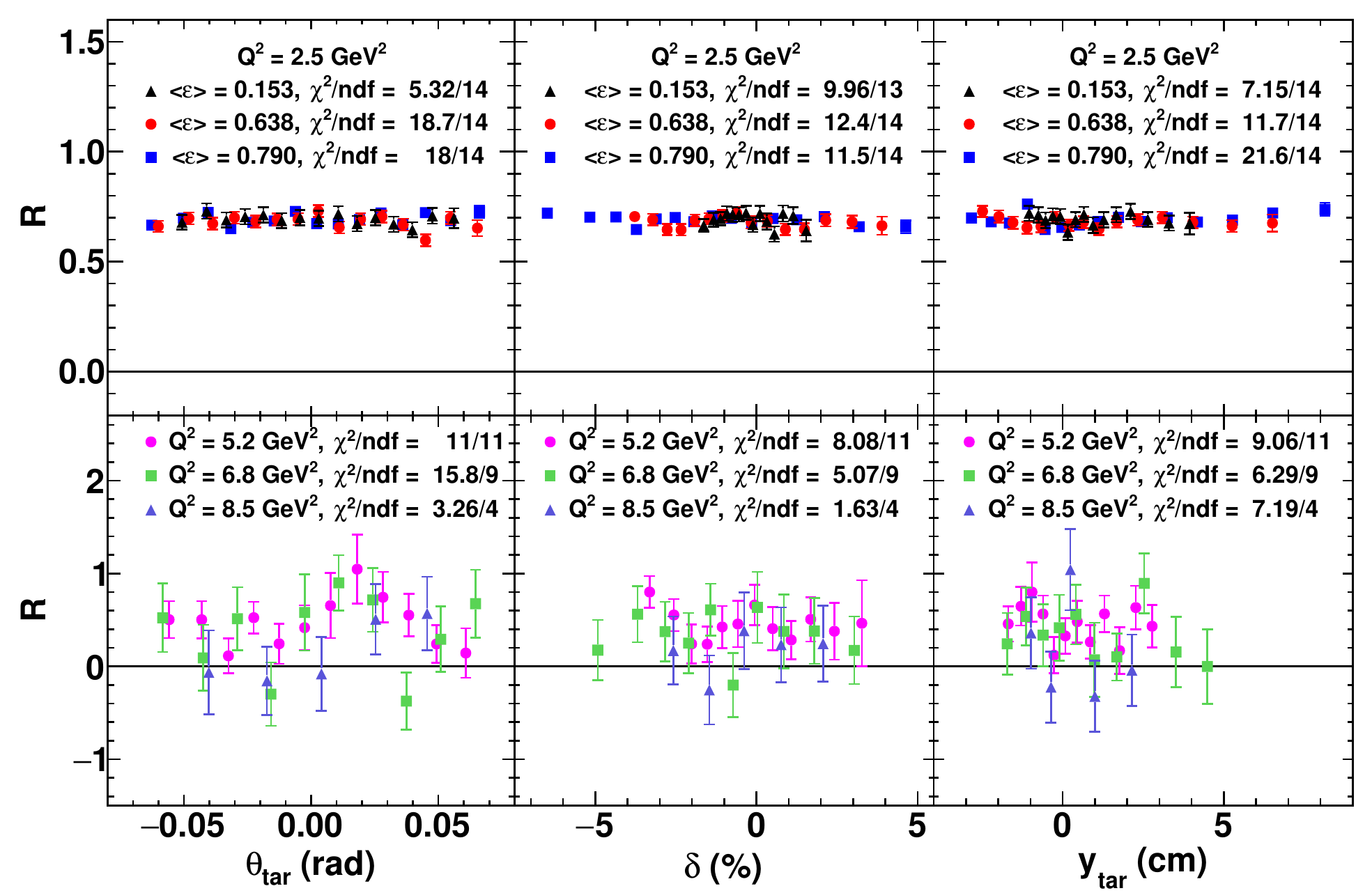}
  \end{center}
  \caption{\label{fig:R_theta_delta_y}  Dependence of $R$ on reconstructed proton trajectory parameters $\theta_{tar}$ (left), $\delta \equiv 100 \times \frac{p-p_0}{p_0}$ (middle), and $y_{tar}$ (right) for GEp-2$\gamma$ (top row) and GEp-III (bottom row). The dependence on $\phi_{tar}$ is not shown, as it is redundant with the $\delta$ dependence, given the kinematic correlation between $\phi_{tar}$ and $\delta$ for elastic $ep$ scattering. The $\chi^2$ values shown here are computed using Eq.~\eqref{R_chi2_definition}; i.e., the $\chi^2$ is computed with respect to the ratio of $R$ to its expected value. No statistically significant deviations of $R$ from a constant or from its expected behavior within the acceptance are observed for any of the six kinematic settings as a function of any of the proton trajectory parameters, confirming the validity of the spin transport calculation and the maximum-likelihood extraction. See text for details.}
\end{figure*}
An important test of the validity of the spin transport calculation using the COSY model of the HMS is that it should not introduce spurious dependence of the extracted values of $P_t$, $P_\ell$ and $R$ on the reconstructed parameters of the proton trajectory at the target, which are the inputs to the calculation. Figure~\ref{fig:R_theta_delta_y} shows the dependence of the ratio $R$ on $\theta_{tar}$, the dispersive plane trajectory angle, $\delta$, the percentage deviation of the reconstructed proton momentum from the HMS central momentum, and $y_{tar}$, the position of the interaction vertex in the TRANSPORT coordinate system, for all six kinematic settings\footnote{The dependence on the non-dispersive plane trajectory angle $\phi_{tar}$ is not shown, since it is redundant with the $\delta$ dependence owing to the kinematic correlation between the proton's scattering angle and its momentum in elastic $ep$ scattering.}. The consistency of $R$ with its ``expected'' behavior was tested by forming a weighted average of the \emph{ratio} of $R$ to its expected value $R_0(Q^2)$, computed from the results of the global proton form factor fit described in the main publication~\cite{Puckett:2017flj}, evaluated at the average $Q^2$ of each kinematic bin, and computing the $\chi^2$ defined as:
\begin{eqnarray}
  \chi^2 &\equiv& \sum_i \frac{\left(\frac{R_i}{R_0(Q_i^2)} - \bar{R} \right)^2}{\sigma_i^2}, \label{R_chi2_definition}\\
  \bar{R} &\equiv& \frac{\sum_i \frac{R_i}{\sigma_i^2 R_0(Q_i^2)} }{\sum_i \frac{1}{\sigma_i^2}}, \nonumber \\
  \sigma_i^2 &\equiv & \left(\frac{\Delta R_i}{R_0(Q_i^2)}\right)^2, \nonumber
\end{eqnarray}
in which $\bar{R}$ is the weighted average ratio of $R$ to its ``expected'' value, and $\sigma_i^2$ is the statistical variance of $R_i/R_0(Q_i^2)$, which acts as a weight in the average $\bar{R}$. As measured by the $\chi^2$ values shown in Fig.~\ref{fig:R_theta_delta_y}, no statistically significant deviations of $R$ from its expected behavior are observed for any of the kinematics as a function of any of the reconstructed proton trajectory parameters. Since $\theta_{tar}$ is mainly sensitive to the azimuthal angle of the reaction plane, it is uncorrelated with $Q^2$ to a good approximation. $R$ is therefore expected to be constant as a function of $\theta_{tar}$, as observed. Since $\delta$ (and $\phi_{tar}$) are both one-to-one correlated with $Q^2$, a weak linear dependence of $R$ on $\delta$ is expected. To within uncertainties, the observed $\delta$ dependence of $R$ within the acceptance is compatible with both the expected $R(\delta)$ and with a constant for all six kinematics. Although no \emph{direct} dependence of $R$ on $y_{tar}$ is expected, the average $Q^2$ ($R(Q^2)$) exhibits a slight negative (positive) correlation with $y_{tar}$ due to acceptance effects and the proton angle-momentum correlation, with the most pronounced $y_{tar}$ dependence of the expected $R$ occurring for $\left<\epsilon\right> = 0.79$ at $Q^2 = 2.5$ GeV$^2$.

\section{Systematic Uncertainties}
\label{sec:systematics}
\subsection{Systematics for $R = \mu_p G_E^p/G_M^p$}
\begin{figure*}
  \begin{center}
    \includegraphics[width=0.85\textwidth]{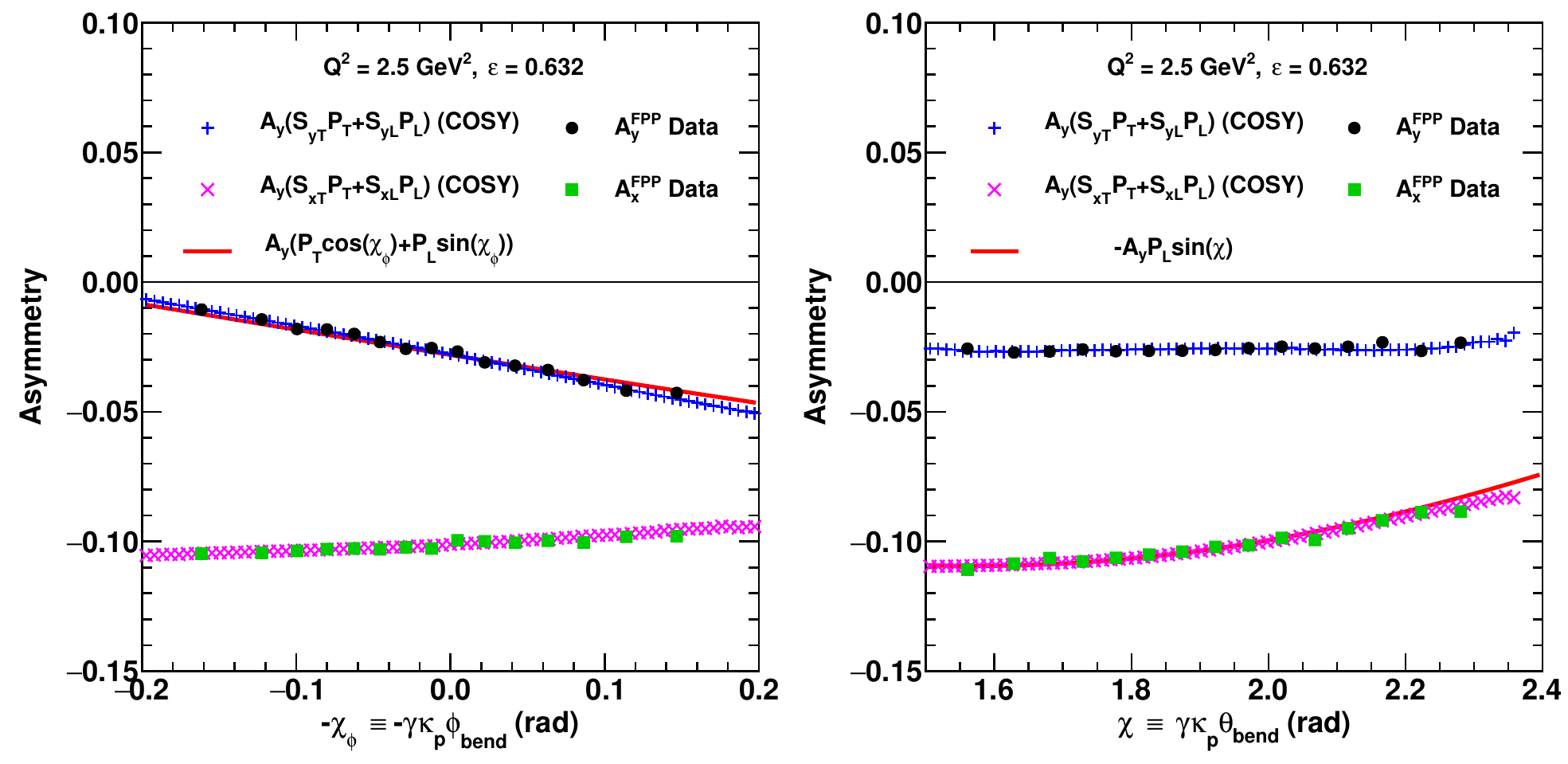}
  \end{center}
  \caption{\label{fig:aychiphiaxchi} Dependence of measured focal-plane asymmetries $A_y^{FPP} \equiv A_y P_y^{FPP}$ (black filled circles) and $A_x^{FPP} \equiv A_y P_x^{FPP}$ (green filled squares) on the non-dispersive-plane precession angle $\chi_\phi$ (left plot) and the dispersive-plane precession angle $\chi$ (right plot), for $Q^2 = 2.5$ GeV$^2$, $\epsilon = 0.632$. The beam-helicity-dependent asymmetry in the azimuthal angle distribution of protons scattered in the FPP is $A(\varphi) \equiv \left[f_+(\varphi)-f_-(\varphi)\right]/\left[f_+(\varphi)+f_-(\varphi)\right] = A_y^{FPP} \cos \varphi - A_x^{FPP} \sin \varphi$ (see Eq.~(20) of Ref.~\cite{Puckett:2017flj}). Measured asymmetries are compared to the approximate expressions $A_y^{FPP} \approx A_y(P_t \cos \chi_\phi + P_\ell \sin \chi_\phi)$ and $A_x^{FPP} \approx A_y(P_t \sin \chi_\phi \sin \chi - P_\ell \cos \chi_\phi \sin \chi) \approx -A_y P_\ell \sin \chi$, as well as the asymmetries predicted by the full COSY calculation. }
\end{figure*}

The polarization transfer method is highly robust against systematic uncertainty, particulary where the determination of the ratio $R$ is concerned. This is a consequence of several exact cancellations, including the cancellation of the polarimeter instrumental asymmetries by the beam helicity reversal, and the cancellation of both the beam polarization and the analyzing power in the ratio $P_t/P_\ell$. An important source of systematic uncertainty for the ratio $R$ is the calculation of the proton spin precession through the HMS magnets. The dominant source of systematic uncertainty in the spin transport calculation is the accuracy of the inputs to the calculation; i.e., the reconstructed proton kinematics at the target. 

The simplicity of the QQQD layout of the HMS magnets (in contrast to the somewhat more complicated QQDQ layout of the HRSs in Hall A~\cite{Punjabi:2005wq,Puckett:2011xg}), leads to a simple and intuitive behavior of the spin transport. To a good approximation, the total rotation of the proton spin through the HMS can be decomposed into two rotations relative to the proton trajectory; a rotation by an angle $\chi_\phi \equiv \gamma \kappa_p (\phi_{fp} - \phi_{tgt}) \equiv \gamma \kappa_p \phi_{bend}$ in the non-dispersive plane, followed by a rotation through an angle $\chi \equiv \gamma \kappa_p (\Theta_0 + \theta_{tgt} - \theta_{fp}) \equiv \gamma \kappa_p \theta_{bend}$ in the dispersive plane, with $\Theta_0 = 25^\circ$ denoting the central vertical bend angle of the HMS. In this approximation, $R$ has the following simple expression:
\begin{eqnarray}
  R &=& -K\frac{P_t}{P_\ell} = K\frac{\tan(\chi_\phi) + \sin(\chi)\frac{P_y^{FPP}}{P_x^{FPP}}}{1 - \tan(\chi_\phi)\sin(\chi) \frac{P_y^{FPP}}{P_x^{FPP}} } \label{eq:Rgeom}             
\end{eqnarray}
Figure~\ref{fig:aychiphiaxchi} shows an illustrative example of the $\chi_\phi$ and $\chi$ dependences of the focal-plane asymmetries $A_y^{FPP}$ and $A_x^{FPP}$, for $(Q^2 = 2.5\text{ GeV}^2,E_e = 2.847\text{ GeV})$. The asymmetries measured at the HMS focal plane behave as expected, and the differences between the full COSY calculation and the simple ``geometric'' approximation described above are small compared to the statistical uncertainties of the asymmetries. It must be noted that this logic is partially circular, as the behavior of the focal plane asymmetries is predicted using the values of $P_t$ and $P_\ell$ extracted from the measured asymmetries, \emph{assuming} validity of the COSY calculation. However, as shown in Fig.~\ref{fig:R_theta_delta_y} above and in Figure 13 of Ref.~\cite{Puckett:2017flj}, the extracted values of $P_t$, $P_\ell$, and $R = \mu_p G_E^p/G_M^p$ based on the COSY spin transport model all closely follow the predictions of the one-photon-exchange or Born approximation within the HMS acceptance, providing strong evidence for the accuracy of the COSY model and the self-consistency of the extraction method for $P_t$ and $P_\ell$. 

When both $\chi_\phi$ and the ratio $P_y^{FPP}/P_x^{FPP}$ are ``small'', as is typically the case in this experiment, the ratio $R$ can be approximated by
\begin{eqnarray}
   R &=& -K \frac{P_t}{P_\ell} \approx K\left[\chi_\phi + \sin(\chi) \frac{P_y^{FPP}}{P_x^{FPP}}\right],                         
\end{eqnarray}
showing that the ratio is highly sensitive to the precession in the
non-dispersive plane, which mixes $P_t$ and $P_\ell$, and is far less
sensitive to $\chi$. To first order, a systematic error $\Delta
\phi_{bend}$ in the non-dispersive-plane trajectory bend angle leads
to a systematic error
\begin{eqnarray}
  \Delta R \approx \gamma \kappa_p K \Delta \phi_{bend}.
\end{eqnarray}
On the other hand, an error $\Delta \theta_{bend}$ in
the dispersive plane trajectory bend angle leads to an error
\begin{eqnarray}
  \Delta R \approx \gamma \kappa_p K \cos(\chi) \frac{P_y^{FPP}}{P_x^{FPP}} \Delta \theta_{bend},
\end{eqnarray}
which is generally much smaller. When the precession angle is \emph{favorable} for the determination of $P_\ell$; i.e., when $\left|\sin(\chi)\right| \rightarrow 1$, $\Delta R / \Delta \theta_{bend}$ vanishes like $\cos \chi$. When the precession angle is \emph{unfavorable} for the determination of $P_\ell$ ($\left|\sin(\chi)\right| \rightarrow 0$), as is the case at $Q^2 = 5.2$ GeV$^2$, the sensitivity of $R$ to $\theta_{bend}$ also tends to vanish. Recalling that $P_x^{FPP} \approx -\sin (\chi) P_\ell$ and $P_y^{FPP} \approx P_t$, the limiting value of the full expression for the geometric approximation~\eqref{eq:Rgeom} is 
\begin{eqnarray}
   \lim_{\chi \rightarrow \pi} R &=& \lim_{\chi \rightarrow \pi} K \frac{P_\ell \tan(\chi_\phi) - P_t}{P_\ell + \tan(\chi_\phi) P_t} \approx R + K\chi_\phi, 
\end{eqnarray}
which lacks any sensitivity to $\chi$, even as the \emph{statistical} uncertainty in the determination of $P_\ell$ diverges in this limit\footnote{The wide $\chi$ acceptance of the HMS allows for an adequate statistical precision on $P_\ell$, and the weighting of events by the spin transport matrix elements in the calculation of the maximum-likelihood estimators for $P_t$ and $P_\ell$ automatically optimizes the statistical precision of the extraction and suppresses the contribution of events with $\chi$ very close to $\pi$, which have vanishing sensitivity to $P_\ell$.}. This somewhat counterintuitive result is borne out by the detailed systematic uncertainy evaluation for the full COSY calculation, in that the $Q^2 = 5.2$ GeV$^2$ setting, for which the central $\chi$ value is close to 180 degrees, is the \emph{least} sensitive to $\Delta \theta_{bend}$ of the six kinematics, and in all cases the contribution of $\Delta \theta_{bend}$ to the total systematic uncertainty $\Delta R$ is small compared to the total $\Delta R$. $\Delta \phi_{bend}$ generally gives the most important precession-related contribution to $\Delta R$ at large $Q^2$.

There are several additional sources of systematic uncertainty beyond those directly related to the spin precession. The uncertainties in the FPP scattering angles $\vartheta$ and $\varphi$ are minimized by the software alignment procedure described above. By analyzing FPP straight-through data obtained in different configurations using a single set of alignment parameters, it was estimated that the systematic uncertainty in the difference between the FPP and HMS track slopes is $\Delta x' = \Delta y' = 0.1$ mrad, which translates to a $\vartheta$-dependent uncertainty $\Delta \varphi \approx 0.14\text{ mrad}/\sin(\vartheta)$ in the azimuthal angle $\varphi$. The inelastic background subtraction also introduces systematic uncertainty. While the correction itself is rather small, the uncertainty associated with the correction ranges from 10-50\% of the size of the correction, and is usually dominated by the \emph{statistical} uncertainty in the background polarization in the region of overlap with the elastic peak in $\delta p_p$. The uncertainty in the beam energy does not directly affect the spin transport or the polarimetry, but does affect the calculation of $\epsilon$ and the kinematic factor multiplying $P_t/P_\ell$ entering the expression for $R$. Uncertainties in $A_y$ and $P_e$ affect $P_\ell$ but do not affect $R$. 

\begin{figure*}
  \begin{center}
    \includegraphics[width=.7\textwidth]{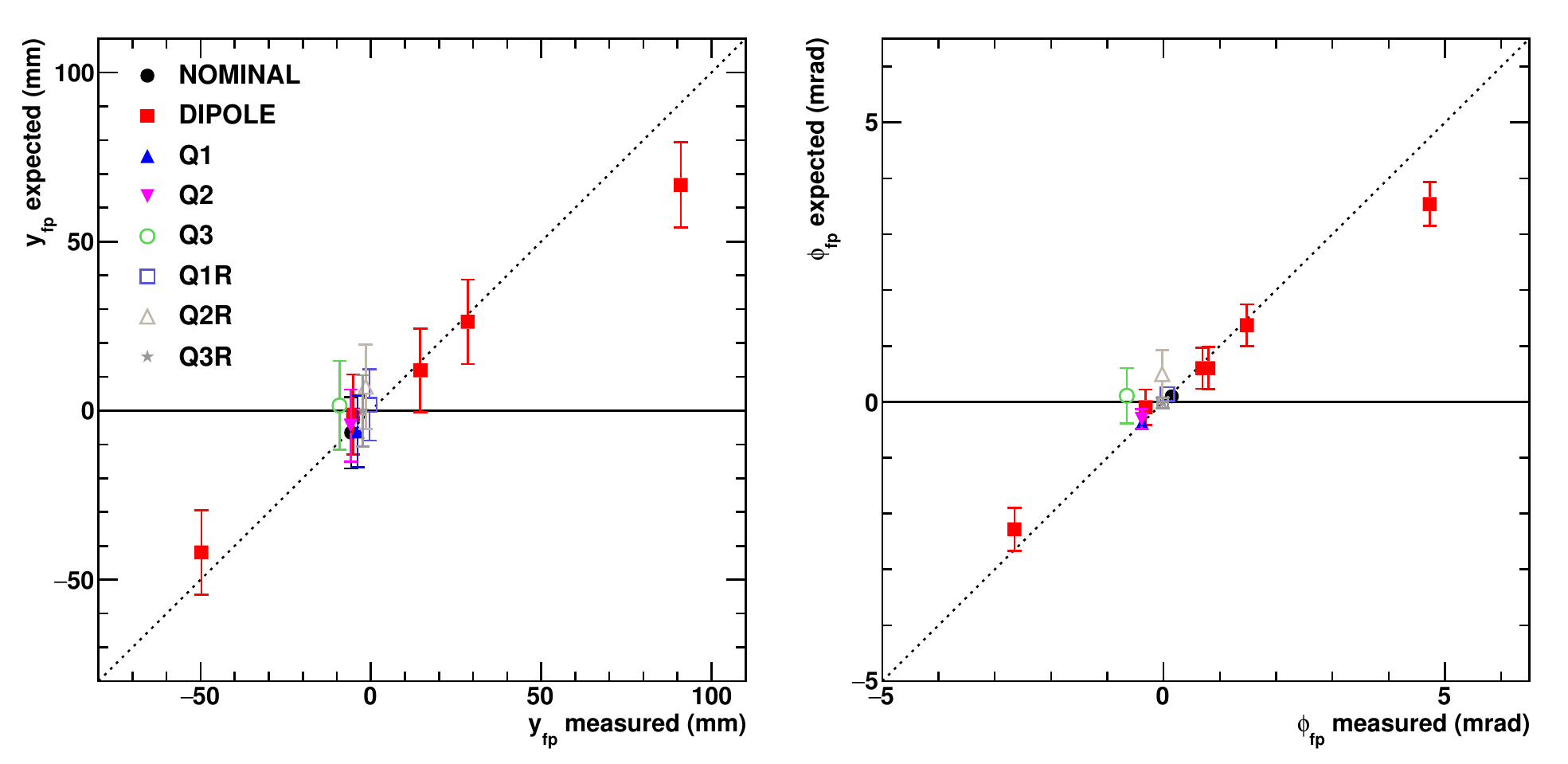}
  \end{center}
  \caption{\label{fig:optical_studies} Results of the HMS optics study in the non-dispersive plane. Correlations between measured and predicted offsets in $y_{fp}$ (left) and $\phi_{fp}$ (right) for rays passing through the central sieve hole for the seven different mistunings of the HMS magnets. The predicted offsets are computed from the best-fit quadrupole offsets and the first-order matrix elements computed for each tune from the HMS COSY model. Since the best-fit result for $\phi_{bend}^{(total)}$ is consistent with zero, the COSY model was not updated and no changes to the spin transport coefficients were made.}
\end{figure*}
Because the ratio $R$ is highly sensitive to the total non-dispersive plane trajectory bend angle $\phi_{bend}$, a dedicated study of the HMS optics in the non-dispersive plane was carried out to reduce the systematic error $\Delta \phi_{bend}$. With the sieve slit collimator in place, scattering of an unrastered electron beam from a thin carbon target foil located at the origin of Hall C was measured for seven deliberate mistunings of the HMS magnets. The resulting displacements at the HMS focal plane of the non-dispersive-plane coordinate $y_{fp}$ and trajectory angle $\phi_{fp}$ of rays passing through the central sieve hole were used to constrain the unknown offsets in the setup that affect $\phi_{bend}$. A systematic error in $\phi_{bend}$ can arise from horizontal misalignments of the HMS quadrupoles relative to the HMS optical axis, or from unknown offsets in $y_{tar}$, $y_{fp}$ and $\phi_{fp}$. The mistunings were chosen for their sensitivities to the various offsets. The first setting, denoted ``DIPOLE'' (for ``dipole-only'') involved turning off and ``degaussing'' all three of the quadrupoles and obtaining data with only the HMS dipole field. At this setting, the horizontal beam position on target was varied in order to vary the $\phi$ angle of scattered electrons passing through the central sieve hole and to center the beam-target intersection point with respect to the HMS optical axis. With no quadrupoles to focus particles in the non-dispersive direction, small displacements in $y_{tar}$ and/or $\phi_{tar}$ lead to large displacements in $y_{fp}$ and $\phi_{fp}$. The results of the horizontal beam position scan for the DIPOLE setting were used to set the final, fixed beam position used for the other six settings, which are as follows:
\begin{description}
\item[``Q1''] Dipole and Q1 at their nominal settings, Q2 and Q3 off. 
\item[``Q2''] Dipole and Q2 at their nominal settings, Q1 and Q3 off. 
\item[``Q3''] Dipole and Q3 at their nominal settings, Q1 and Q2 off.
\item[``Q1R''] Q1 set at 70\% of its nominal current, all other magnets at their nominal setpoints.
\item[``Q2R''] Q2 set at 70\% of its nominal current, all other magnets at their nominal setpoints.
\item[``Q3R''] Q3 set at 70\% of its nominal current, all other magnets at their nominal setpoints.
\end{description}
For each setting, the first-order forward transport coefficients $(y_{fp}|y_{tar})$, $(y_{fp}|\phi_{tar})$, $(\phi_{fp}|y_{tar})$, and $(\phi_{fp}|\phi_{tar})$, as well as the coefficients $(y_{fp}|s_i)$ and $(\phi_{fp}|s_i)$ describing the first-order deflections in $y$ and $\phi$ due to horizontal displacements $s_i$ in quadrupole $i$, were computed using COSY. The procedure for isolating events passing through the central sieve hole is described at length in Ref.~\cite{Puckett:2015soa}. Restricting the analysis to the central sieve hole minimizes deviations from the central ray and the effects of higher-order coefficients. The coordinate $y_{tar}$ of the interaction vertex was computed from the target foil position and the horizontal beam position on target measured by the BPMs, accounting for the slight mispointing of the HMS optical axis with respect to the ``ideal'' target position. The ray from the vertex to the central sieve hole defines $\phi_{tar}$. The foil position, the HMS pointing angle, and the horizontal spatial mispointing of the HMS were all determined from a survey performed on the HMS at the location used for the study. The known values of $y_{tar}$ and $\phi_{tar}$, the measured displacements $y_{fp}$ and $\phi_{fp}$, and the first order HMS COSY coefficients for each setting were used to determine the quadrupole misalignments $(s_1, s_2, s_3)$ and the zero offsets $y_0^{tar}$ and $\phi_0^{fp}$. $y_0^{tar}$ represents a zero offset in the $y_{tar}$ position of the intersection of the beam with the thin carbon foil, and is treated as a free parameter in the fit due to the uncertainty in the horizontal beam position; the target foil and sieve hole positions are both known quite accurately. $\phi_0^{fp}$ represents a possible angular offset of the HMS track relative to the HMS optical axis. No explicit offsets in $\phi_{tar}$ or $y_{fp}$ were included in the fit. A $\phi_{tar}$ offset would be redundant with $y_0^{tar}$ and $\phi_0^{fp}$. No $y_0^{fp}$ offset was allowed because the study is insufficiently sensitive to a zero offset in $y_{fp}$ to provide a more stringent constraint than even the most conservative estimate of the accuracy with which $y_{fp}$ is already known from surveys and previous optics calibration studies. The lack of sensitivity to $y_0^{fp}$ is due to the large magnification in $y_{fp}$ of small offsets in $y_{tar}$ and (especially) $\phi_{tar}$. For example, in the ``DIPOLE'' setting, the first order coupling $(y_{fp}|\phi_{tar}) = 25.6$ mm/mrad.

Instead, $y_0^{fp}$ was fixed at $y_0^{fp} = 0$ mm in the fit. A 10-mm uncertainty was assigned to $y_0^{fp}$ as a very conservative estimate; the surveyed drift chamber positions in the HMS detector hut have a nominal accuracy of about $\pm$1 mm. The uncertainty assigned to $y_0^{fp}$ only affects the fit result via the relative weighting of the measured $y_{fp}$ and $\phi_{fp}$ displacements in the $\chi^2$ calculation, and $\Delta y_0^{fp} = \pm10$ mm gives a fit result with a $\chi^2/ndf$ close to one. Figure~\ref{fig:optical_studies} summarizes the results of the study. The ``DIPOLE'' setting was studied for five different horizontal beam positions, producing large variations in $y$ and $\phi$ as the beam was scanned across the target foil. The point at $(y_{fp}, \phi_{fp}) \approx (91\text{ mm}, 4.7\text{ mrad})$ corresponds to a fairly extreme ray with $(y_{tar},\phi_{tar}) \approx (-4.6\text{ mm},2.6\text{ mrad})$ passing through the central sieve hole, and is the only measurement which deviates significantly from the prediction of the first-order optics model using the best-fit offsets. The fit results are not particularly sensitive to this point in any case so it is included in the fit nonetheless. 

\begin{table*}
  \caption{\label{tab:quadoffsets} Results of the HMS non-dispersive optics study, for two different uncertainties assigned to $y_0^{fp}$. Fig.~\ref{fig:optical_studies} shows the results for $\Delta y_0^{fp} = \pm 10$ mm. $\phi_{bend}^{(s)}$ is the total offset in the non-dispersive bend angle for the nominal HMS tune due to the best fit quadrupole misalignments $s_{1,2,3}$, while $\phi_{bend}^{total}$ also includes the contributions of $\phi_0^{fp}$ and $y_0^{tar}$. See text for details.}
  \begin{center}
    \begin{tabular}{ccc}
      \hline \hline 
      $y_0^{fp} \pm \Delta y_0^{fp}$ (mm) & $0 \pm 10$  & $0 \pm 2$ \\ \hline 
      $\phi_0^{fp} \pm \Delta \phi_0^{fp}$ (mrad) & $-0.05 \pm 0.18$ & $-0.03 \pm 0.07$ \\
      $y_0^{tar} \pm \Delta y_0^{tar}$ (mm) & $-0.3 \pm 0.2$ & $-0.3 \pm 0.1$ \\
      $s_1 \pm \Delta s_1$ (mm) & $0.8 \pm 0.3$ & $0.7 \pm 0.1$ \\
      $s_2 \pm \Delta s_2$ (mm) & $1.0 \pm 0.7$ & $1.1 \pm 0.2$ \\
      $s_3 \pm \Delta s_3$ (mm) & $2.7 \pm 1.3$ & $3.1 \pm 0.8$ \\ \hline 
      $\phi_{bend}^{(s)} \pm \Delta \phi_{bend}^{(s)}$ (mrad) & $0.16 \pm 0.18$ & $0.13 \pm 0.07$ \\
      $\phi_{bend}^{(total)} \pm \Delta \phi_{bend}^{(total)}$ (mrad) & $0.12 \pm 0.14$ & $0.13 \pm 0.08$ \\
      $\chi^2/ndf$ & $22.2/21$ &  $35.1/21$ \\ \hline \hline 
    \end{tabular}
  \end{center}
\end{table*}
Table~\ref{tab:quadoffsets} shows the fit results for two different choices of the uncertainty $\Delta y_0^{fp}$. In both cases, small, positive offsets are favored for all three quadrupoles, including a noticeable offset of about 3 mm for Q3. As shown in Fig.~\ref{fig:optical_studies}, this Q3 offset is mainly driven by the deviation of the measured $y$ and $\phi$ positions for the Q3 setting from the observed values for the ``DIPOLE'' setting at the same horizontal beam position. Of more relevance than the individual offsets, however, is the implication of the results for $\phi_{bend}$. Table~\ref{tab:quadoffsets} shows the total offset in $\phi_{bend}$ for the nominal HMS tune corresponding to the best-fit values of the quadrupole offsets and the zero offsets $\phi_0^{fp}$ and $y_0^{tar}$: 
\begin{eqnarray}
  \phi_{bend}^{(s)} &=& \sum_i (\phi_{fp}|s_i)s_i \nonumber \\
  \phi_{bend}^{(total)} &=& \phi_{bend}^{(s)} + \phi_0^{fp} + (\phi_{fp}|y_{tar}) y_0^{tar}  \nonumber \\
                    & & + \left[(\phi_{fp}|\phi_{tar})-1\right] \frac{y_{sieve}-y_0^{tar}}{z_{sieve}},
\end{eqnarray}
where $y_{sieve}$ and $z_{sieve}$ are the $y$ and $z$ positions of the central sieve hole in TRANSPORT coordinates, respectively. The quantity $\phi_{bend}^{(total)}$ represents the total trajectory bend angle in the non-dispersive plane for the central ray due to the quadrupole misalignments and the offsets $y_0^{tar}$ and $\phi_0^{fp}$. The significant correlations that exist among the best-fit parameters are accounted for in the calculation of the uncertainties $\Delta \phi_{bend}^{(s)}$ and $\Delta \phi_{bend}^{(total)}$. Because the best-fit quadrupole offsets are all in the same direction, and because the first-order couplings $(\phi_{fp}|s_i)$ are positive for Q1 and Q3 but negative for Q2, the resulting cumulative deflection of the central ray due to these offsets is nonetheless quite small. Because the central value of $\phi_{bend}^{(total)}$ was found to be consistent with zero, the COSY spin transport model was not modified. The effect of the final uncertainty $\Delta \phi_{bend} = 0.14$ mrad on the polarization transfer observables was measured by shifting $\phi_{tar}$ in the analysis by an amount $\Delta \phi_{tar} = \tfrac{\Delta \phi_{bend}}{|\left(\phi_{bend}|\phi_{tar}\right)|} = 0.1$ mrad, where $\left(\phi_{bend}|\phi_{tar}\right) \approx -1.4$ is the first order coupling between $\phi_{bend}$ and $\phi_{tar}$ for the nominal tune.

\begin{figure}
  \begin{center}
    \includegraphics[width=0.75\columnwidth]{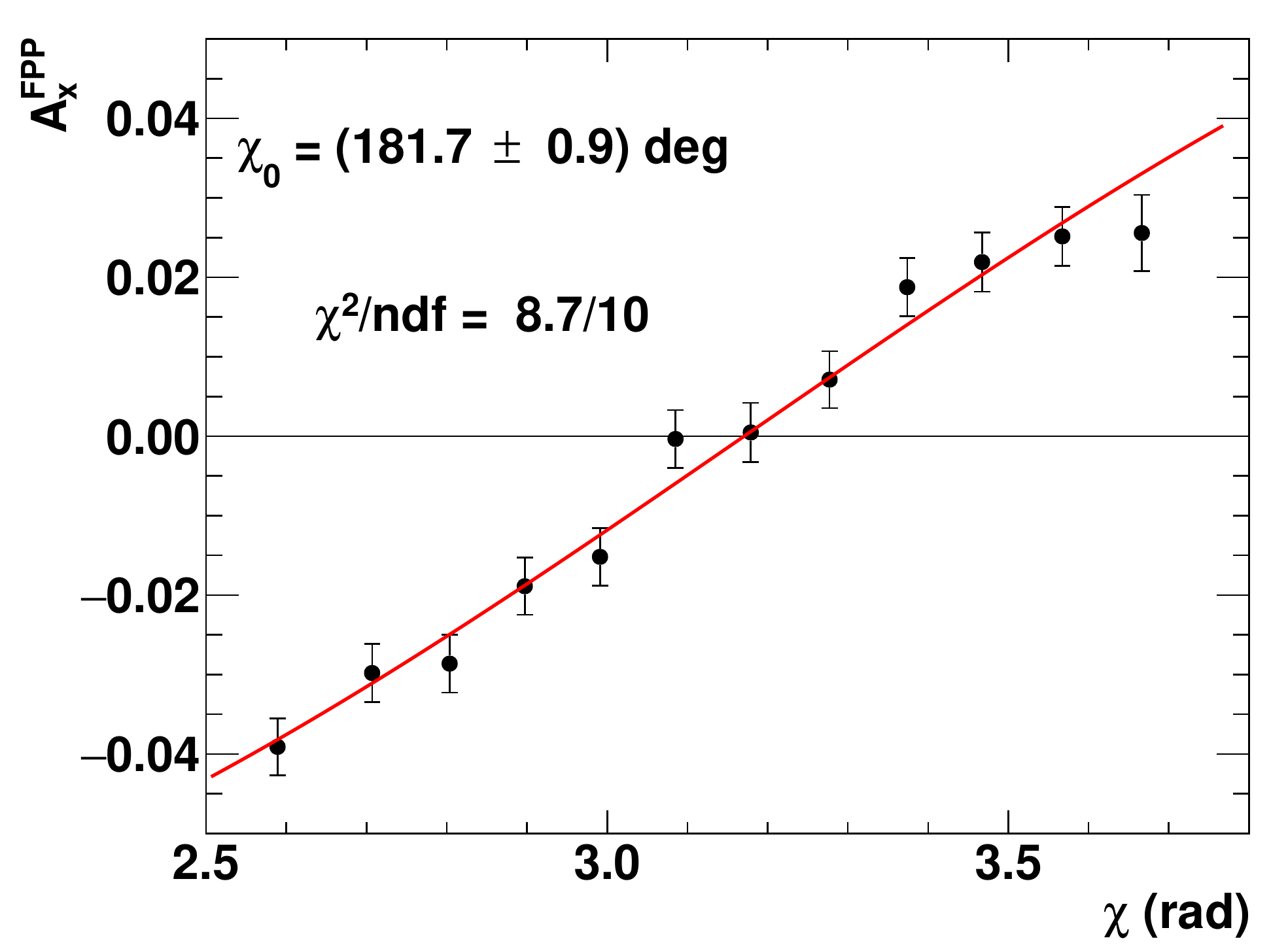}
  \end{center}
  \caption{\label{fig:zerocrossing} Focal-plane normal asymmetry $A_x^{FPP}$ vs. $\chi = \gamma \kappa_p \theta_{bend}$, for $Q^2 = 5.2$ GeV$^2$. The fit function is $A_x^{FPP} = -A_0 \sin(\chi - \delta)$. The zero-crossing angle in radians is $\chi_0 \equiv \pi + \delta$. The expected zero-crossing angle based on the values of $P_t$ and $P_\ell$ and the COSY spin transport is $(180.42 \pm 0.02)^\circ$.}
\end{figure}
The uncertainty in the dispersive bend angle $\theta_{bend}$ was estimated using the $Q^2 = 5.2$ GeV$^2$ data. In the ideal dipole approximation, the asymmetry $A_x^{FPP} \propto -\sin(\chi)$ has a zero crossing at exactly 180 degrees. At $Q^2 = 5.2$ GeV$^2$, the precession angle corresponding to the central momentum setting for the central ray is 177.2$^\circ$, and the asymmetry crosses zero near the center of the acceptance. The actual expected location of the zero crossing is slightly different from 180$^\circ$ because of the slight mixing of $P_t$ and $P_\ell$ in $A_{x}^{FPP} = -P_e A_y\left( S_{x\ell} P_\ell + S_{xt} P_t\right)$. The expected zero crossing angle $\hat{\chi}_0 = (180.42 \pm 0.02)^\circ$ was computed from the COSY spin transport matrix elements and the extracted values of $P_t$ and $P_\ell$. Figure~\ref{fig:zerocrossing} shows the measured zero crossing of $\chi_0 = (181.7 \pm 0.9)^\circ$. The difference between the expected and measured zero-crossing angles, while not statistically signficant, provides for a conservative estimate of the systematic uncertainty $\Delta \theta_{bend} \equiv \frac{\Delta \chi_0}{\gamma\kappa_p} = 3.2$ mrad. The systematic effect of $\Delta \theta_{bend}$ on $P_t$, $P_\ell$ and $R$ was measured by shifting $\theta_{tar}$ in the analysis by $\Delta \theta_{tar} \equiv \frac{\Delta\theta_{bend}}{\left|(\theta_{bend}|\theta_{tar})\right|} = 2.4$ mrad, with $(\theta_{bend}|\theta_{tar}) \approx 1.33$ being the first-order coupling between $\theta_{bend}$ and $\theta_{tar}$ for the nominal tune.  

The systematic uncertainty in the percentage deviation $\delta$ of the reconstructed proton momentum from the HMS central momentum was estimated to be $\Delta \delta = 0.14\%$, based on the observed variations of the offset of the elastic peak position from zero in $\delta p_p$ among the various kinematics, after accounting for the uncertainties in the beam energy, the HMS central angle, the corrections for energy loss and radiative effects, and all other contributions to the observed ``zero offset'' of the elastic peak. The contribution of $\Delta \delta$ to $\Delta R$ is quite small except at $Q^2 = 2.5$ GeV$^2$, $\epsilon = 0.15$, for which it is comparable to the other contributions. The systematic uncertainty in $y_{tar}$ was estimated to be $\Delta y_{tar} = 0.4$ mm based on the results of the non-dispersive optical studies of the HMS described above. The systematic uncertainties $\Delta y_{tar}$ and $\Delta \phi_{tar}$ are partially correlated due to the uncertainty in the horizontal beam position during the optics calibration. The estimated correlation coefficient is $\rho_{\Delta \phi \Delta y} = -0.43$. Because the correlation between $\Delta y$ and $\Delta \phi$ is negative, but the slopes $\frac{dR}{dy}$ and $\frac{dR}{d\phi}$ always have the same sign (see Tab.~\ref{tab:Rsystematics}), the effect of the correlation is to slightly reduce the magnitude of $\Delta R$: $(\Delta R)^2 = (\frac{dR}{d\phi}\Delta \phi)^2 + (\frac{dR}{dy}\Delta y)^2 + 2\rho_{\Delta \phi \Delta y} \frac{dR}{d\phi}\frac{dR}{dy}\Delta \phi \Delta y$. Similarly, the uncertainties $\Delta \theta_{tar}$ and $\Delta \delta$ are positively correlated ($\rho_{\Delta \theta \Delta \delta} = +0.26$). The derivatives $\frac{dR}{d\delta}$ and $\frac{dR}{d\theta}$ are opposite in sign for all of the $Q^2 = 2.5$ GeV$^2$ data, leading to a slight reduction in $\Delta R$, but have the same sign for the GEp-III kinematics\footnote{The change in relative sign of $\frac{dR}{d\delta}$ and $\frac{dR}{d\theta}$ between $Q^2 = 2.5$ GeV$^2$ and the GEp-III data is likely related to the sign change of $\sin(\chi)$.}, leading to a slight increase in $\Delta R$.

\begin{table*}
  \caption{\label{tab:Rsystematics} Systematic uncertainty contributions for $R = -K\frac{P_t}{P_\ell} = \mu_p \frac{G_E^p}{G_M^p}$. The total systematic uncertainty includes the effects of partial correlations among the various systematic contributions, including $\Delta \phi_{tar}$ and $\Delta y_{tar}$ (correlation coefficient $\rho_{\Delta \phi \Delta y} \approx -0.43$), and $\Delta \theta_{tar}$ and $\Delta \delta$ (correlation coefficient $\rho_{\Delta \theta \Delta \delta} \approx +0.26$). $\Delta R_{syst}^{total}$ is the total systematic uncertainty, while $\Delta R_{syst}^{ptp}$ is the ``point-to-point'' systematic uncertainty for $Q^2 = 2.5$ GeV$^2$ \emph{relative} to the $\epsilon = 0.79$ setting.}
  \begin{center}
    \begin{tabular}{ccccccc}
      \hline \hline Nominal $Q^2$ (GeV$^2$) & 2.5 & 2.5 & 2.5 & 5.2 & 6.8 & 8.5 \\
      $\left<\epsilon\right>$ & 0.153 & 0.638 & 0.790 & 0.38 & 0.52 & 0.24 \\ \hline
      $\frac{dR}{d\phi_{tar}} \Delta \phi_{tar}$ & $-3.4 \times 10^{-3}$ & $-2.1\times 10^{-3}$ & $-2.0 \times 10^{-3}$ & $-4.8 \times 10^{-3}$ & $-5.7 \times 10^{-3}$ & -0.010 \\ 
      $\frac{dR}{dy_{tar}} \Delta y_{tar}$ & $-2.0 \times 10^{-3}$ & $-1.2 \times 10^{-3}$ & $-1.2 \times 10^{-3}$ & $-2.9 \times 10^{-3}$ & $-3.9 \times 10^{-3}$ & $-7.7 \times 10^{-3}$ \\
      $\frac{dR}{d\theta_{tar}} \Delta \theta_{tar}$ & $-2.2 \times 10^{-3}$ & $-2.5 \times 10^{-3}$ & $-2.5 \times 10^{-3}$ & $1.4 \times 10^{-3}$ & $-5.0 \times 10^{-3}$ & $3.0 \times 10^{-3}$ \\
      $\frac{dR}{d\delta} \Delta \delta $ & $5.8 \times 10^{-3}$ & $1.2 \times 10^{-3}$ & $9.0 \times 10^{-4}$ & $1.2 \times 10^{-3}$ & $-3.3 \times 10^{-6}$ & $2.5 \times 10^{-4}$ \\
      $\frac{dR}{d\varphi_{FPP}} \Delta \varphi_{FPP}$ & $4.1 \times 10^{-3}$ & $2.5 \times 10^{-3}$ & $2.4 \times 10^{-3}$ & $4.6 \times 10^{-4}$ & $-6.0 \times 10^{-3}$ & $-0.017$ \\
      $\frac{dR}{dE_e} \Delta E_e$ & $-1.8 \times 10^{-3}$ & $-1.1 \times 10^{-4}$ & $-5.6 \times 10^{-5}$ & $-1.9 \times 10^{-4}$ & $-8.3 \times 10^{-5}$ & $-1.4 \times 10^{-4}$ \\
      $\Delta R_{syst}($background$)$ & $3.5 \times 10^{-4}$ & $9.6 \times 10^{-5}$ & $9.9 \times 10^{-5}$ & $2.4 \times 10^{-3}$ & $1.6 \times 10^{-3}$ & $0.012$ \\ \hline 
      $\Delta R_{syst}^{total}$ & $7.9 \times 10^{-3}$ & $4.0 \times 10^{-3}$ & $3.9 \times 10^{-3}$ & $5.5 \times 10^{-3}$ & $9.7 \times 10^{-3}$ & $0.024$ \\
      $\Delta R_{syst}^{ptp}$ & $4.3 \times 10^{-3}$ & $2.3 \times 10^{-4}$ & $1.1 \times 10^{-4}$ & N/A & N/A & N/A \\ \hline \hline 
    \end{tabular}
  \end{center}
\end{table*}
Table~\ref{tab:Rsystematics} shows the important contributions to the systematic uncertainty in the ratio $R$, which include those related to the HMS optics and spin transport, the uncertainty of the FPP scattering angle reconstruction, the beam energy uncertainty, and the inelastic background subtraction. For $Q^2 = 2.5$ GeV$^2$, the ``point-to-point'' systematic uncertainties are also shown. The contributions to $\Delta R_{syst}$ can be classified as either \emph{independent}, meaning the systematic errors in the underlying variables are totally uncorrelated from one measurement to the next, or \emph{correlated}, meaning that the uncertainties in the underlying variables are global and independent of kinematics. The beam energy uncertainty and the background subtraction-related uncertainty are independent  by this definition; neither the errors in the variables themselves nor their effects on $R$ are the same for different kinematics. The uncertainties related to the reconstructed proton kinematics are assumed to be the same for all kinematics, though their effects on $R$ can differ from point to point. This is a good assumption in particular for $Q^2 = 2.5$ GeV$^2$, which used the same HMS central momentum setting, and thus the same magnetic field, for all three $\epsilon$ values. The three measurements at $Q^2 = 2.5$ GeV$^2$ differ only in terms of the HMS central angle, which affects neither the spin transport nor the calculation of the event kinematics (since $Q^2$ is calculated from the proton momentum). Similarly, the FPP alignment uncertainty is assumed to be the same for all kinematics in terms of the plane-angle differences $\Delta \theta_x$ and $\Delta \theta_y$, but its effect on the FPP azimuthal angle reconstruction increases with $Q^2$ as the accepted range of $\vartheta$ shifts toward smaller angles. At $Q^2 = 2.5$ GeV$^2$, the FPP angle reconstruction systematics are the same for all three $\epsilon$ values. 

The shifts $\Delta R$ resulting from the correlated systematic contributions have the same sign for all three $\epsilon$ values at 2.5 GeV$^2$, but somewhat different magnitudes as a result of the different kinematic factors involved at each $\epsilon$ value. The total systematic uncertainties in $R$ are small for all kinematics and comparable to the statistical uncertainties at $Q^2 = 2.5$ GeV$^2$. The ``point to point'' systematic uncertainty shown in Tab.~\ref{tab:Rsystematics} is defined as the quadrature sum of the \emph{independent} contributions and the \emph{differences} in each correlated contribution between the point in question and the chosen ``reference'' point\footnote{The partially correlated $\Delta \phi$/$\Delta y$ and $\Delta \theta$/$\Delta \delta$ contributions are combined internally at each point before taking the differences with the reference point.} ($\left<\epsilon\right> = 0.79$ in Tab.~\ref{tab:Rsystematics}). The point-to-point systematic uncertainty for the relative variation of $R$ with $\epsilon$ is quite small (about half the total systematic uncertainty in the worst case at $\left<\epsilon\right> = 0.15$).     

\subsection{Systematics for $P_\ell/P_\ell^{Born}$}
The spin transport systematics affect the determination of $P_\ell$ quite a bit differently; in this case  $\Delta \theta_{bend}$ makes an appreciable contribution to the total systematic uncertainty $\Delta P_\ell$, while the effect of $\Delta \phi_{bend}$ is negligible. In contrast to $R$, $P_\ell$ has no direct sensitivity to the beam energy, but is directly sensitive to the product $P_e A_y$ of the beam polarization and the analyzing power. For the relative variation of $P_\ell/P_\ell^{Born}$ with $\epsilon$, the beam polarization uncertainty $\Delta P_e = \pm 0.5\%$ (point to point, relative) is the dominant contribution. 
\begin{table*}
  \caption{\label{tab:PLsystematics} Systematic uncertainty contributions for $P_\ell$ and the ratio $P_\ell/P_\ell^{Born}$ at $Q^2 = 2.5$ GeV$^2$. The point-to-point systematic uncertainty is calculated \emph{relative} to the $\left<\epsilon\right> = 0.153$ setting. The total systematic uncertainties in $P_\ell$ do not include the global uncertainty of $\Delta P_e \approx 1\%$ in the beam polarization measurement. This is because any global overestimation (underestimation) of $P_e$ is exactly compensated by an equal and opposite underestimation (overestimation) of the polarimeter analyzing power $A_y$. See text for details.}
  \begin{center}
    \begin{tabular}{cccc}
      \hline \hline     $Q^2$ (GeV$^2$) & 2.5 & 2.5 & 2.5 \\
      $\left<\epsilon\right>$ & 0.153 & 0.638 & 0.790 \\ \hline
      $\frac{dP_\ell}{d\phi_{tar}} \Delta \phi_{tar}$ & $1.3 \times 10^{-4}$ & $1.6 \times 10^{-4}$ & $1.3 \times 10^{-4}$ \\
      $\frac{dP_\ell}{d\theta_{tar}} \Delta \theta_{tar}$ & $4.2 \times 10^{-3}$ & $3.2 \times 10^{-3}$ & $2.5 \times 10^{-3}$ \\
      $\frac{dP_\ell}{dy_{tar}}\Delta y_{tar}$ & $8 \times 10^{-5}$ & $9 \times 10^{-5}$ & $8 \times 10^{-5}$ \\
      $\frac{dP_\ell}{d\delta} \Delta \delta$ & $-2.5 \times 10^{-4}$ & $-1.8 \times 10^{-4}$ & $-1.4 \times 10^{-4}$ \\
      $\frac{dP_\ell}{d\varphi_{FPP}} \Delta \varphi_{FPP}$ & $-1.6 \times 10^{-4}$ & $-2.0 \times 10^{-4}$ & $-1.7 \times 10^{-4}$ \\
      $\Delta P_\ell$ (background) & $8 \times 10^{-5}$ & $3 \times 10^{-5}$ & $2\times 10^{-5}$ \\
      $\frac{dP_\ell}{dA_y} \Delta A_y$ & N/A & $-1.5 \times 10^{-3}$ & $-1.2 \times 10^{-3}$ \\
      $\frac{dP_\ell}{dP_e} \Delta P_e$ & N/A & $-3.7 \times 10^{-3}$ & $-2.9 \times 10^{-3}$  \\ \hline
      Total $\Delta P_\ell^{syst}$ & $4.2 \times 10^{-3}$ & $5.1 \times 10^{-3}$  & $4.0 \times 10^{-3}$ \\
      Total $\Delta_{syst} \left(\frac{P_\ell}{P_\ell^{Born}}\right)$ & N/A & $7.0 \times 10^{-3}$ & $7.1 \times 10^{-3}$ \\
      $\Delta_{syst}^{ptp} \left(\frac{P_\ell}{P_\ell^{Born}}\right)$ & N/A & $5.3 \times 10^{-3}$ & $6.1 \times 10^{-3}$  \\ \hline \hline 
    \end{tabular}
  \end{center}
\end{table*}  
Table~\ref{tab:PLsystematics} shows the important contributions to the systematic uncertainties in $P_\ell$ and $P_\ell/P_\ell^{Born}$. The ratio $P_\ell/P_\ell^{Born}$ is not a meaningful quantity for the measurement at $\left<\epsilon\right> = 0.153$, since it is used to extract $A_y$ under the assumption that $P_\ell = P_\ell^{Born}$. The quoted systematic uncertainty $\Delta P_\ell^{syst}$ at the lowest $\epsilon$ therefore includes only the contributions from the HMS optics/spin transport, the FPP azimuthal angle reconstruction, and the inelastic background. The analyzing power is subject to a global normalization uncertainty  $\Delta A_y/A_y = 0.2\%$ equal to the relative \emph{statistical} uncertainty in $P_\ell$ at $\left<\epsilon\right> = 0.153$. The quoted systematic uncertainties in $P_\ell$ do not include the global uncertainty $\Delta P_e \approx \pm1\%$ in the beam polarization measurement. This is because a global uncertainty in $P_e$ is exactly compensated by the analyzing power calibration; to the extent that the beam polarization is globally overestimated (underestimated), the analyzing power is underestimated (overestimated) by the same amount. All systematic uncertainty contributions other than the beam polarization measurement and the inelastic background subtraction are strongly correlated among the three $\epsilon$ points, such that their contribution to the relative $\epsilon$ dependence of $P_\ell/P_\ell^{Born}$ is very small. Since the systematic uncertainty associated with the inelastic background is essentially negligible, the beam polarization measurement dominates the point-to-point systematic uncertainty of $P_\ell/P_\ell^{Born}$, as mentioned above.  

\begin{figure}
  \begin{center}
    \includegraphics[width=0.9\columnwidth]{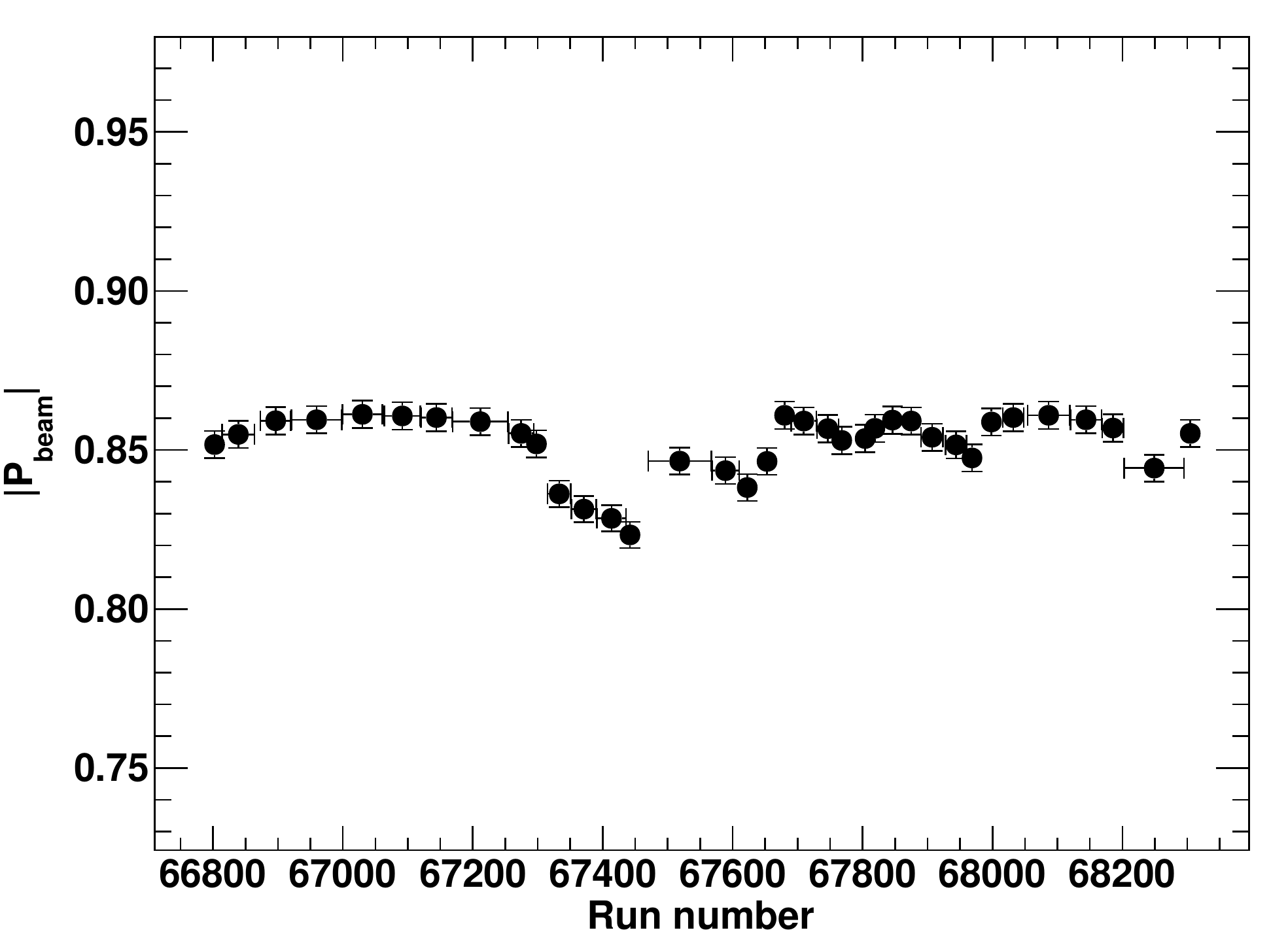}
  \end{center}
  \caption{\label{fig:BeamPol_runnumber} Beam polarization database used for the final GEp-2$\gamma$ analysis, including the corrections described in Ref.~\cite{GEP3_PRC_Erratum}. Vertical error bars indicate the point-to-point systematic uncertainty $\Delta P_e = \pm 0.5\%$. Horizontal ``error bars'' indicate the run ranges for which the indicated value of the beam polarization is used. The horizontal axis coordinate of each point represents the midpoint of the associated run range. See text for details.}
\end{figure}

\begin{figure}
  \begin{center}
    \includegraphics[width=0.9\columnwidth]{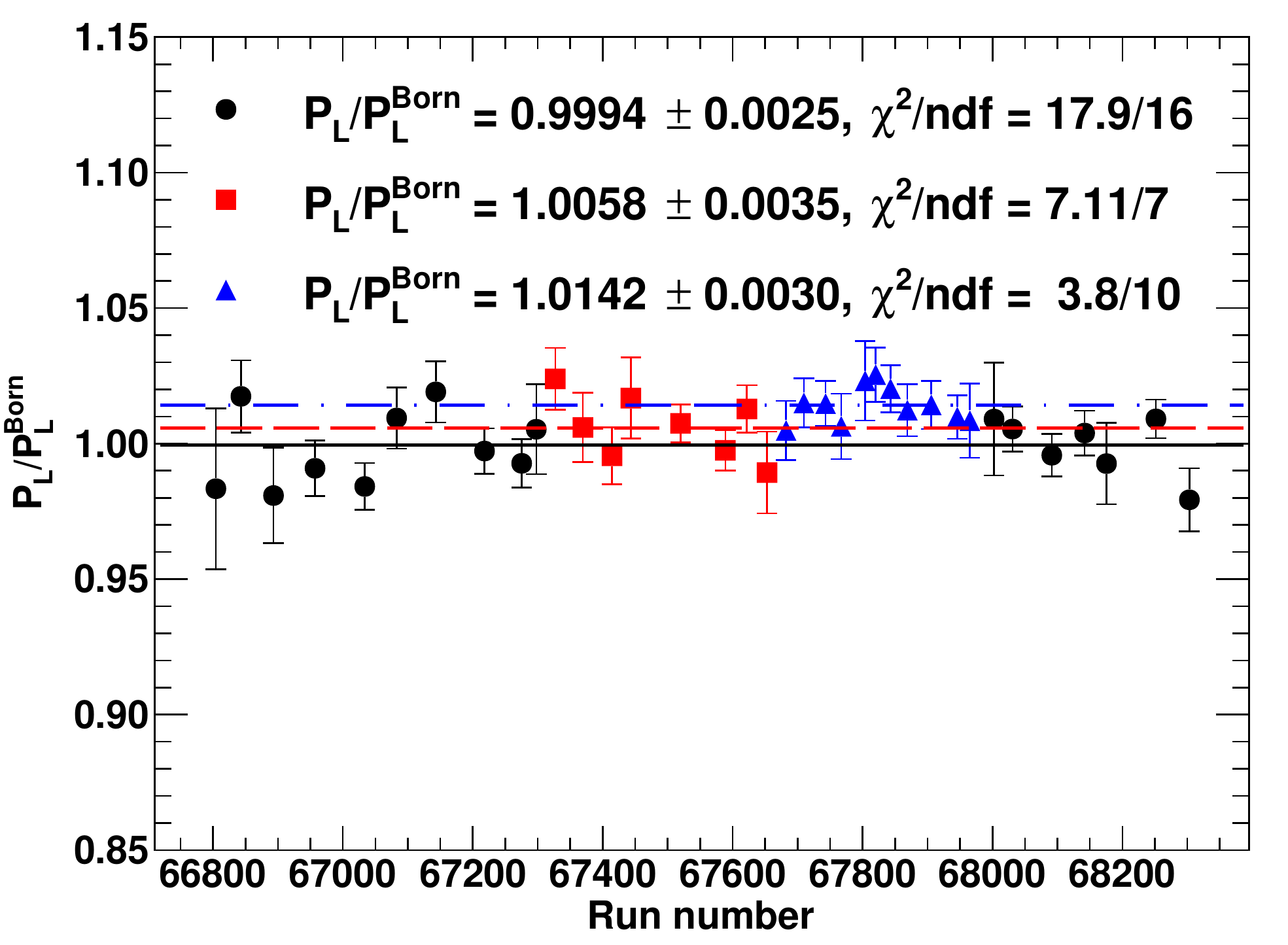}
  \end{center}
  \caption{\label{fig:PLBornRatio_stability} Time dependence of $P_\ell/P_\ell^{Born}$ during the GEp-2$\gamma$ experiment. Data are $\left<\epsilon\right> = 0.153$ (black circles), $\left<\epsilon\right> = 0.638$ (red squares), and $\left<\epsilon\right> = 0.790$ (blue triangles). Data at each kinematic setting are divided into bins based on the run ranges corresponding to the unique beam polarization assignments shown in Fig.~\ref{fig:BeamPol_runnumber}. In contrast to Fig.~\ref{fig:BeamPol_runnumber}, each point is plotted at the statistics-weighted average run number of all events in the corresponding run range.}
\end{figure}
On average, the beam polarization was measured using the Hall C M{\o}ller polarimeter~\cite{Hauger:1999iv} roughly once every two days during the GEp-2$\gamma$ experiment, and always after any change in accelerator operating conditions affecting the polarized beam delivery to Hall C. Because the beam polarization measurement is invasive, it was not possible to continuously monitor the beam polarization directly. However, the stability of the beam polarization could be monitored indirectly via the FPP asymmetry magnitude. Figure~\ref{fig:BeamPol_runnumber} shows the beam polarization database used for the final analysis of the GEp-2$\gamma$ data as a function of time, including the correction of the typographical error discovered subsequent to the original publication, detailed in Ref.~\cite{GEP3_PRC_Erratum}. Each M{\o}ller measurement performed during the GEp-2$\gamma$ experiment was assigned to an appropriate range of data acquisition runs, after correcting for any interceding changes to accelerator operating conditions that affected the beam polarization, including, for example, changes in the configuration of the Wien filter at the injector that determines the initial and final orientation of the electron spin, and changes in the quantum efficiency of the accelerator photocathode resulting from changes in the position of the laser spot on the photocathode. Typically, the beam polarization during GEp-2$\gamma$ was 85-86\%. Figure~\ref{fig:PLBornRatio_stability} shows the time dependence of the ratio $P_\ell/P_\ell^{Born}$ during the $Q^2 = 2.5$ GeV$^2$ running at all three $\epsilon$ values, extracted using the final beam polarization database corrected as discussed in Ref.~\cite{GEP3_PRC_Erratum}. The data from each kinematic setting are divided into run ranges corresponding to the unique beam polarization assignments shown in Fig.~\ref{fig:BeamPol_runnumber}. The $\chi^2/ndf$ values shown in Fig.~\ref{fig:PLBornRatio_stability} are based on the quadrature sum of the statistical uncertainties of the data and the point-to-point systematic uncertainty of the M{\o}ller measurement, $\Delta P_e/P_e = \pm 0.5\%$. The extracted ratio $P_\ell/P_\ell^{Born}$ is compatible with a constant at each $\epsilon$ value. The stability of the extracted $P_\ell/P_\ell^{Born}$ as a function of time confirms the stability of the beam polarization between M{\o}ller measurements and the overall accuracy of the database. 

\section{Summary and Conclusions}
\label{sec:conclusion}
This technical note has presented details of the detector
performance and the data analysis of the GEp-III and GEp-2$\gamma$ experiments that go beyond the scope of the main body of the recent archival publication of both experiments~\cite{Puckett:2017flj}. This detailed documentation, including the performance of the detectors that were newly constructed for these measurements, the lessons learned during the experiment, and the details of the final systematic uncertainty evaluation, will serve as a useful reference for future Hall C data analyses in general, and for the planning and optimization of future experiments using the polarization transfer method and/or the High Momentum Spectrometer in Hall C. 

\section{Acknowledgments}

The collaboration thanks the Hall C technical staff and the Jefferson
Lab Accelerator Division for their outstanding support during the
experiment. This material is based upon work supported by the U.S. Department of Energy, Office of Science, Office of Nuclear Physics, under Award Number DE-SC-0014230 and contract Number(s) DE-AC02-06CH11357 and DE-AC05-06OR23177, the U.S.
National Science Foundation, the
Italian Institute for Nuclear Research, the French Commissariat
\`a l'Energie Atomique and Centre National de la Recherche Scientifique
(CNRS), and the Natural Sciences and Engineering
Research Council of Canada.

\section*{References}
\bibliographystyle{elsarticle-num}
\bibliography{gep3_prc_master_references}

\end{document}